\documentclass[11pt,twoside,a4paper,openright]{report}

\usepackage{amsmath}
\usepackage{graphicx,lineno}
\usepackage{graphics}   
\usepackage{dcolumn,enumerate}    
\usepackage{bm,bbm}         
\usepackage{mathrsfs}
\usepackage{mathtools}
\usepackage{color}

\usepackage{rotating}
\usepackage{times}
\usepackage{charter}
\numberwithin{equation}{chapter}

\usepackage{geometry}
\usepackage{fancyhdr}
\usepackage[small,bf]{caption}
\usepackage{setspace}
\usepackage{hyperref}
\usepackage{fancyhdr}
\usepackage{tikz}
\newcommand*\circled[1]{\tikz[baseline=(char.base)]{
            \node[shape=circle,draw,inner sep=2pt] (char) {#1};}}
\makeatletter
\renewcommand{\paragraph}{\@startsection{paragraph}{4}{0ex}%
   {-3.25ex plus -1ex minus -0.2ex}%
   {1.5ex plus 0.2ex}%
   {\normalfont\normalsize\bfseries}}
\makeatother

\stepcounter{secnumdepth}
\stepcounter{tocdepth}
\begin{document}

\begin{titlepage}
\begin{center}



{\LARGE \bfseries Quantum Simulations with}\\
{\LARGE \bfseries Unitary \& Nonunitary Controls: }\\
{\LARGE \bfseries    
NMR implementations}\\[1.5cm]

A thesis\\
Submitted in partial fulfillment of the requirements\\
Of the degree of\\
Doctor of Philosophy\\[0.8cm]
By\\[1cm]
{\large Swathi S Hegde}\\
20103089 \\[2.5cm]

\includegraphics[width=3cm]{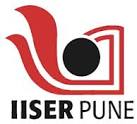}\\    
[0.5cm]
{INDIAN INSTITUTE OF SCIENCE EDUCATION AND RESEARCH PUNE}\\[1cm]
{July, 2016}

\end{center}
\end{titlepage}
\thispagestyle{empty}
\cleardoublepage

\pagenumbering{roman}
\chapter*{Certificate}\addcontentsline{toc}{section}{\textbf{Certificate}}
Certified that the work incorporated in the thesis entitled 
``\textit{Quantum Simulations with Unitary and Nonunitary Controls: NMR implementations}'',
submitted by \textit{Swathi S Hegde} was carried out by the candidate, under my supervision. The work presented here or any part of it has not been included in any other thesis submitted previously for the award of any degree or diploma from any other University or institution.\\[3cm]
\textit{Date} \hspace{10cm} \textbf{Dr. T. S. Mahesh}
\cleardoublepage

\chapter*{Declaration}\addcontentsline{toc}{section}{\textbf{Declaration}}
I declare that this written submission represents my ideas in my own words and where others' ideas have been included, I have adequately cited and referenced the original sources. I also declare that I have adhered to all principles of academic honesty and integrity and have not misrepresented or fabricated or falsified any idea/data/fact/source in my submission. I understand that violation of the above will be cause for disciplinary action by the Institute and can also evoke penal action from the sources which have thus not been properly cited or from whom proper permission has not been taken when needed.\\[3cm]
\textit{Date} \hspace{10.4cm} \textbf{Swathi S Hegde}
\begin{flushright}
Roll No.- 20103089
\end{flushright}
\cleardoublepage

\chapter*{Acknowledgements}\addcontentsline{toc}{section}{\textbf{Acknowledgements}}
All through the years of my PhD life, I have been very fortunate for having met and worked with a lot of wonderful people. 

I am immensely grateful to my supervisor Dr. T S Mahesh for guiding and teaching
me so patiently and consistently ever since I joined his lab. His spontaneous creative
ideas during almost every discussion have always motivated me to learn to think outside
the box. I am lucky to have worked with him as his student. His expertise in the field
of NMR QIP has been a great source to students like me, and this thesis would not have
been possible without his support. 
 
Collaboration with Dr. Arnab Das was a very fruitful  experience and I thank him for introducing me to many  exciting theoretical results. I am inspired by his deep  love for physics and enthusiasm in explaining new results.

I thank my RAC members - Dr. Arijit Bhattacharyay and Dr. T. G. Ajitkumar - for monitoring my yearly work progress and for suggesting new ideas. I also thank the IISER community, both academic and administration, for the excellent lab facilities and  financial support.

My first NMR experiments in the lab were done with the help of Soumya. His experience and his ability to respond quickly to many NMR related questions have helped me greatly during the early part of my PhD. I thank him for teaching me the basics of NMR experiments. I was  also lucky to have Abhishek as my senior whose patience has always inspired me. The never ending arguments with him that ranged from physics to politics were quite refreshing.  
Working with the matlab expert Hemant was a very pleasant experience. I learnt a good deal by discussing with my other collaborators too - Koteswar, Anjusha and Ravishankar -  and I thank them for involving me actively. I am glad to have Sudheer as my labmate - he was always available to answer and discuss  any difficult problem with a great clarity (don't be surprised if you find him in lab at 3am!).  I also thank my other juniors - Anjusha, Deepak and Soham - who are super cool, both personally and academically, and who are responsible for a very healthy lab environment. Outside the lab, but within the academic circle, it was a great relief to share  our mutual  miseries (and of course mutual happiness sometimes!) with my batchmates - Mubeena, Snehal, Sunil, Koushik and Arindam.

  Bhavani, my friend since seven years (and counting), has always been very dear and kind to me. The saturday outings, movies, treks, walks, and discussions over ``anything" have made my years in IISER extremely enjoyable. Miss B, I owe you for sharing these amazing activities - I will certainly cherish these moments in the future. Cheers to our eternal friendship!

The love and encouragement from my family has been the greatest strength and inspiration to me. It is with a profound  love  (infinite I wish!) that I dedicate this thesis to my Amma, Appa and Ruthu.

\cleardoublepage
\chapter*{Publications}\addcontentsline{toc}{section}{\textbf{Publications}}
\begin{enumerate}
\item Ravi Shankar, Swathi S. Hegde, and T. S. Mahesh,\\
\textit{Quantum simulations of a particle in one-
dimensional potentials using NMR}, \\
Physics Letters A 378, 10 (2014).

\item Swathi S. Hegde and T. S. Mahesh, \\
\textit{Engineered Decoherence: Characterization and Suppression},\\
Phys. Rev. A 89, 062317 (2014).
\item Swathi S. Hegde, Hemant Katiyar, T. S. Mahesh, and Arnab Das, \\
\textit{Freezing a Quantum Magnet by Repeated Quantum Interference: An Experimental Realization}, \\
Phys. Rev. B 90, 174407(2014).
\item T. S. Mahesh, Abhishek Shukla, Swathi S. Hegde, C. S. Sudheer Kumar, Hemant Katiyar, Sharad Joshi, and K. R. Koteswara Rao, \\
\textit{Ancilla assisted measurements on quantum ensembles: General protocols and applications in NMR quantum information processing},\\
Current Science, 109, 1987 (2015).
\item Anjusha V. S., Swathi S. Hegde, and T. S. Mahesh, \\
\textit{NMR simulation of the Quantum Pigeonhole Effect}, \\
Phys. Lett. A, 380, 577 (2016).
\item Swathi S. Hegde, K. R. Koteswara Rao, and T. S. Mahesh, \\
\textit{Pauli Decomposition over Commuting Subsets: Applications in Gate Synthesis, State Preparation, and Quantum Simulations},\\
arXiv:1603.06867 (2016).
\end{enumerate}

\cleardoublepage

\fancyhf{}
\fancyhead[RO]{\nouppercase{\emph{\rightmark}}}
\fancyhead[LE]{\nouppercase{\emph{\leftmark}}}
\fancyfoot[C]{\thepage}
\setlength{\headheight}{15pt} 
\pagestyle{fancy}

\tableofcontents
\clearpage{\pagestyle{empty}\cleardoublepage}
\addcontentsline{toc}{section}{\textbf{List of Figures}}
\listoffigures
\clearpage{\pagestyle{empty}\cleardoublepage}
\thispagestyle{plain}

\newcommand{\ket}[1]{\vert{#1}\rangle}
\newcommand{\bra}[1]{\langle{#1}\vert}
\newcommand{\outpr}[2]{\vert{#1}\rangle\langle{#2}\vert}

\pagenumbering{arabic}

\thispagestyle{empty}
\part{Background}

\chapter{Overview  \label{chap1}}

\begin{center}
\begin{minipage}[t]{4in}
\textit{``Nature isn't classical, 
dammit, and if you want to make a 
simulation of nature, you'd better make it 
quantum mechanical, and by golly it's a 
wonderful problem, because it doesn't look 
so  easy''.
\begin{flushright}
- Richard Feynman, 1982 \cite{fey82}.
\end{flushright}}
\end{minipage}
\end{center}

\section{Quantum simulation  \label{chap1_1}}

The origin of the quantum physics dates back to the year 1900 when Max Planck tried to give an explanation for the properties of the black-body radiation \cite{planck}. This  quantum theory was further developed by Schr\"{o}dinger, Dirac and other eminent physicists leading to the understanding of quantum mechanics as we now know \cite{shankar,sakurai,dirac}. More than a century since its inception, we still believe that quantum mechanics is the correct description of the present understanding of nature. 
Yet this subject is so counter-intuitive that it has never ceased to surprise us even now.

Quantum mechanics has a lot of applications in the present day science and technology. For example, it is an indispensable tool  to understand the structure of atoms, molecules and their interactions; the invention of magnetic resonance imaging 
has revolutionized the field of medicine; lasers are heavily used in medicine, communication, industries, etc; 
and the list goes on. This thesis deals with one other application of quantum physics, i.e, quantum information processing (QIP) and quantum computation (QC).

Quantum computers are believed to be capable of solving certain physical and mathematical
problems much more efficiently than the classical computers \cite{chuang,preskill,nuclearspins}. The main reason for this efficiency is the phenomenon of
quantum superposition that offers computational parallelism and is beyond the classical paradigm.

Coupled quantum particles that can be precisely addressed, controlled and measured form the basic hardware of a quantum computer. Moreover, in order to implement quantum computation, Di Vincenzo gave certain criteria that the quantum computer should possess \cite{divincenzo1}. These requirements are as follows: 
\begin{enumerate}
\item Scalable and well defined quantum systems. 
\item Ability to initialize the quantum systems to a desired initial state.
\item Long coherence times of the quantum systems so as to implement specific gate operations.
\item A set of quantum gates which are universal.
\item Ability to perform a qubit-specific measurement.
\end{enumerate}

Three different classes of quantum algorithms are believed to be solvable on a quantum computer much more efficiently than on a classical computer. The first class of algorithm is based on quantum Fourier transform such as the Deutsch-Jozsa algorithm and Shor's algorithm \cite{deutsch1992rapid,shor1}.  The quantum computer uses only $n^2$ steps to Fourier transform $2^n$ numbers but a classical computer uses $n2^n$ steps for the same. The second class of algorithm is  based on quantum search algorithm \cite{grover1996fast}. Suppose the goal is to search an specific element in the search space of size  $2^n$. In these cases, a classical computer  requires about $2^n$ operations while a quantum computer does the job by using only about $\sqrt{2^n}$ operations. Finally, another class of algorithm is the quantum simulation \cite{fey82,qsimu14}. This field of quantum simulations is the primary subject of this thesis and is described below.

\begin{figure}[b!]
\begin{center}
\includegraphics[trim=0cm 1cm 0cm 1cm, clip=true, width=10cm]{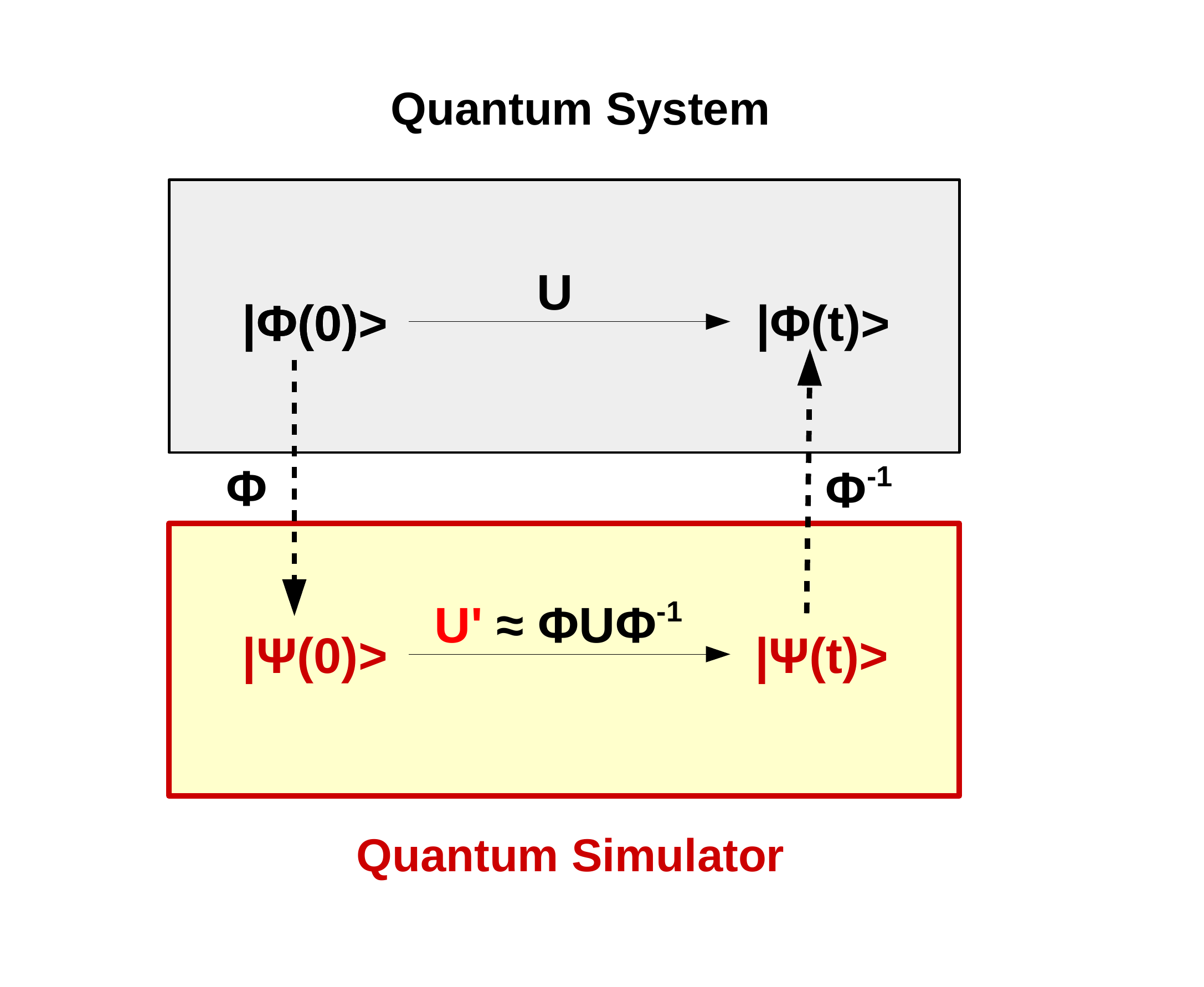}
\caption{Quantum simulation protocol.}
\label{qs_fig}
\end{center}
\end{figure}

A quantum computer that can simulate the dynamics of other quantum systems is  a quantum simulator \cite{fey82}.   
A typical quantum simulation protocol is explained in Fig. \ref{qs_fig} \cite{tseng}.  
The upper box represents the dynamics of a quantum system that we wish to study. Here the quantum system in the initial state $|\phi(0)\rangle$  evolves to a final state  $|\phi(t)\rangle$ under the action of an operator $U$.  The lower box corresponds to an accessible and controllable quantum simulator that is used to simulate the above evolution. The way to implement quantum simulation protocol is by encoding $|\phi(0)\rangle$ into the initial state $|\psi(0)\rangle$  of the quantum simulator via a linear map $\phi$  followed by the application of $U'$ on $|\psi(0)\rangle$.  The operator $U'$ has a one-to-one correspondence with $U$ and is related by the transformation $U' \approx \phi U \phi^{-1}$. The read-out of the final state 
$|\psi(t)\rangle$ of the quantum simulator encodes the information corresponding to $|\phi(t)\rangle$.

In most of the cases, the problem of interest is the final state or the expectation value of an operator in the corresponding state. Even when the required output is the expectation value of an operator, a classical computer has to calculate the state of the quantum system as a prerequisite step. However, the memory required to store the probability amplitudes of the basis states of the quantum system grows exponentially  with the number ($n$) of the quantum systems \cite{qsimu14} (see section. \ref{multiple_qubits}). 
For example, for $n$  2-level coupled quantum systems, a computer has to store $2^n$ complex numbers in a vector and multiply it by a unitary matrix consisting of $2^{2n}$ complex numbers. Although most of the quantum systems can be efficiently simulated using classical computers for small $n$, the same class of problems become intractable for large $n$. For example when $n = 50$, a classical computer has to store  $2^{50}$ parameters and has to be multiplied by a unitary matrix consisting of $2^{100}$ complex numbers which is beyond the reach of present day supercomputers.
Thus owing to this huge memory requirement, simulating quantum systems using a classical computer is a challenging problem.
  As a possible solution to this limitation of classical computers, Feynman  in 1982   proposed the concept of quantum simulator to perform quantum simulations \cite{fey82}:
\begin{center}
\begin{minipage}[t]{4in}
\textit{``Let the computer itself be built of quantum mechanical elements which obey quantum mechanical laws.''}
\end{minipage}
\end{center} 
\vskip 0.3cm
The major advantage of using quantum simulators is that the Hilbert space of the composite quantum systems with $n$ qubits is inherently capable of storing all the $2^n$ complex amplitudes simultaneously. Thus the quantum simulation of an $n$ qubit system can be simulated using only $n$ qubit quantum simulator.

\section{Implementations \label{1_implement}}

Since a couple of decades various quantum devices are believed to be promising candidates for quantum simulations. Among them are the nuclear spins \cite{nuclearspins1,nuclearspins2}, electron spins in quantum dots \cite{dots}, neutral atoms \cite{atoms}, trapped ions \cite{iontrap}, superconducting circuits \cite{squid}, etc, each with strengths and challenges as shown in table \ref{qs_table} \cite{qsimu14}. 
\begin{table}[b!]
\centering
\begin{tabular}{|c| c| c|} 
 \hline
 Quantum simulators & Strength & Challenges  \\ \hline\hline
 Nuclear spins & Well established,   & Scaling,\\  &readily available technology &individual control  \\ \hline
 Electron spins & Individual control, readout   & Scaling \\ \hline
Neutral atoms & Scaling  & Individual control,\\& & readout  \\ \hline
Trapped ions &  Individual control, readout & Scaling  \\ \hline
Superconducting circuits &  Individual control, readout &  Scaling \\
  \hline
\end{tabular}
\caption{Strengths and challenges of a few quantum simulators.}
\label{qs_table}
\end{table}
As of now, the number of small-scale quantum simulation problems that are experimentally implemented or are proposed to be implemented  is almost exhaustive \cite{suter2005,suter2010,nori,cirac,kota,ravi,anjusha,cavity,atoms1}. However, large-scale quantum simulators are yet to become a reality. The main obstacles for this are the scalability, precise control of the dynamics and decoherence.

In this thesis, I will explain our work on quantum simulations using both unitary and nonunitary controls.  While these works indicate successful implementations of the quantum simulations, they also  address the problems of quantum control and decoherence. Although we used nuclear spin 1/2 systems in a liquid state NMR setup as our quantum simulators, most of the concepts are general and are applicable elsewhere. The experimental implementations of these aspects that are a part of this thesis are briefly explained below:   
\begin{enumerate}
\item Chapter \ref{chap4} describes the unitary control, the methodology, and one particular quantum simulation realized using the advanced optimal quantum control techniques. 
I will first describe the phenomenon of  dynamical quantum many-body localization, wherein a spin-chain freezes its dynamics for certain specific frequencies of external drive \cite{arnab}.
Unlike classical systems, the quantum systems freeze and respond non-monotonically with the frequency of the external drive.  Here I will describe the first experimental observation of quantum exotic freezing using an NMR system consisting of three mutually interacting spin 1/2 nuclei \cite{freeze2}.  I will also describe the importance of robust unitary control over spin-dynamics.  Particularly, I will describe the implementation of GRadient Ascent Pulse Engineering (GRAPE) protocol for robust unitary control.

\item Chapter \ref{chap6} addresses the problem of decomposition of an arbitrary unitary operator in terms of simpler unitaries. 
Here we propose a general numerical algorithm, namely Pauli Decomposition over Commuting Subsets (PDCS), to decompose an arbitrary unitary operator in terms of simpler \textit{rotors} \cite{pdcs}.  Each rotor is expressed as a generalized rotation over a mutually commuting set of Pauli operators.   Using PDCS, we decomposed several quantum gates and circuits and also showed its application in designing quantum circuits for
state preparation. We hypothesize the decomposition method to scale efficiently with the size of the system, and propose its application in quantum simulations.  As an example, I will describe quantum simulation of three-body interaction using a three-spin NMR system and monitor the dynamics with the help of overall magnetization.  

\item In practice, quantum systems are affected by their interactions with the environment leading to an undesirable nonunitary process  known as decoherence. This process is accompanied by the loss of information in the quantum processors and is a major obstacle in experimental quantum information processing and computation. One of the ways to fight this process is to understand decoherence. Teklemariam
 {\em et al.}, in 2003  \cite{cory_dec}, described a way of introducing the artificial decoherence on a closed quantum system by randomly perturbing an ancillary system.  Recently, in a different context, Alvarez
 {\em et al.} and  Yuge {\em et al.}, have independently proposed noise spectroscopy to characterize the noise acting on a quantum system \cite{suter_ns,yuge}.   In Chapter \ref{chap5}, I will describe the experimental implementation of such an engineered noise introduced by random RF pulses on an ancillary spin using   an NMR spin-system.  I will also describe the characterization of the engineered noise  by both noise spectroscopy and quantum process tomography. Further, we suppressed this induced noise using dynamical decoupling (DD) which is a  process of suppression of decoherence by systematic modulation of system state. 
Chapter \ref{chap5} also describes the first experimental study of competition between the engineered decoherence and DD \cite{dec}.

\section{Thesis structure}
Fig. \ref{thesis_structure_fig} gives the pictorial representation of the thesis structure. 
 \begin{figure}[h]
\begin{center}
\includegraphics[trim=2.5cm 5cm 2.0cm 1.5cm, clip=true, width=11.2cm]{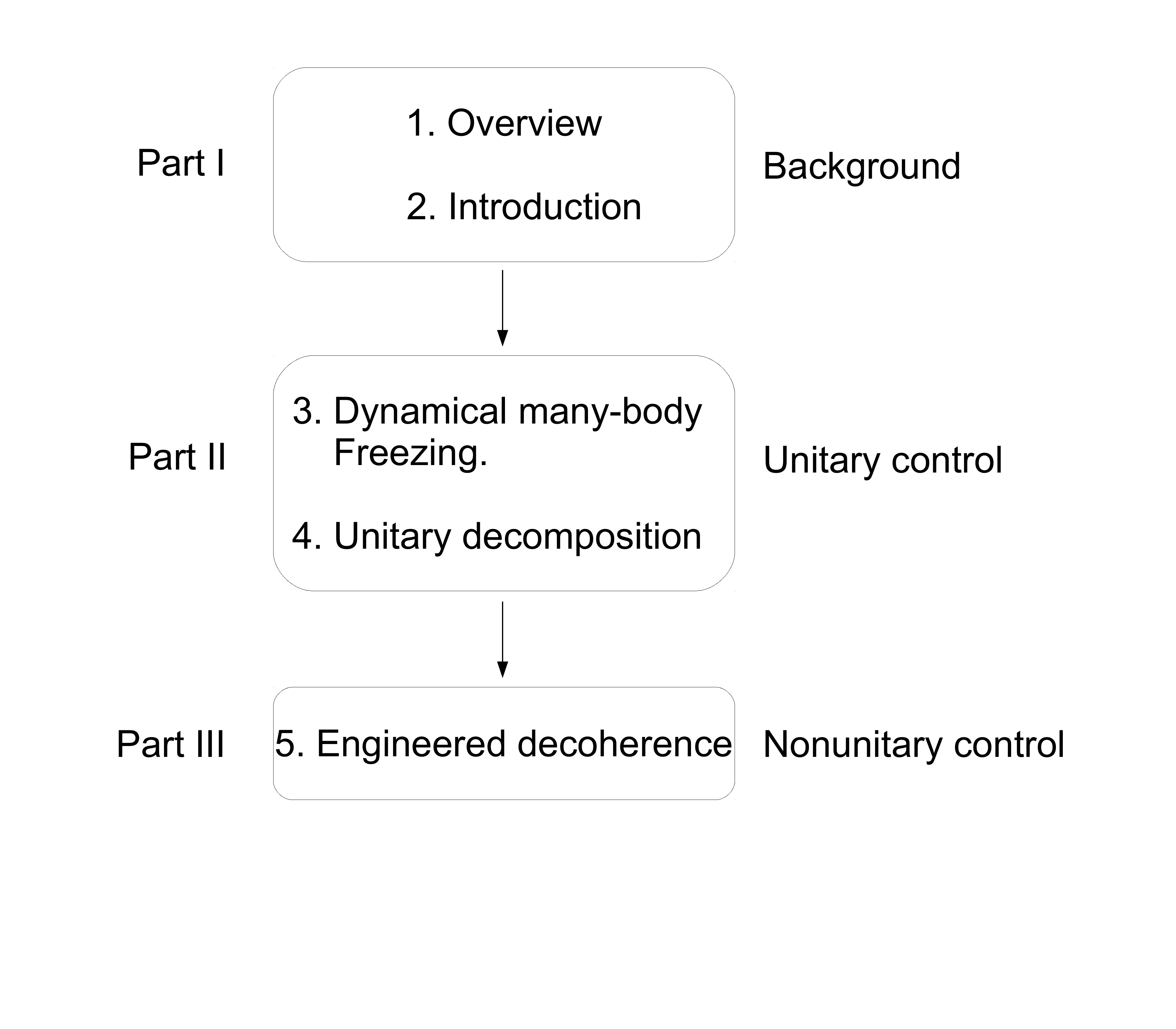}
\caption{Structure of the thesis.}
\label{thesis_structure_fig}
\end{center}
\end{figure}

The thesis consists of the following parts: 
\begin{itemize}
\item The part I consists of chapter \ref{chap1} and \ref{chap2}. Chapter \ref{chap1} gives a brief overview of this thesis. Chapter \ref{chap2} deals with the basic terminology and theory of quantum information processing. It includes the description of quantum states, their evolution and measurement schemes. These form the platform to understand a quantum simulation protocol as mentioned in Fig. \ref{qs_fig}. Chapter \ref{chap2} also explains the basics of nuclear magnetic resonance (NMR) and how nuclear spins in NMR can be used as quantum simulators. 

\item The part II is about the implementations of quantum simulations using unitary control. It consists of two related works and are explained in chapters \ref{chap4} and \ref{chap6}. 
\item Finally, part III deals with the implementations of non-unitary dynamics and the related work is explained in chapter \ref{chap5}. 
\end{itemize}

To summarize, three different works that  are explained in detail in chapters \ref{chap4} to \ref{chap5} form the backbone of this thesis  \cite{freeze2,pdcs,dec}. The abstracts of these three chapters were given in the section \ref{1_implement}.

\end{enumerate}

\thispagestyle{empty}
\chapter{Introduction \label{chap2}}

This chapter gives a brief introduction to the theory of  quantum information processing (QIP) and nuclear magnetic resonance (NMR) QIP with a goal to provide a few useful techniques for implementing quantum simulations. A typical quantum computation algorithm consists of an input, processing and an output. Below is a brief summary of these three major steps:
\begin{enumerate} 
\item State initialization:  A \textit{quantum state} $|\psi(t)\rangle$ contains the entire description of the quantum system.  As an input of any quantum algorithm, it is required that any given quantum system is initialized to a known state $|\psi(0)\rangle$. 
\item Gate implementation: Processing of the information is done using quantum gates. A quantum gate is realized by the \textit{unitary operator} $U(t)$ that evolves the initial state $|\psi(0)\rangle$ to the final state $|\psi(t)\rangle$.  
\item Measurements:  The final state $|\psi(t)\rangle$ or the  expectation value of any hermitian operators $\hat{A}$ in the state $|\psi(t)\rangle$ that encodes the solution to the algorithm is obtained by a \textit{measurement} process. 
\end{enumerate}
With reference to the above steps, this chapter first concentrates on the theory of state description, gate operation and measurements. The latter parts of the chapter deal with the same for a specific quantum device, namely nuclear spins in liquid state NMR set up. 
For extensive details about these topics it is recommended to refer to \cite{chuang,preskill,cavanagh,olivera,nmr_review}. 

\section{Quantum information processing and computation}
\subsection{Quantum States \label{chap2_1}}
 This section explains the basic terminology and properties of the quantum states. 

\subsubsection{Single qubit}\label{S_single_qubit}

A qubit is a quantum counterpart of a classical bit.  Physically, any two level quantum system is a qubit. Mathematically, the most general state of a qubit is represented as
\begin{equation}
|\psi\rangle = \alpha|0\rangle+\beta|1\rangle,
\label{psi}
\end{equation}
where $\alpha$, $\beta$ are the probability amplitudes with $|\alpha|^2+|\beta|^2=1$, and $|0\rangle$, $|1\rangle$ are the orthogonal states and form a computational basis. 

\begin{figure}[h]
\begin{center}
\includegraphics[trim=1cm 1cm 1cm 1cm, clip=true, width=7cm]{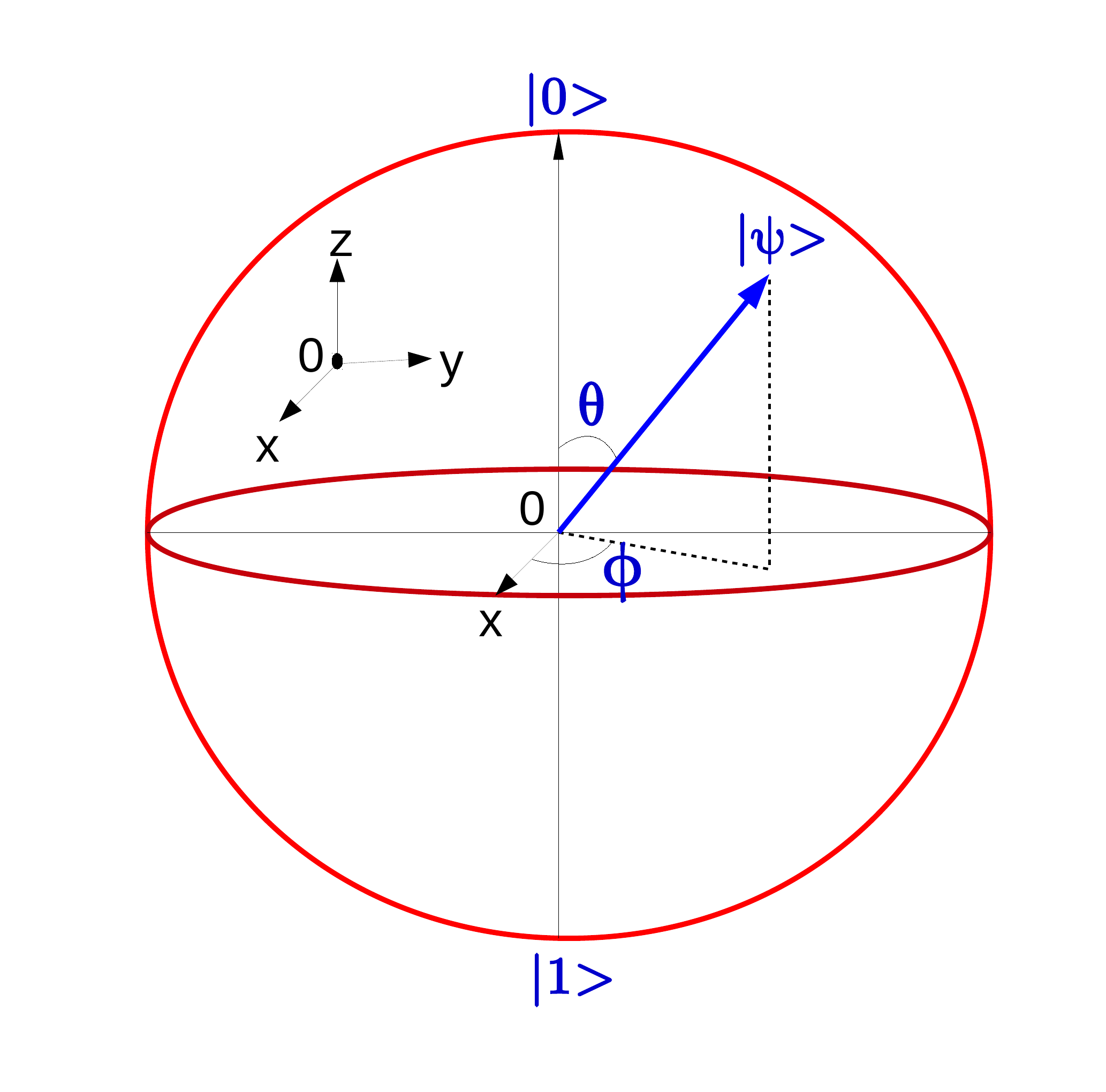}
\caption{Bloch sphere representation of the single qubit state $|\psi\rangle$. }
\label{bloch}
\end{center}
\end{figure}

The geometric representation of a single qubit state (Eq. \ref{psi}) is visualized by Bloch sphere as shown in Fig. \ref{bloch}. Here $\alpha=\cos(\frac{\theta}{2})$ and $\beta=e^{i\phi}\sin(\frac{\theta}{2})$ where $\theta=[0,\pi]$ and $\phi=[0,2\pi]$ are the points on the unit sphere. The state $|\psi\rangle$ can exist anywhere in the sphere.

Thus as seen from Eq. \ref{psi}, a qubit can exist in a linear superposition of $|0\rangle$ and $|1\rangle$. The complex numbers $\alpha$ and $\beta$ have the information of the basis states and thus a qubit can store infinite amount of information until measured. This is in contrast with the classical bits that can be either 0 or 1 and can have only one bit of information at a time. This property of superposition enables quantum parallelism that renders quantum computers more powerful than classical computers in terms of the computational speed and storage capacity. 

\subsubsection{Multiple qubits \label{multiple_qubits}}
As discussed in section \ref{chap1_1}, a typical quantum computer requires multiple interacting qubits. Apart from the phenomenon of quantum superposition, such quantum systems exhibit one of the most powerful properties called entanglement. 

Suppose there are two qubits described by the states $|\psi_1\rangle = \alpha_1|0\rangle+\beta_1|1\rangle$ and $|\psi_2\rangle = \alpha_2|0\rangle+\beta_2|1\rangle$ where $|\alpha_1|^2+|\beta_1|^2=1$ and $|\alpha_2|^2+|\beta_2|^2=1$. The state of the composite system is represented by 
\begin{equation}
|\psi\rangle = |\psi_1\rangle \otimes |\psi_2\rangle = |\psi_1\psi_2\rangle,
\end{equation}
where $\otimes$ is the tensor product. Hence $|\psi\rangle = \alpha_1 \alpha_2|00\rangle+\alpha_1\beta_2|01\rangle+\beta_1\alpha_2|10\rangle+\beta_1\beta_2|11\rangle$ with $|\alpha_1\alpha_2|^2+|\alpha_1\beta_2|^2+|\beta_1\alpha_2|^2+|\beta_1\beta_2|^2=1$ and is described by $2^2=4$ complex numbers. The states $\{|00\rangle,|01\rangle,|10\rangle, |11\rangle\}$ form the computational basis of this two qubit system.

In a similar way, an $n$-qubit system is represented by the state
\begin{equation}
|\psi\rangle^{\otimes n} = |\psi_1\psi_2  \cdots \psi_n\rangle,
\end{equation}
One can observe that a total of $2^n$ basis states are required to describe an $n$-qubit state. Thus in order to describe an $n$-qubit state, one requires $2^n$  probability amplitudes indicating an  exponential growth  with  $n$. 

\subsubsection{Density operator formalism}
Quantum state for an ensemble of quantum systems is generally described by using the density operators \cite{landau}. In this section, I will introduce density operator formalism.

 A density operator of an $n$-qubit  system is defined as

\begin{equation}
\rho = \sum_i^n p_i|\psi_i\rangle \langle\psi_i|,
\label{density}
\end{equation}
where $|\psi_i\rangle$ is the state of the $i^{\mathrm{th}}$ sub-system and $p_i$'s are the  probabilities of finding the $i^{\mathrm{th}}$ sub-system in the state $|\psi_i\rangle$ such that $\sum_i^n p_i=1$.

The geometrical description of a single qubit density operator expressed in Pauli operator basis is given by 
\begin{equation}
\rho = \frac{1}{2}(\mathbbm{I}+r\cdot\sigma),
\label{density1}
\end{equation}
where $\mathbbm{I}$ is the Identity operator, $r$ is the 3-dimentional unit vector and $\sigma\in\{X,Y,Z\}$ are the Pauli operators defined by:
\begin{equation}
 X = \left[ \begin{array}{cc}
0 & 1\\
1 & 0\\
\end{array}\right];
\qquad
Y = \left[ \begin{array}{cc}
0 & -i\\
i & 0\\
\end{array}\right]; \qquad
Z = \left[ \begin{array}{cc}
1 & 0\\
0 & -1\\
\end{array}\right];
\end{equation} 
Also, in the matrix representation
\begin{equation}
\rho = \left(
\begin{array}{cc}
\rho_{00} & \rho_{01} \\
\rho_{10} & \rho_{11} \\
\end{array}
\right). 
\label{rho_pop_coh}
\end{equation}
It is important to note that the diagonal elements $\rho_{00}, \rho_{11}$ correspond to the populations and the off-diagonal elements $\rho_{01}, \rho_{10}$ correspond to the coherences of the state. It should be noted that the populations add up to one and $\rho_{01} = \rho_{10}^\star$ since $\rho$ is hermitian. 

 One can also express the density operator of the composite system  as $\rho = \rho_1\otimes\rho_2\otimes\cdots\otimes\rho_n$.

Most importantly, any  operator $\rho$ should satisfy the following properties:
\begin{itemize}
\item $\mathrm{Tr}[\rho]=1$.
\item $\rho$ should be a positive operator (i.e., it should have non-negative eigenvalues).
\item $\rho$ should be hermitian. i.e.,  $\rho=\rho^\dagger$.
\end{itemize}

\subsubsection{Reduced density operator}
A reduced density operator describes the state of the sub-system when the density operator of the composite system is known.  

Suppose the composite system is in the state $\rho_{12}$ which contains two sub-systems namely $1$ and $2$. Then the sub-system states are given by
\begin{equation}
\rho_1=\mathrm{tr}_2(\rho_{12}),
\end{equation}
\begin{equation}
 \rho_2=\mathrm{tr}_1(\rho_{12}),
\end{equation}
where the operation tr$_i$, with $i=1,2$, is called as partial trace. For example, when $\rho_{12} = |\psi_1\rangle \langle\psi_1| \otimes |\psi_2\rangle \langle\psi_2|$, the partial trace over the sub-system 1 is defined as
\begin{equation}
\rho_2 = \mathrm{tr}_1(\rho_{12}) = \mathrm{tr}_1(|\psi_1\rangle \langle\psi_1| \otimes |\psi_2\rangle \langle\psi_2|) = |\psi_2\rangle \langle\psi_2|\mathrm{tr}(|\psi_1\rangle \langle\psi_1|) = |\psi_2\rangle \langle\psi_2| \langle\psi_1|\psi_1\rangle.
\nonumber
\end{equation}

\subsubsection{State types}
A state can be either pure, mixed, separable or entangled depending on the following properties. 

When all the sub-systems are in the same state $|\psi\rangle$, the composite system is known to be in pure state.  It is a required assumption that the individual sub-systems in Eq. \ref{density} are pure but the composite system may not always be pure. When different sub-systems have different states, the composite system is known to be in a mixed state.  
 The condition for the composite state $\rho$ to be either pure or mixed is defined as follows:
\begin{itemize}
\item Pure state: $\mathrm{Tr}[\rho^2]=1$. 
\item Mixed state: $\mathrm{Tr}[\rho^2]<1$.
\end{itemize}
Geometrically, the states on the surface of the bloch sphere of Fig. \ref{bloch} are  pure states and any other states inside the surface of the bloch sphere are the mixed states.

An interesting consequence of ensemble quantum systems is the property of entanglement. 
If an $n$-qubit density matrix is expressed as 
\begin{equation}
\rho = \rho_1\otimes\rho_2\otimes\cdots\otimes\rho_n,
\end{equation}
then such a state is known to be separable state. And if
\begin{equation}
\rho \ne \rho_1\otimes\rho_2\otimes\cdots\otimes\rho_n,
\end{equation}
then such a state is known as entangled state.

It should be noted that suppose the composite system is described by a separable state then its reduced density operator will be a pure state and if the composite system is described by an entangled state then its reduced density operator will be a mixed state.

\subsection{{Quantum gates}\label{chap2_2}}

A quantum gate is an operation that evolves the quantum state from a specific initial state to a final state. 

\subsubsection{State evolution \label{section_state_evolution}}
Any closed quantum system with initial state $|\psi(0)\rangle$ evolves under a time dependent Hamiltonian  ${\cal H}(t)$ according to
\begin{equation}
|\psi(t)\rangle = U(t) |\psi(0)\rangle, 
\label{state_evolution}
\end{equation}
where $U(t) = {\cal T}e^{-i\int_0^{t}{\cal H}(t') dt'}$  is a unitary operator. Here $\hbar$ is set to unity and ${\cal T}$ is the time ordering operator.

Similarly, for time independent Hamiltonian, the evolution of the state in terms of $n-$qubit density operator $\rho(0)$ is obtained by combining equations \ref{density} and \ref{state_evolution} and  is described as
\begin{equation}
\rho(t) = \sum_i^n p_i[U(t)|\psi_i(0)\rangle][ \langle\psi_i(0)|U(t)^\dagger],
\nonumber
\end{equation}
\begin{equation}
\rho(t) = U(t)\rho(0)U(t)^\dagger.
\label{density_evolution}
\end{equation}

One of the main features of unitary operators is that they preserve the purity of the quantum states over time. In other words, unitarity imposes reversibility criteria which means that one should be able to get back the initial state $\rho(0)$ starting from $\rho(t)$:
\begin{equation}
U(t)^\dagger\rho(t)U(t) =  U(t)^\dagger[U(t)\rho(0)U(t)^\dagger]U(t)  = \rho(0),
\nonumber
\end{equation}
since $U U^\dagger = U^\dagger U = \mathbbm{I}$.

In the language of quantum computation, a unitary operator $U(t)$ corresponding to the transformation
\begin{equation}
\rho(0) \xrightarrow{U(t)} \rho(t)
\nonumber
\end{equation} 
is  a quantum gate. Below, I will explain the quantum gates with reference to the circuit model of quantum computation.

\subsubsection{Single qubit gates \label{section_single_gates}}

Any unitary $U$ transforms the quantum system from one state to another. 
Geometrically, $U$ rotates any  state vector $|\psi(0)\rangle$ to $|\psi(t)\rangle$ in the bloch sphere. Thus each single qubit $U$ corresponds to a rotation  about an axis $\hat{n}$  and is given by
\begin{equation}
R_{\hat{n}}^{\theta} = e^{-i\theta\hat{n}\cdot\vec{\sigma}/2}
= \cos\left(\frac{\theta}{2}\right)\mathbbm{I}-i\sin\left(\frac{\theta}{2}\right)(n_x X+n_y Y+ n_z Z),
\label{rotation}
\end{equation}
where $\hat{n}=\{n_x,n_y,n_z\}$ is the $3$-dimensional unit vector, $\sigma=\{X,Y,Z\}$ is the Pauli operator and $\theta$ is the rotation angle.

Any single qubit operator can be constructed using Eq. \ref{rotation}. Some standard quantum gates like Hadamard ($H$) and phase gate ($S$) are listed below:
\begin{equation}
H =  \frac{1}{\sqrt{2}}\left[ \begin{array}{cc}
1 & 1\\
1 & -1\\
\end{array}\right]; \qquad
S = \left[ \begin{array}{cc}
1 & 0\\
0 & i\\
\end{array}\right];
\end{equation} 

\begin{figure}[h]
\begin{center}
\includegraphics[trim=1cm 1cm 3.5cm 1cm, clip=true, width=5cm]{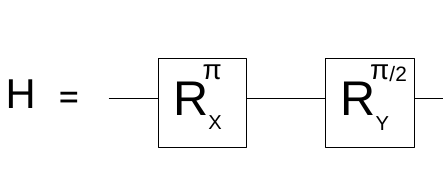}
\caption{Single qubit Hadamard gate. Here the rotations are of the form of Eq. \ref{rotation} and are implemented from left to right.}
\label{hadamard}
\end{center}
\end{figure}

Quantum operators with multiple non-commuting rotations should be carefully implemented in a specific time order. For convenience, the operators are acted from left to right in a quantum circuit. For example,  as shown in Fig. \ref{hadamard}, $H$ corresponds to the rotation about  $X$-axis with $\theta = \pi$  followed by a rotation about  $Y$-axis with $\theta=\pi/2$. Thus, $H = R_y^{\pi/2} R_x^\pi$. 

\subsubsection{Two-qubit gates \label{section_multi_gates}}

\begin{figure}[h]
\begin{center}
\includegraphics[trim=0cm 1cm 0cm 1cm, clip=true, width=7cm]{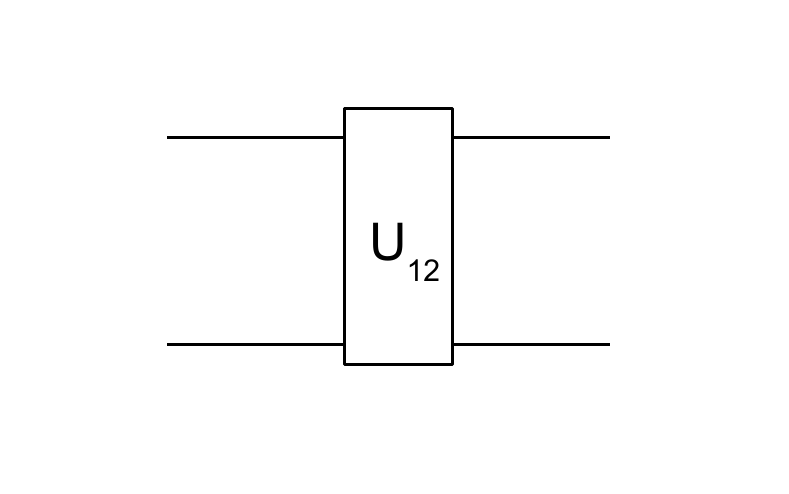}
\caption{A general two qubit gate.}
\label{U10}
\end{center}
\end{figure}

A two qubit gate $U_{12}$ exploits the knowledge of single qubit gates as well as the interaction between the two qubits. Such gates play an important role in quantum computation as they can entangle the qubits. The circuit representation of $U_{12}$ is shown in figure \ref{U10}.

For example, a standard two qubit gate is a controlled-NOT (CNOT) gate and is given by 
\begin{equation}
U_{CNOT} = \left[ \begin{array}{cccc}
1 & 0 & 0 & 0 \\
0 & 1 & 0 & 0 \\
0 & 0 & 0 & 1 \\
0 & 0 & 1 & 0 \\
\end{array}\right]. 
\label{Ucnot}
\end{equation} 
\begin{figure}[b!]
\begin{center}
\includegraphics[trim=0cm 1cm 0cm 1cm, clip=true, width=10cm]{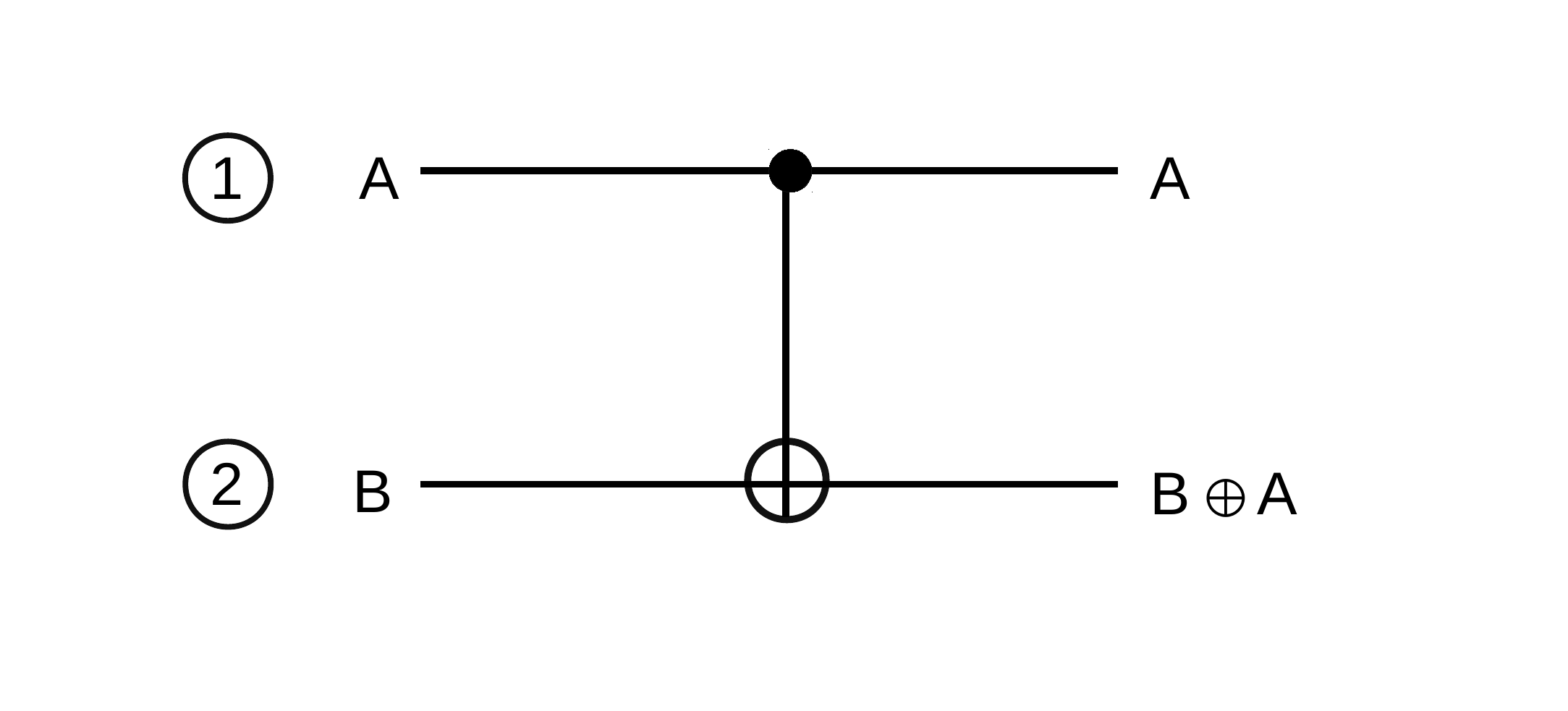}
\caption{Two qubit CNOT gate.}
\label{cnot}
\end{center}
\end{figure}

Figure \ref{cnot} gives the circuit representation of $U_{CNOT}$.
 Qubit \circled{1} is the control and the qubit \circled{2} is the target with A and B as inputs. In convention, a filled circle indicates control and the plus indicates the target.
The action of $U_{CNOT}$ is written as
\begin{equation}
|A,B\rangle \longrightarrow |A,B\oplus A\rangle,
\nonumber
\end{equation}
where $A,B\in\{0,1\}$ and $\oplus$ is the addition modulo 2. The effect of CNOT gate is to flip the state of the target qubit  when the control qubit is in state $|1\rangle$ and to do nothing when the control qubit is in state $|0\rangle$.

\subsubsection{Universal gates}
In order to realize arbitrary computation, one needs  a universal set of gates.  Just like a combination of NAND gates is universal in classical computation, there exists a set of quantum gates which are universal. 

\textit{Any arbitrary single qubit gates along with CNOT gates form a universal set of quantum gates.}

Specifically, one can consider Hadamard, phase gate, CNOT and $pi/8$ gates as a set of universal quantum gates. A more general observation is that any arbitrary single and two qubit gates can form universal quantum gates. 

\subsection{Measurements}
Measurements are an important part of any algorithm and is a necessary step to extract any useful information. This step requires that the measuring device interacts with the quantum system  and thus treats the quantum system as an open quantum system.  In general, measurements operations are nonunitary.

Suppose $\{M_m\}$ is the set of measurement operators that act on the state space of the system being measured. Here $m$ is the measurement outcome of the operator that is measured. Let $|\psi\rangle$ be the state just before the measurement and the action of the measurement operators on  $|\psi\rangle$ is defined as
\begin{equation}
|\psi\rangle^{'} = \frac{M_m|\psi\rangle}{\sqrt{p}},
\end{equation}
where $|\psi\rangle^{'}$ is the state after the measurement and $p = \langle\psi|M_m^\dagger M_m|\psi\rangle$   is the probability of obtaining the outcome $m$. Here  $\sum_m p(m) =1$, and $\sum_m M_m^\dagger M_m = \mathbbm{I}$.

\subsubsection{Projective measurements}
Another class of measurements are projective measurements which is described by Hermitian operator $M$ as
\begin{equation}
M = \sum_m mP_m,
\end{equation} 
where $P_m = |m\rangle \langle m|$ with $\{|m\rangle\}$ being the eigenstates of $M$ and $m$ are its eigenvalues. 

The probability of obtaining the outcome $m$ after $|\psi\rangle$ is measured is given by
\begin{equation}
p(m) = \langle\psi|P_m|\psi\rangle,
\end{equation}
and thus the post-measurement state has the form
\begin{equation}
\frac{P_m|\psi\rangle}{\sqrt{p(m)}}.
\end{equation}

Further it should be noted that for projective measurements, $P_m$ should satisfy the following conditions:
\begin{itemize}
\item $\sum_m P_m^\dagger P_m = \mathbbm{I}$.
\item $P_m P_{m'} = \delta_{m, m'} P_m$.
\end{itemize}
Measurements are non-unitary operations. For example, the measurement operators for  single qubit are $|0\rangle\langle 0|$ and $|1\rangle\langle 1|$. One can verify that each of these operators is Hermitian but  not unitary.

\subsubsection{Ensemble average measurements \label{ensemble_measurement}}
In many cases, one is interested in obtaining the expectation value of an arbitrary operator $A$. The way to measure such an operator is to prepare a large number of quantum systems in the same initial states and the outcome corresponds to the probability weighted eigenvalues of  $A$ in some final state. It is defined as follows:
\begin{equation}
\langle A (t)\rangle = \mathrm{Tr}[A(t)\rho(t)],
\end{equation}
where $\rho(t)$ is the normalized state at time $t$.    It is important that the operator $A$ is hermitian since one expects that the measurement outcomes are real.

\section{NMR QIP}
The previous sections dealt with the mathematical descriptions of the quantum states, quantum gates and measurements. In this section, I will describe the same but with reference to their physical realization using nuclear spins in liquid state NMR. 
Below, I will introduce to the phenomenon of NMR and how this phenomenon can be exploited to realize quantum simulators \cite{cavanagh,olivera,nmr_review}.

\subsection{Nuclear magnetic resonance\label{nmr}}

When a quantum particle with non-zero nuclear spin angular momentum 
 is placed in an external static magnetic field ($B_0$), there is an interaction between the particle and the field. This interaction leads to the splitting of the spin energy levels of the quantum particle, a phenomenon known as ``Zeeman effect''. Thus in the presence of $B_0$ along the $z-$axis, the splitting of the levels correspond to the following quantized energy levels:
\begin{equation}
E_m =  -\gamma \hbar m B_0. 
\label{E_nmr}
\end{equation}
Here  $\gamma$ is the gyromagnetic ratio of the nuclei and $m=[-I, -I+1,\cdots,I-1,I]$ is the magnetic quantum number that takes $2I+1$ values where $I$ is the nuclear spin quantum number.

The energy difference between the states $m$ and $m+1$ can be obtained by calculating $\Delta E = E_{m+1}-E_m$ using Eq. \ref{E_nmr}
and the corresponding frequency $\omega_0 = \Delta E/\hbar$ is given by
\begin{equation}
\omega_0=  - \gamma  B_0.
\label{omega_0}
\end{equation}
This frequency is known as the \textit{Larmor frequency} and plays a major role in addressing different nuclear spin species. 
A resonant absorption of energy is achieved when such  nuclear  spins with definite $\omega_0$ is perturbed by an external electromagnetic field with same frequency as $\omega_0$. This phenomenon is called as nuclear magnetic resonance. 

Nuclei which exhibit this phenomenon are called as NMR active nuclei. Some common examples include $^1$H,$^{13}$C,$^{14}$N,$^{19}$F, etc and their intrinsic properties are listed in table \ref{nmr_nuclei}. A few examples of NMR samples are chloroform, trifluoroiodoethylene, 1-bromo-2,4,5-trifluorobenzene, crotonic acid, aspirin, etc.  
\begin{table}[h!]
\centering
\begin{tabular}{|c| c| c|} 
 \hline
 Nucleus & $I$ & $\gamma$ $(Ts)^{-1}$  \\ \hline
 $^1$H & 1/2   & $2.6752 \times 10^8$  \\ 
 $^{13}$C & 1/2 & $6.728 \times 10^7$   \\
 $^{14}$N & 1 & $1.934 \times 10^7$   \\
 $^{19}$F & 1/2 & $2.5181 \times 10^8$  \\
$^{31}$P & 1/2 & $1.0841 \times 10^8$  \\
  \hline
\end{tabular}
\caption{NMR active nuclei and their intrinsic properties.}
\label{nmr_nuclei}
\end{table}

\subsection{NMR qubits}
This thesis  deals with nuclear spins corresponding to $I=1/2$. In the following, I will explain the physical realization of single and multiple qubits using NMR.
\subsubsection{Single qubits}
\begin{figure}[h]
\begin{center}
\includegraphics[trim=0cm 0.5cm 0cm 0.5cm, clip=true, width=8.5cm]{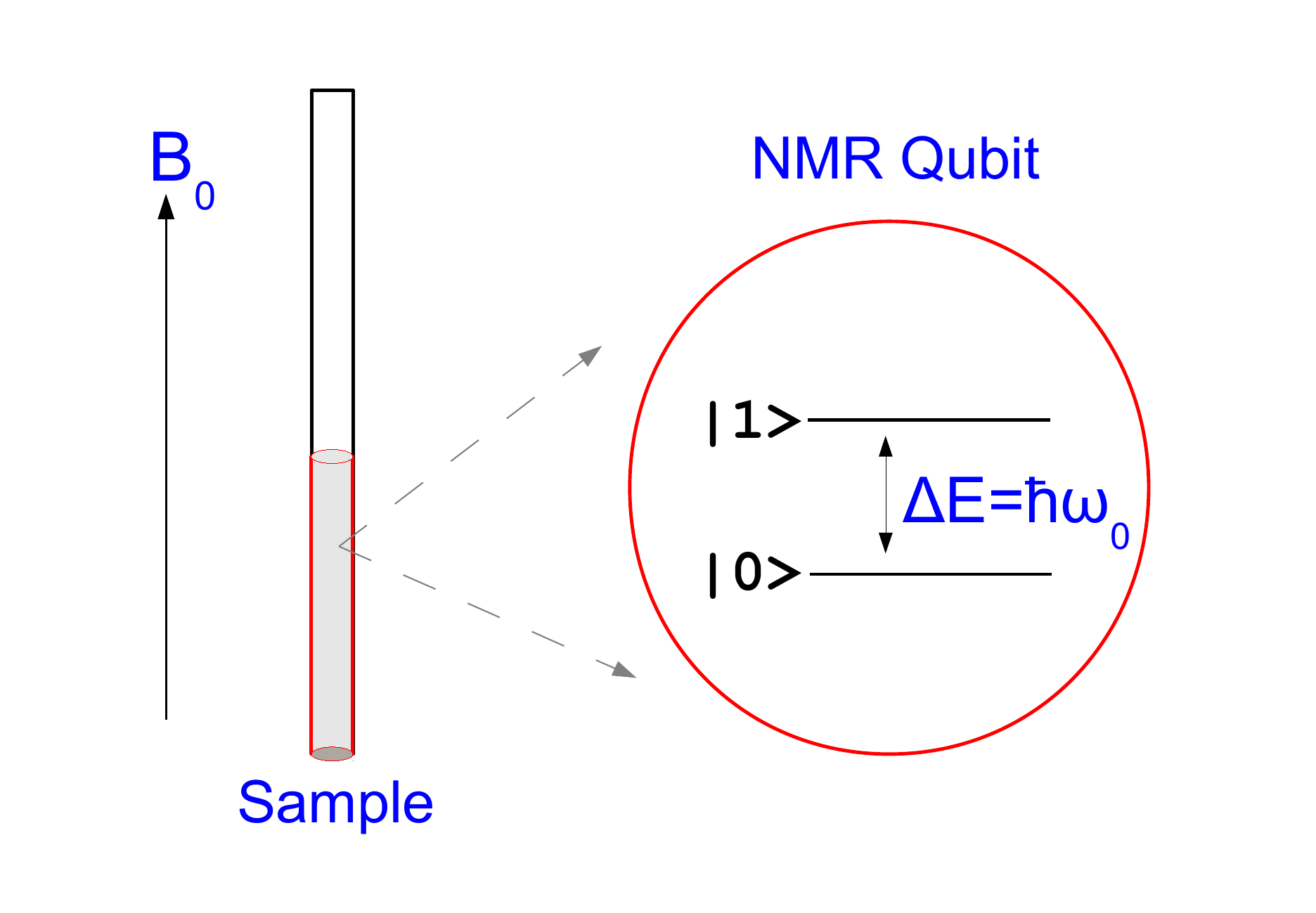}
\caption{Zeeman splitting of a spin-1/2 nuclei. Here $|0\rangle$ corresponds to $m=1/2$ and $|1\rangle$ corresponds to $m=-1/2$.}
\label{zeeman}
\end{center}
\end{figure}

A single NMR active nuclei with $I=1/2$ in a molecule  that is placed in $B_0$ has a unique $\omega_0$ and represents a qubit  as shown in figure \ref{zeeman}. 
The internal Hamiltonian of such a single qubit system is given by
\begin{equation}
H_0 = -\omega_0 I_z,
\end{equation}
where the spin operator $I_z = Z/2$. 
The eigenstates and eigenvalues of $H_0$ are given by $\{|0\rangle, |1\rangle\}$  and $\{\omega_0/2,  -\omega_0/2\}$ respectively. This corresponds to  the energy difference of $\Delta E = \hbar \omega_0$.

Typically, in liquid state NMR, a sample consisting of NMR active molecules are dissolved in an NMR silent solvent. In a dilute solution the intermolecular interactions are negligible and hence one can treat the sample as an ensemble of single spin systems.

\subsubsection{Multiple qubits}
\begin{figure}[h]
\begin{center}
\includegraphics[trim=0cm 0.5cm 0cm 0.5cm, clip=true, width=8.5cm]{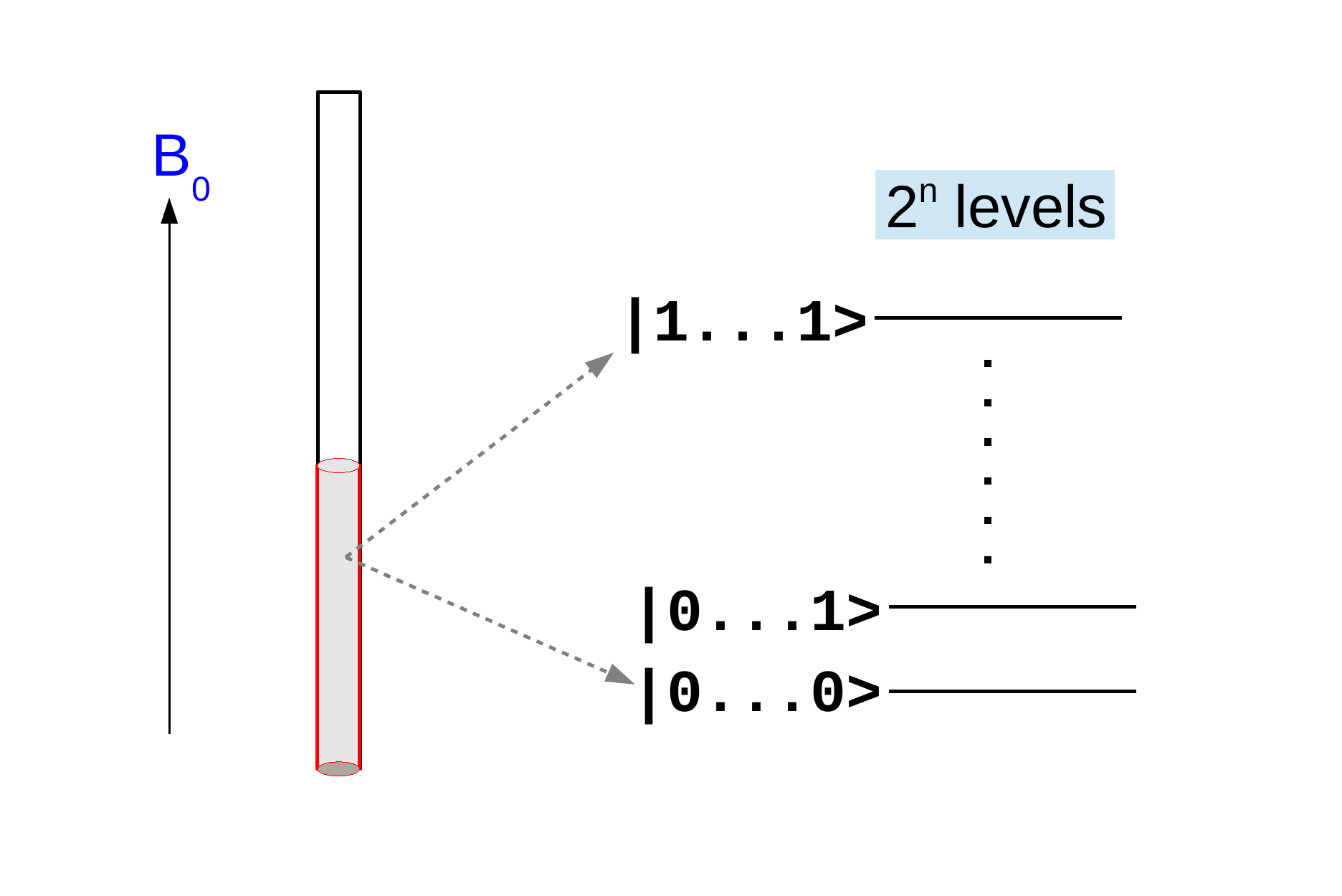}
\caption{Energy levels of an $n$ qubit spin-1/2 nuclei.}
\label{nqubits}
\end{center}
\end{figure}
A molecule may contain multiple coupled spin-1/2 nuclei. These spins in a molecule can be either of the same or different  species  and are categorized as homonuclear or heteronuclear molecules respectively. In isotropic liquids state, the inter-molecular and intra-molecular dipolar couplings are averaged out due to the rapid, random and isotropic  motions of the molecules and thus only the scalar couplings that are mediated by the electrons in the intra-molecular nuclear bonds survive. Thus the internal Hamiltonian for multiple qubits in the lab frame is given by
\begin{equation}
H_0 = \sum_{i=1}^n \omega_i I_z^i + 2\pi \sum_{i<j}^n J_{ij} I^i\cdot I^j,
\label{H_strong}
\end{equation}
where $n$ is the number of qubits, $J_{ij}$ is the scalar coupling  and $\omega_i$ is the larmor frequency.  

If $|\omega_i-\omega_j|\gg 2\pi |J_{ij}|$ then the NMR qubits are weakly coupled, otherwise they are strongly coupled. Under the weak coupling limit, Eq. \ref{H_strong} reduces to 
\begin{equation}
H_0 = \sum_{i=1}^n \omega_i I_z^i + 2\pi \sum_{i<j}^n J_{ij} I_z^i I_z^j.
\label{H_weak}
\end{equation}

\subsection{NMR States \label{section_nmr_states}}
Physical qubits are not perfectly isolated from the environment and the NMR qubits are surrounded by the lattice which is at a temperature $T$. Their interactions with the lattice lead to a thermal equilibrium state at $T$ and such a state is given by
\begin{equation}
\rho_{eq} = \frac{e^{-H_0/k_B T}}{\mathrm{Tr}[e^{-H_0/k_B T}]},
\end{equation}  
where $H_0$ is the internal Hamiltonian. For example, for a single qubit case,  the above equation in the eigenbasis of $Z$ is equivalent to
 \begin{equation}
 \frac{1}{\sum_{i=0}^{1}e^{-E_i/k_B T}}\left(
\begin{array}{cc}
e^{-E_0/k_BT} & 0 \\
0 & e^{-E_1/k_BT} \\
\end{array}
\right),
\label{rho_1q_eq}
\end{equation}
where $E_0 =  \omega_0/2$ and $E_1 = - \omega_0/2$. The diagonal elements indicate the populations and thus, from Eq. \ref{rho_1q_eq}, it can be observed that the populations follow the Boltzmann distribution. 

At high temperature, $E_i\ll k_BT$ and $\rho_{eq}$ can be expanded using first order Taylor series:

\begin{align}
\rho_{eq} & \approx \frac{\mathbbm{I}-H_0/k_BT}{\mathrm{Tr}[{\mathbbm{I}-H_0/k_BT}]}
\approx \frac{\mathbbm{I}-H_0/k_BT}{\mathrm{Tr}[{\mathbbm{I}}]}\nonumber \\
&\approx \frac{\mathbbm{I}}{2^n}-\frac{H_0}{2^nk_BT},
\label{rhoeq1}
\end{align}
where $2^n$ is the dimensionality of the $n$ qubit system. 
The first term neither contributes to the NMR signal nor evolves under unitary operations. The traceless second part is known as the deviation density matrix and is given by

\begin{equation}
\rho_{I} \approx -\sum_{i=0}^{2^n-1}\frac{\hbar\omega_{0i}}{2^nk_BT}I_{z}^i.
\label{rho_eq}
\end{equation}  
Here, $\epsilon = \frac{\hbar\omega_{0i}}{2^nk_BT}$ is the spin polarization. At room temperature and at typical $B_0$ strengths, $\epsilon \approx 10^{-5}$ for $n=1$. This suggests that the population difference between the energy levels is extremely low. Certain consequences due to the small $\epsilon$ values in liquid state NMR are summarized as below:
\begin{enumerate}
\item NMR states are highly mixed. However, it was shown that one can prepare pseudo-pure states from $\rho_I$ that can effectively mimic the pure states. Several methods like spatial averaging, logical labeling, temporal averaging have been proposed to realize pseudo-pure states \cite{pps1,pps2,pps3,pps4,nmrqst,pps5}.
\item A typical NMR signal is proportional to $\rho_{I}$. Since $\epsilon$ is inversely proportional to $2^n$, the signal intensity drops exponentially with $n$. This limits the NMR qubits to the small scale quantum simulators.
\item  As seen in Eq. \ref{rhoeq1}, the NMR state has a dominant contribution from the mixed state. 
It was shown by Peres that the  state is entangled if the eigenvalues of its partial trace are negative \cite{peres1996separability}. Also a more general study of $n$-qubit pseudo-entalged state was given by Braunstein {\em et. al.} \cite{braunstein1999separability}. They showed that  a pseudo-pure states can be non-seperable if $\epsilon > 1/(1+2^{n/2})$. However at room temperature, NMR states do not reach this non-seperable region. An entangled state obtained from the NMR pseudo-pure state is always a pseudo-entalged state. 
\end{enumerate}

NMR states are not limited to the thermal equilibrium states.
Various states can be prepared be the application of suitable quantum gates on $\rho_I$. 

\subsection{NMR gates \label{sec_gates_nmr}}
The mathematical description of quantum gates is given in section \ref{chap2_2}. In this section, I will explain the physical implementation of a unitary operator $U$ in the context of NMR. 

\subsubsection{Single qubit gates}
A typical NMR nuclei has  energy differences in the radio frequency (RF) range and hence any  single qubit gate can be realized by an RF pulse. Such an RF pulse is defined by its amplitude, phase and duration. 

The single qubit Hamiltonian in the lab frame under the action of an RF field is
\begin{align}
H(t) &= H_0+H_{rf} \nonumber \\
&=\omega_0I_z+ \omega_1[I_x\cos(\omega_{rf}t+\phi)+I_y\sin(\omega_{rf}t+\phi)],
\end{align}
where $\omega_1 = -\gamma B_1$ is the RF amplitude with $B_1$ being the amplitude of the applied field, $\omega_{rf}$ is the RF frequency and $I_x, I_y,I_z$ are the spin operators.

In many cases, it is customary to transform the time dependent $H(t)$ to a time independent Hamiltonian.  This transformation is obtained by the operator $U = e^{-i\omega_{rf}I_zt}$ and the effective time independent Hamiltonian is given by 
\begin{equation}
H_e = \Omega I_z+ \omega_1[I_x\cos(\phi)+I_y\sin(\phi)],
\end{equation}
where $\Omega = \omega_0-\omega_{rf}$ is the offset frequency.

An initial state $\rho(0)$ evolves under this $H_e$ as 
\begin{equation}
\rho(\tau) = e^{-iH_e \tau} \rho(0) e^{iH_e \tau}.
\end{equation}
For example, an on-resonant RF pulse with $\Omega=0$ and for $\phi=0$ corresponds to an operator $e^{-i\omega_1 I_x \tau}$. Here the amplitude and duration of the pulse is $\omega_1$ and $\tau$ respectively. The pulse can also be represented as $e^{-i\theta I_x}$ where $\theta = \omega_1 \tau$ is the rotation angle and thus has the form of a general rotation operator given by Eq. \ref{rotation}.

A convenient way to represent the evolution of any density operator is given by product operator formalism \cite{cavanagh}. Any state, such as the following, evolves for a time $t$ under $\Omega I_z$ as follows
\begin{align}
I_z \longrightarrow I_z.\nonumber \\
I_x \longrightarrow I_x\cos(\Omega t)+I_y\sin(\Omega t). 
\nonumber \\
I_y \longrightarrow I_y\cos(\Omega t)-I_x\sin(\Omega t).  
\end{align}
Similarly, the states under the action of the RF pulse given by $\theta I_{x}$ evolve as
\begin{align}
I_x \longrightarrow I_x.\nonumber \\
I_y \longrightarrow I_y\cos(\theta)+I_z\sin(\theta).  
\nonumber \\
I_z \longrightarrow I_z\cos(\theta)-I_y\sin(\theta).  
\label{state_rf}
\end{align}
and under the action of $\theta_1 I_y$ evolve as
\begin{align}
I_x \longrightarrow I_x\cos(\theta_1)-I_z\sin(\theta_1).  \nonumber \\
I_y \longrightarrow I_y.  
\nonumber \\
I_z \longrightarrow I_z\cos(\theta_1)+I_x\sin(\theta_1).  
\label{state_rf1}
\end{align}
A simple NOT gate corresponds to an on-resonant RF pulse with $\theta_1=\pi$ in Eq. \ref{state_rf1}. Similarly, any single qubit gate can be realized by various values of $\theta$, $\theta_1$ and $\Omega$.   

\subsubsection{Two qubit gates}
The internal Hamiltonian of a two qubit system is shown in Eq. \ref{H_weak}. A two qubit gate is realized by the evolution of the qubits under the action of the coupling strength $J$ as well as the external RF pulses. The action of RF pulses is the same as in Eqs. \ref{state_rf} and \ref{state_rf1}. However, the evolution of the density operators under the action of the two qubit coupling Hamiltonian $2\pi J I_z S_z$ (where $I$ and $S$ are the spin angular momentums of two different qubits) is given by
\begin{align}
I_x \longrightarrow I_x\cos(\Theta)+2I_yS_z\sin(\Theta),  
\nonumber \\
I_y \longrightarrow I_y\cos(\Theta)-2I_xS_z\sin(\Theta),  
\nonumber \\
I_z \longrightarrow I_z, \nonumber \\
2I_xS_z \longrightarrow 2I_xS_z\cos(\Theta)+I_y\sin(\Theta),  
\nonumber \\
2I_yS_z \longrightarrow 2I_yS_z\cos(\Theta)-I_x\sin(\Theta), 
\nonumber \\
2I_zS_z \longrightarrow 2I_zS_z,
\end{align}
where the rotation angle $\Theta=\pi J t$. While $J$ is fixed in any NMR system, any effective rotation angle can be realized by changing the pulse time $t$. 

\begin{figure}[h]
\begin{center}
\includegraphics[trim=1cm 1cm 1cm 1cm, clip=true, width=14cm]{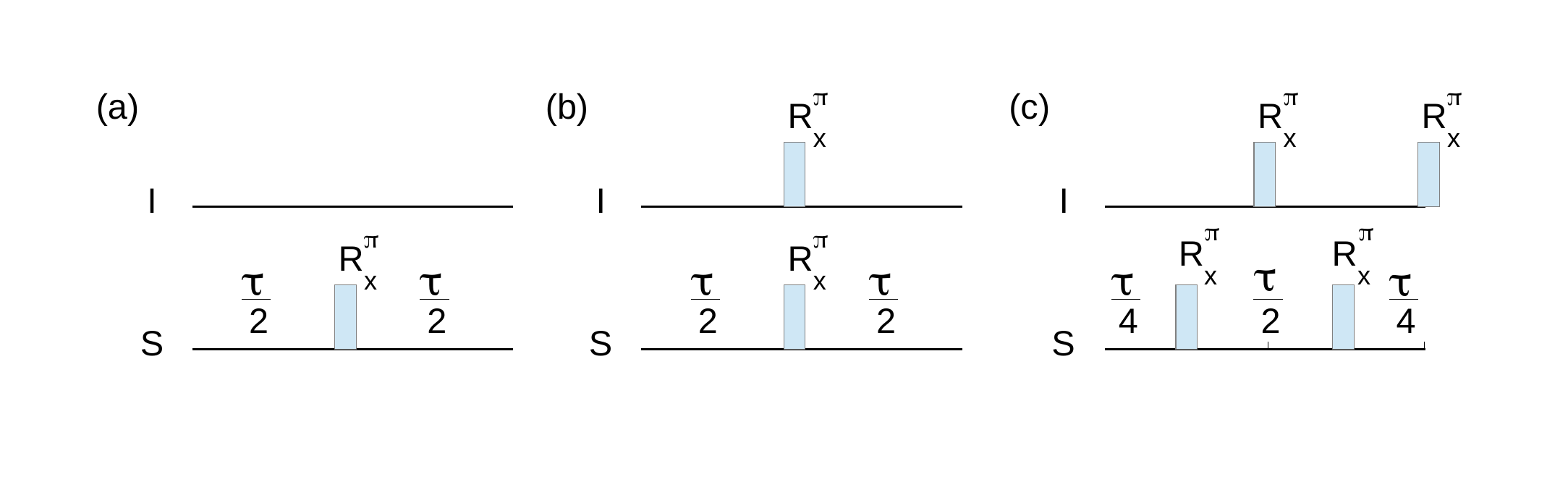}
\caption{Pulse sequences for refocusing (a) coupling strength  (b) chemical shifts (c) both chemical shifts and coupling strength over a time $\tau$. The RF pulses with rotation angles $\pi$ about $x$-axis are represented by $R^\pi_x$.}
\label{refocus}
\end{center}
\end{figure}

In addition, it is also possible to effectively cancel the evolution of the states under the chemical shift Hamiltonian and the coupling Hamiltonian. This technique is called as refocusing scheme and the corresponding pulse sequences are shown in Fig. \ref{refocus}. Fig. \ref{refocus}(a) is the standard Hahn echo sequences (also see section \ref{hahn_echo}). 

A standard two qubit gate is the CNOT gate as represented in Eq. \ref{Ucnot}. The corresponding NMR pulse sequence is given by
\begin{equation}
U_{CNOT} = (R^{\pi/2}_z)_{I}(R^{-\pi/2}_z)_{S}(R^{\pi/2}_x)_{S}U(\frac{1}{2J})(R^{\pi/2}_y)_{S},
\end{equation} 
where $U(t) = e^{-i2\pi J I_z S_zt}$ and the operator $R_{\hat{n}}^{\theta}$ is the rotaion of the spins about the axis $\hat{n}$ with rotation angle $\theta$.

Generally, the pulses that we normally implement are hard pulses. These correspond to short duration pulses and hence cover a larger frequency range. However in many cases, selection of a specific spin or its transition with a precise frequency is important. 
Although,  it might be possible to use long duration shaped pulses (e.g. Gaussian), they are not universally applicable and are prone to RF inhomogeinity. 
In such cases, it is recommended to use optimal control algorithms  to design selective and robust quantum gates.  
Such algorithms maximize the fidelity between the desired opererator and the RF operator by optimizing the control parameters such as rf amplitudes, phases,durations, delays, etc.  Fidelity is the overlap between the two operators and is a measure of how close the operators are. The fidelity F between the operators $U_1$ and $U_2$ is defined as 
\begin{equation}
F = |\mathrm{Tr}[U_1 U_2^\dagger]|/2^n.
\end{equation}
 Among many such optimal control algorithms are the strongly modulated pulses \cite{smp1,smp2,smp3}, GRAPE \cite{grape}, bang-bang control \cite{gaurav} etc. In this thesis, we have used GRAPE and bang-bang control.

In principle, any effective Hamiltonian can be realized using the internal Hamiltonian and RF Hamiltonian. 

\subsection{NMR measurements}

The average of the magnetic moments $\mu$ of the nuclei in thermal equilibrium at static magnetic field in the sample produces a bulk magnetization. Upon the perturbation of the nuclei by an external RF field, the bulk magnetization precess about the static magnetic field $B_0$ with a frequency $\omega_0$ and produces a time varying magnetic field.  This time varying magnetic field produces an electromotive field in the coils placed near the sample in accordance with the Faraday's law of induction. In the conventional NMR set up, the orientation of $B_0$ and the coils allows the detection of the transverse magnetization precessing about $B_0$. The signal that we observe  corresponds to the bulk transverse magnetization of the sample and thus NMR quantum computers are regarded as ensemble average quantum computers. 

The bulk magnetization that is recorded in the NMR experiment is given by
\begin{equation}
M(t) \propto \mathrm{Tr}[\rho(t)D],
\label{M_det}
\end{equation}
where $M(t)=M_x(t)+iM_y(t)$, $\rho(t)$ is the instantaneous state of the qubits and $D = \sum_{j=1}^{n}(I^j_x+iI^j_y)$ is the detection operator. 
The fourier transform of Eq. \ref{M_det} gives the signal in the frequency domain.

\begin{figure}[h]
\begin{center}
\includegraphics[trim=1cm 1cm 2cm 1cm, clip=true, width=14cm]{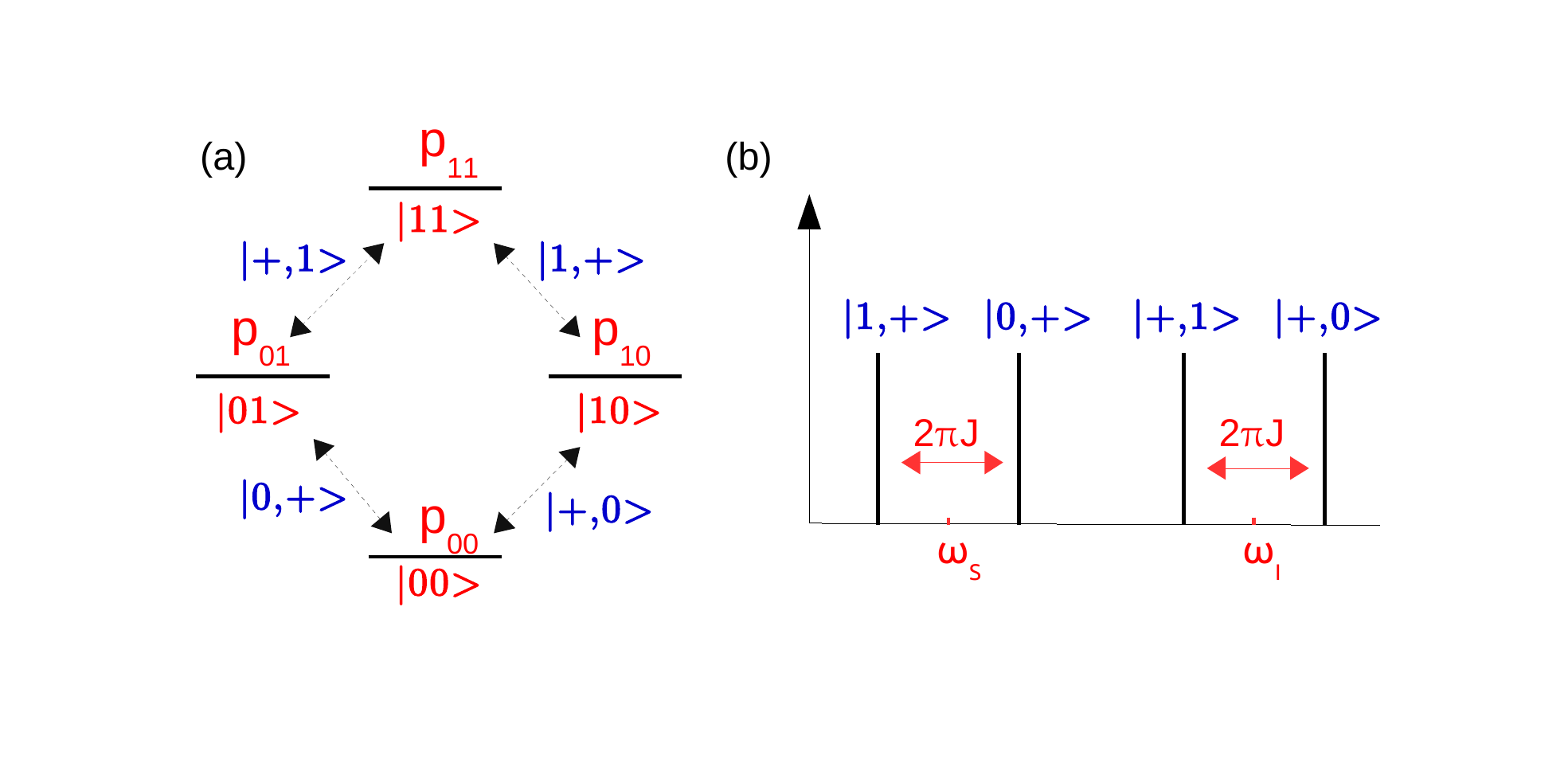}
\caption{(a) Energy lelvel diagram of a weakly coupled two qubit system at thermal equilibrium. (b) NMR read-out of the transitions corresponding to (a).}
\label{twoq_spectra}
\end{center}
\end{figure}

As an example, Fig. \ref{twoq_spectra} shows the read-out of the equilibrium spectra of a two qubit weakly coupled NMR system represented by spins $I, S$. The Hamiltonian $H_0$ of such a system is given by Eq. \ref{H_weak} with $n=2$. Fig. \ref{twoq_spectra}(a) shows the eigenstates of $H_0$ that are given by $|00\rangle,|01\rangle,|10\rangle,|11\rangle$ and the corresponding populations are given by $p_{00},p_{01},p_{10},p_{11}$ respectively. Here, the first and the second spins are represented by   $I$ and $S$ respectively. Only the transitions for which $\Delta m= \pm1$ are allowed and thus only four different transitions can be observed in this system. The labels $|0,+\rangle,|1,+\rangle$ indicate the transitions of spin $S$ when spin $I$ is in state $|0\rangle$, $|1\rangle$ respectively and  $|+,0\rangle,|+,1\rangle$ indicate the transitions of spin $I$ when spin $S$ is in state $|0\rangle$, $|1\rangle$ respectively. 
Fig. \ref{twoq_spectra}(b) shows the NMR signal with $D = \sum_jI_x^j$ for the transitions shown in Fig. \ref{twoq_spectra}(a). The positions of the peaks are given by the eigenvalues of $H_0$ in frequency units. $\omega_I, \omega_S$ are the larmor frequencies of spins $I,S$ respectively and $J$ is the coupling between them.

In general, an $n$ qubit weakly coupled system will have $2^n$ energy levels and a total of $n2^{n-1}$ transition lines can be observed. The area under the spectra gives the bulk transverse magnetization $M^x(t)$ of the corresponding spins. Although, the conventional NMR signal gives $M^{x/y}(t)$, it is also possible to measure the expectation values of any Hermitian operators, e.g, Moussa protocol \cite{moussa,sharad}.

By systematically measuring the transverse magnetization, the  state at any instant can be reconstructed using the quantum state tomography (QST) \cite{nmrqst,nmrqst1,nmrqst2}. As an explict demonstration, QST for a single qubit is explained below. A general single qubit state is represented by
\begin{equation}
\rho = \left(
\begin{array}{cc}
a & c+id \\
c-id & 1-a \\
\end{array}
\right), 
\label{rho_qst}
\end{equation}
where $a,(1-a)$ are the populations and $c,d$ are real numbers. Here, the single quantum coherence terms are the off-diagonal terms that can be directly observed.
The goal of QST is to recontruct \ref{rho_qst} and it involves obtaining the values of $a,c$ and $d$. This requires two experiments:
\begin{enumerate}
\item 
The real part of the spectra gives the value of $c$ and the imaginary part (that is obtained by a changing the spectrum phase by $\pi/2$) gives the value of $d$. 
\item Application of the pulse field gradient that destroys the coherence terms followed by a $R_x^{\pi/2}$ pulse give the values of $a$.
\end{enumerate}

This method can be generalized to $n$ qubits \cite{aqst}.



\section{NMR quantum computers vs Di Vincenzo criteria}

As mentioned in chapter \ref{chap1}, it is important for any quantum computer to follow Di Vincenzi criteria. In the following, I will briefly explain how well NMR quantum computers follow these criteria.  
\begin{enumerate}
\item The NMR signal intensity  exponentially decreases with the number of qubits. This poses a severe challenge in realizing a scalable quantum computer. As of present, the maximum number of NMR qubits that are realized in the lab is 12 \cite{mahesh_prl}. 
Although an estimate of about one hundred qubits is necessary to realize a large scale quantum computer, the problem of scalability persists in most of the present day quantum technologies.
\item   Due to very low spin polarization, NMR states are highly mixed as shown in Eq. \ref{rho_eq}. However, it is possible to mimic the pure states by preparing pseudo pure states \cite{pps1,pps2,pps3,pps4,nmrqst,pps5}.

\item The NMR qubit life times are characterized by the time scales called spin-lattice relaxation ($T_1$) and spin-spin relaxation ($T_2$). While $T_1$ refers to the energy relaxation, $T_2$ refers to the coherence decay of the qubits. Typically, these decay constants are of the orders of a few seconds. Since the gates are realized by RF pulses, the typical gate implementation times range from a few microseconds to milli seconds. This allows to implement hundreds of gates before the coherences decay.

\item Any effective Hamiltonian can be realized by the system Hamiltonian, and RF Hamiltonian, and thus NMR gates are universal.

\item The average transverse magnetization can be directly measured. From this data, one can reconstruct the quantum state (QST) or even measure the expectation values of other operators.   However, it is difficult to perform projective measurements in NMR. 
\end{enumerate}

Despite the challenges, NMR systems are widely used as efficient test beds for small scale quantum computers. Various algorithms have been successfully implemented using NMR since the concept of pseudo pure states has been put forth. The first quantum algorithm that was experimentally demonstrated using NMR was Deutsch algorithm \cite{Deutsch}. Later  Deutsch $–$ Jozsa algorithm was implemented in \cite{dj1,dj2,dj3}. Grover's algorithm was implemented for the first time in \cite{grover1}. Other related works were carried out by \cite{grover2,grover3,grover4}. One of the famous experiments was the Shor’s factorization algorithm  that factored the number 15 using a 7 qubit NMR system \cite{shor}. The experimental quantum simulations were performed for the first time by \cite{qs1} in 1999. Since then a large number of quantum simulations were experimentally performed using NMR \cite{tseng,qs3,qs4,qs5}.

\thispagestyle{empty}
\part{Unitary Control}

\chapter{Experimental realization  of ``dynamical many-body freezing''\label{chap4}}

In this chapter, I will explain the experimental work on the quantum simulation of a new phenomenon known as dynamical many-body freezing (DMF) using unitary controls \cite{freeze2}.

\section{Introduction}

Consider a classical system perturbed by external periodic drive with frequencies much higher than the characteristic frequencies of the system. In this high frequency regime, it is intuitive to note that the system does not get sufficient time to adjust itself within the drive period and hence does not respond to the external drive. This phenomenon of no-response is known as freezing and this reason behind freezing 
has already been adopted in various important results in classical as well as quantum physics \cite{kibble1976topology,zurek1985cosmological,damski2005simplest,zurek2005dynamics,dziarmaga2005dynamics,landau1991quantum,zener1932non,chakrabarti1999dynamic,damski2006adiabatic}.

However, recent theoretical studies have shown that this intuitive mechanism of freezing of quantum many-body systems, which implies strong freezing effects for higher drive frequencies, may fail in certain cases  due to the quantum interference of excitation amplitudes \cite{arnab,mondal2013dynamics,bhattacharyya2012transverse,das2012switching,bukov2015universal,suzuki2012quantum}.  
It was shown by A. Das in 2010 \cite{arnab} that when a 1-dimensional spin chain was driven by high drive frequencies, the spin chain exhibited a peculiar response behaviour as opposed to the classical case: while the classical systems showed a monotonic response to the drive, the quantum systems showed a peak-valley response behaviour indicating a non-monotonic response. 
Further, it was shown that for specific drive parameters, the spin chain froze for all times and for arbitrary initial states. This phenomenon is known as dynamical many-body freezing (DMF) \cite{freeze2}. 

\begin{figure}[b!]
\begin{center}
\includegraphics[trim = 20mm 20mm 20mm 10mm, clip, width=14.5cm]{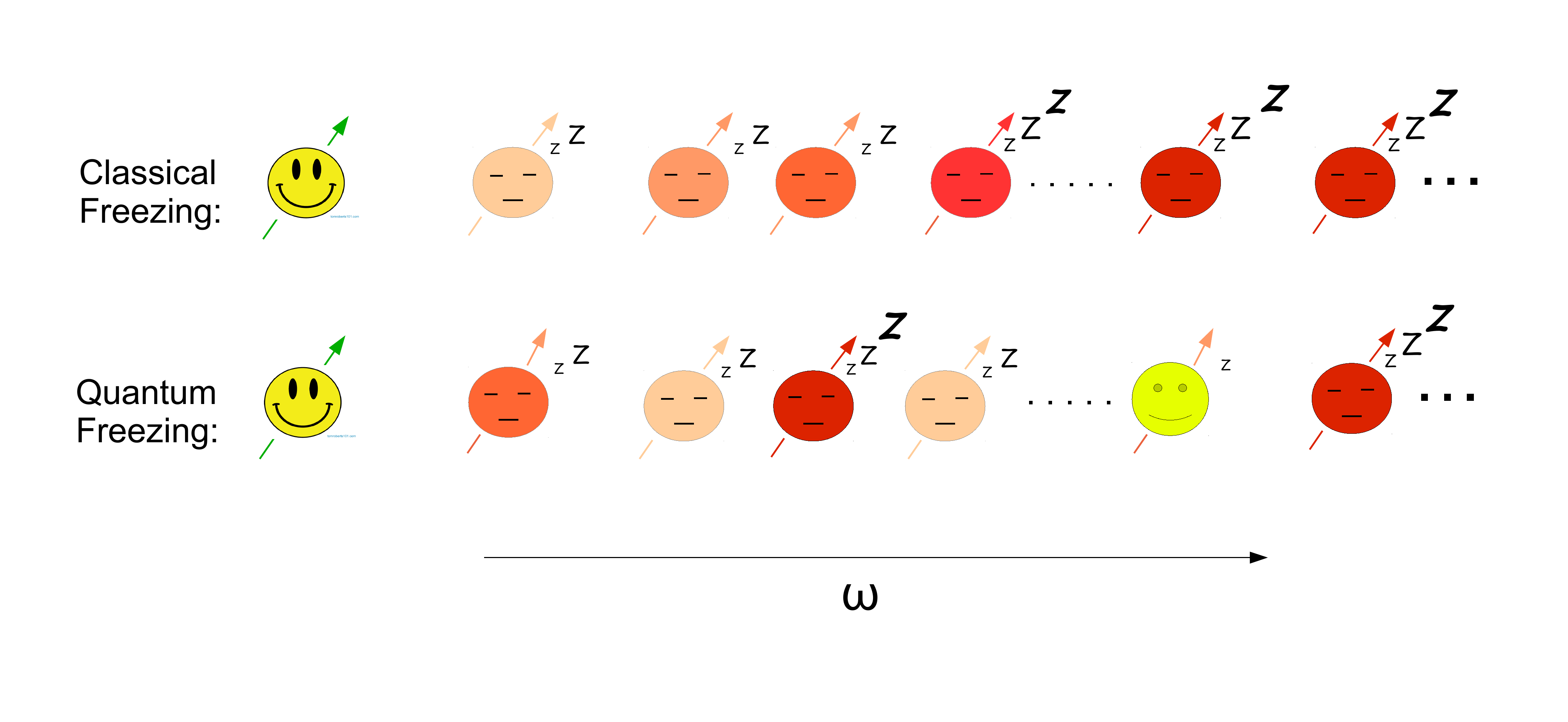}
\caption{Classical (top row) and quantum (bottom row) systems under the influence of external periodic drive corresponding to high frequency ($\omega$) regime.  Each face represents the composite many-body system. The horizontal axis corresponds to the drive frequency ($\omega$). The superscripts (`z's) refer to the response - the strength of freezing increases with the number of 'z's. 
The classical systems freeze for all high frequencies but the quantum systems freeze and respond non-monotonically.	
}
\label{fr_cartoon}
\end{center}
\end{figure}

The comparison between the classical and quantum case in high frequency regime is shown in Fig. \ref{fr_cartoon}. 

Freezing of the particle under the action of external periodic drive was previously observed. Examples include dynamical localization of a single particle \cite{dl} and coherent destruction of tunneling of a single particle \cite{cd}. However the phenomenon of dynamical many-body freezing differs from the above as follows: (1) it is a quantum many-body extension of dynamical localization and (2) the freezing occurs for all times and for arbitrary initial states for specific drive parameters.


In this chapter, I will explain the  experimental demonstration of this phenomenon that was carried out in our lab \cite{freeze2}.
The motivation to  demonstrate this phenomenon is two fold:
\begin{enumerate}
\item The field of driven quantum many-body systems is still largely unexplored. 
Despite the   experimental challenges, we  successfully simulated this phenomenon using a 3-qubit NMR simulator for the first time. 
\item The experimental feasibility of controlling the quantum systems by tuning the drive parameters opens up the possibilities of a novel quantum control technique.
\end{enumerate}

  Below, I will explain the basic outline of the phenomenon of DMF, and present theoretical, and  numerical results that encodes the non-monotonicity in the response of the driven quantum many-body systems. I will also numerically show how the main quantity of interest deviates in the presence of experimental errors and how it can be overcome. 


\section{Quantifying freezing}
This section gives the necessary details required to quantify the amount of freezing. 
We consider a quantum many-body system in one dimension that is evolving under a specific Hamiltoninan starting from an arbitrary initial state. By monitoring the magnetization corresponding to the instantaneous states at regular intervals, we quantify the amount of freezing for specific Hamiltonian parameters by calculating the long time average of the magnetization, called as dynamical order parameter $Q$ \cite{arnab}. We see that the non-monotonic response of the driven quantum many-body system by a periodic field is captured by $Q$.  

Consider an infinite one dimensional Ising spin chain subjected to a transverse periodic field. Such a system is described by the Hamiltonian
\begin{equation}
{\cal H}(t) = -\frac{1}{2}[{\cal J}\sum_i^{n-1} Z_iZ_{i+1}+h_0\cos(\omega t)\sum_i^n X_i],
\label{H_tic}
\end{equation}
where $n=\infty$ is the number of spins, ${\cal J}$ is the coupling between the nearest neighbouring spins, $h_0$ is the drive amplitude and $\omega$ is the drive frequency.

Starting from an initial state $\rho(0)$, the infinite one dimensional spin chain  evolves under the action of the Hamiltonian ${\cal H}(t)$. The final state is $\rho(t)$ and we study the response of the system in terms of its transverse magnetization $m^x(t)$. As previously mentioned, the quantity that characterizes the strength of freezing is  $Q$ which is defined as a long time average of $m^x(t)$ and is given by
\begin{equation}
Q = \lim_{{ T} \to \infty}\frac{1}{ T}\int\limits_{0}^{ T} m^x(t) dt,
\end{equation}
where ${ T}$ is the total evolution time. 

The freezing case requires that  $m^x(t)$ remains the same as $m^x(0)$ for all times $t$. Thus it implies  that $Q=1$ for the freezing case. However, when $m^x(t)$ oscillates, $Q<1$ and thus corresponds to the non-freezing case.

A closed form for $Q$ was analytically derived  by A. Das \cite{arnab} for an infinite spin Ising chain under the periodic boundary condition and is given by
\begin{equation}
Q_\infty = \frac{1}{1+|J_0(2h_0/\omega)|},
\label{Q}
\end{equation}
where $J_0$ is the zeroth order Bessel's function. Thus the non-monotonic feature of $J_0$ imposes non-monotonicity in $Q$.

\begin{figure}[h]
\begin{center}
\includegraphics[width=13.5cm]{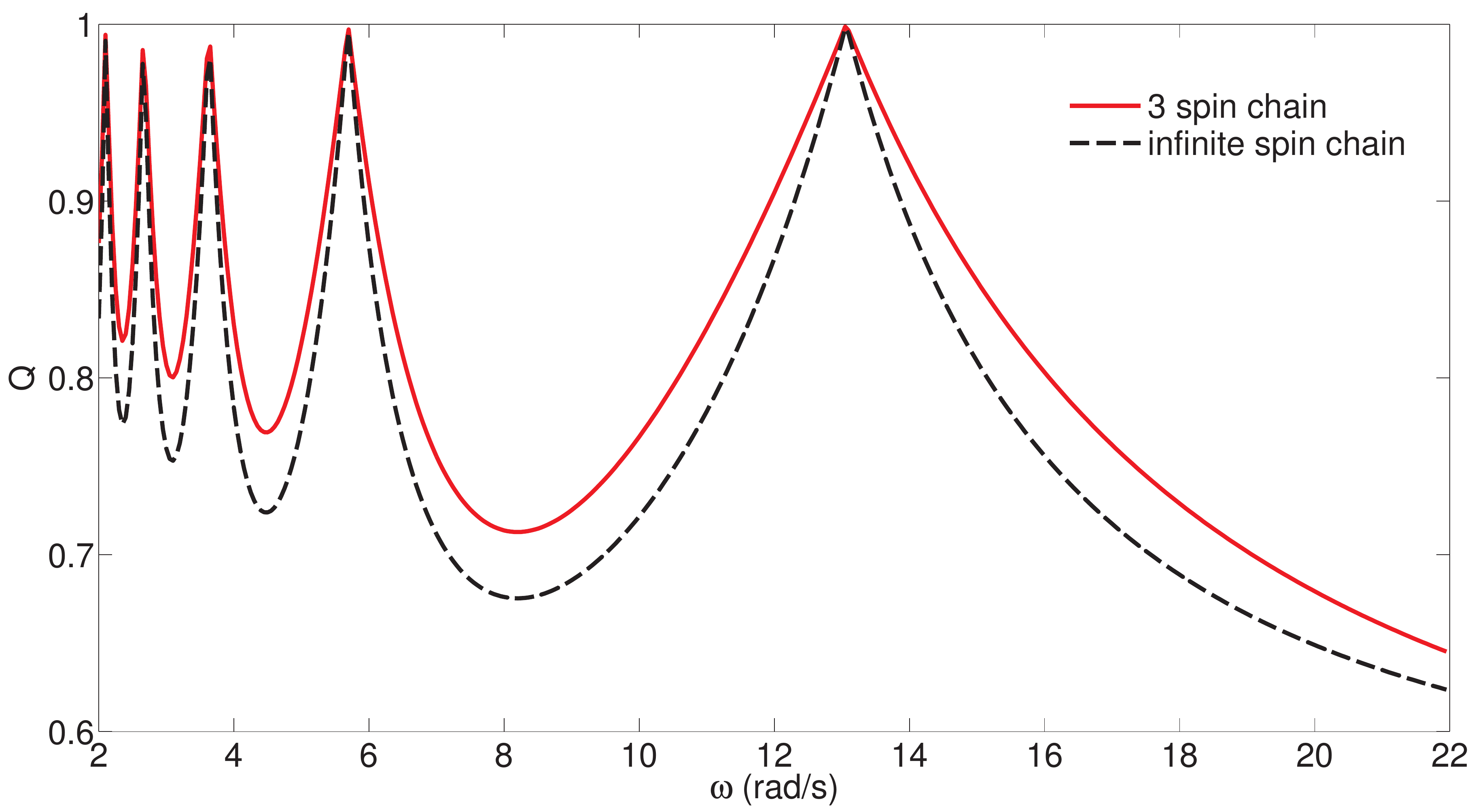}
\caption{The non-monotonic behaviour of $Q$ with $\omega$ for finite and infinite spin chain in the high frequency regime. The simulation is done for the parameters corresponding to $h_0 = 5\pi$ and ${\cal J} = h_0/20$, both in units of rad/s, that are consistent with strong ($h_0\gg {\cal J}$) and fast ($\omega\gg {2\cal J}$) drive scenario. 
}
\label{fr_Q}
\end{center}
\end{figure}

Fig. \ref{fr_Q} shows the numerical plot for $Q$ vs $\omega$. The dotted line corresponds to the $n=\infty$ case. This plot considers the high frequency regime where $\omega$ values are much higher than the maximum characteristic frequency  of the system that is given by $2{\cal J}$.

Similarly, the analytical form for $Q$ for a finite spin chain ($n=3$) was shown to be \cite{freeze2}
\begin{equation}
Q_3 = \frac{1+|J_0(2h_0/\omega)|}{1+3|J_0(2h_0/\omega)|}.
\label{Q3}
\end{equation}
As seen from Fig. \ref{fr_Q}, the  $Q$ vs $\omega$ plot for $n=3$ is  similar to that of the infinite spin chain.  
This feature of $Q$ being independent of $n$ as is reflected in Eq. \ref{Q} and \ref{Q3} allowed us to study this phenomenon using a small scale 3-qubit NMR quantum simulator.

\subsection{Experimental challenges}

The heart of quantum simulation protocol lies in the efficient implementation of the dynamics corresponding to a specific Hamiltonian.  
Here, the Hamiltonian of interest is given by Eq. \ref{H_tic} and the corresponding unitary operator is $U(t) = {\cal T} e^{-i\int_0^t{\cal H}(t') dt'}$ where ${\cal T}$ is the time ordering operator. 
In NMR setup, this $U(t)$ is realized by RF pulses with definite amplitudes and phases. However, in practice, realizing $U(t)$ for a specific implementation time $t$ is a challenging problem due  to the external  imperfect pulses caused by RF inhomogeinity and inherent decoherence. These lead to the faster decay of magnetization and is undesirable.

\begin{figure}[t!]
\begin{center}
\includegraphics[width=12cm]{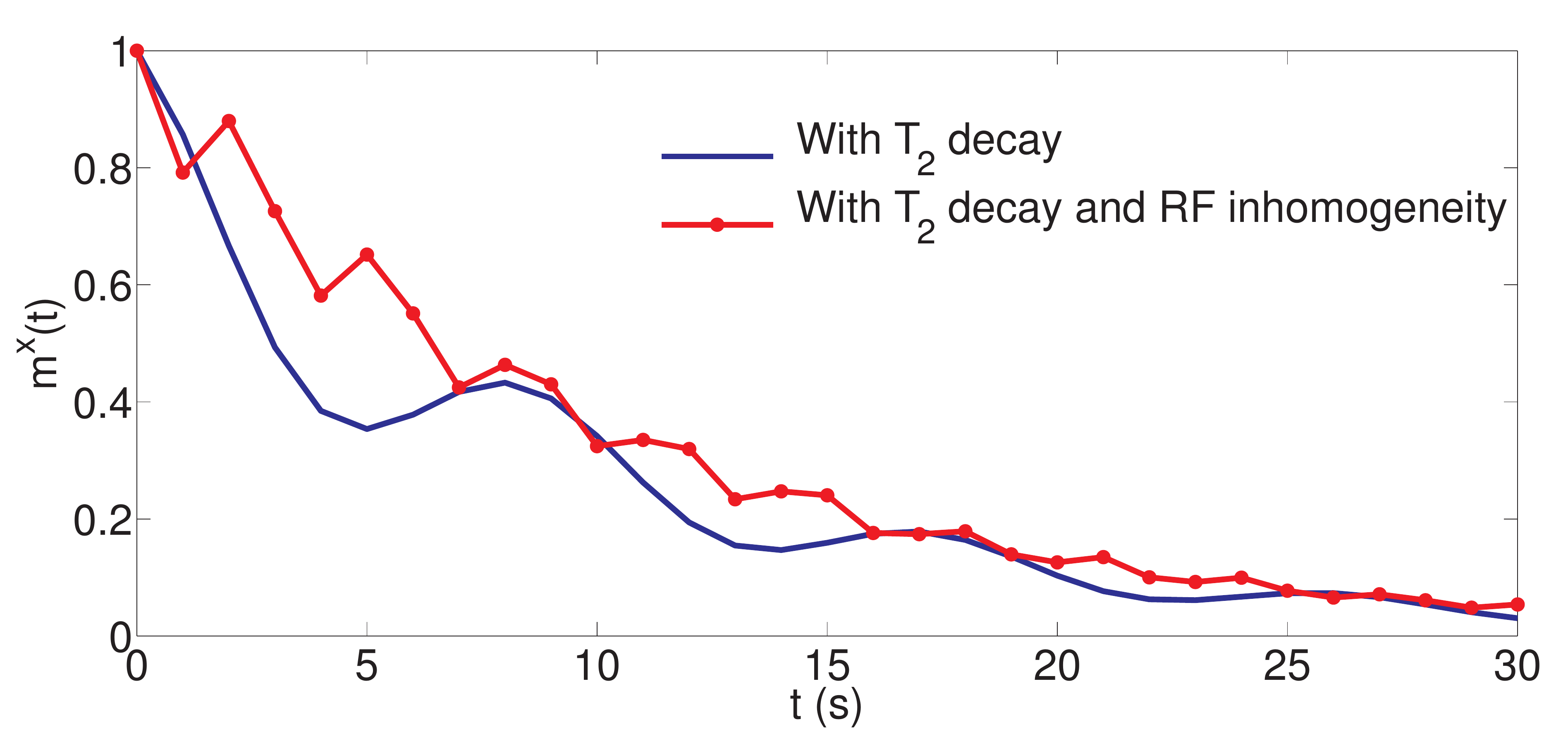}
\caption{Numerical simulation of the evolution of magnetization for $\omega=8.4$ and in the presence of errors for $h_0 = 5\pi$ and ${\cal J} = h_0/20$ starting from an initial state $\rho(0) = \sum_i^3 X_i/2$. 
}
\label{m_rfa}
\end{center}
\end{figure}

Fig. \ref{m_rfa} shows the numerical simulation of $m^x(t)$ in the presence of errors.  
Typically, NMR systems experience RF inhomogeneities that vary from $10\%$ to  $20\%$ depending on the type of  spectrometer probe.  Specific to our probe, on which the experiments were performed, the RF inhomogeneity was measured to be upto $20 \%$. The standard protocol to measure such probe specific RF inhomogeneities is done by measuring the Rabi oscillations - this is done by measuring the magnetization intensities for various pulse durations. These oscillations do not decay in the ideal scenario, however, with RF inhomogeneities the oscillations tend to decay. The fourier transform of the decayed oscillation  corresponds to the RF inhomogeneity distribution.  Also as shown in \cite{shukla2012characterization}, RF inhomogeneities can be mapped quantitatively with the help of Torrey oscillations. This means that quantum operations on some percent of the sample will differ from the ideal intended operations.  
 Thus by incorporating $20\%$  RF  inhomogeneity and a  decay constant with $T_2 = 10s$, we see that the non-freezing point corresponding to $\omega = 8.4$ rad/s 
\begin{figure}[b!]
\begin{center}
\includegraphics[width=12cm]{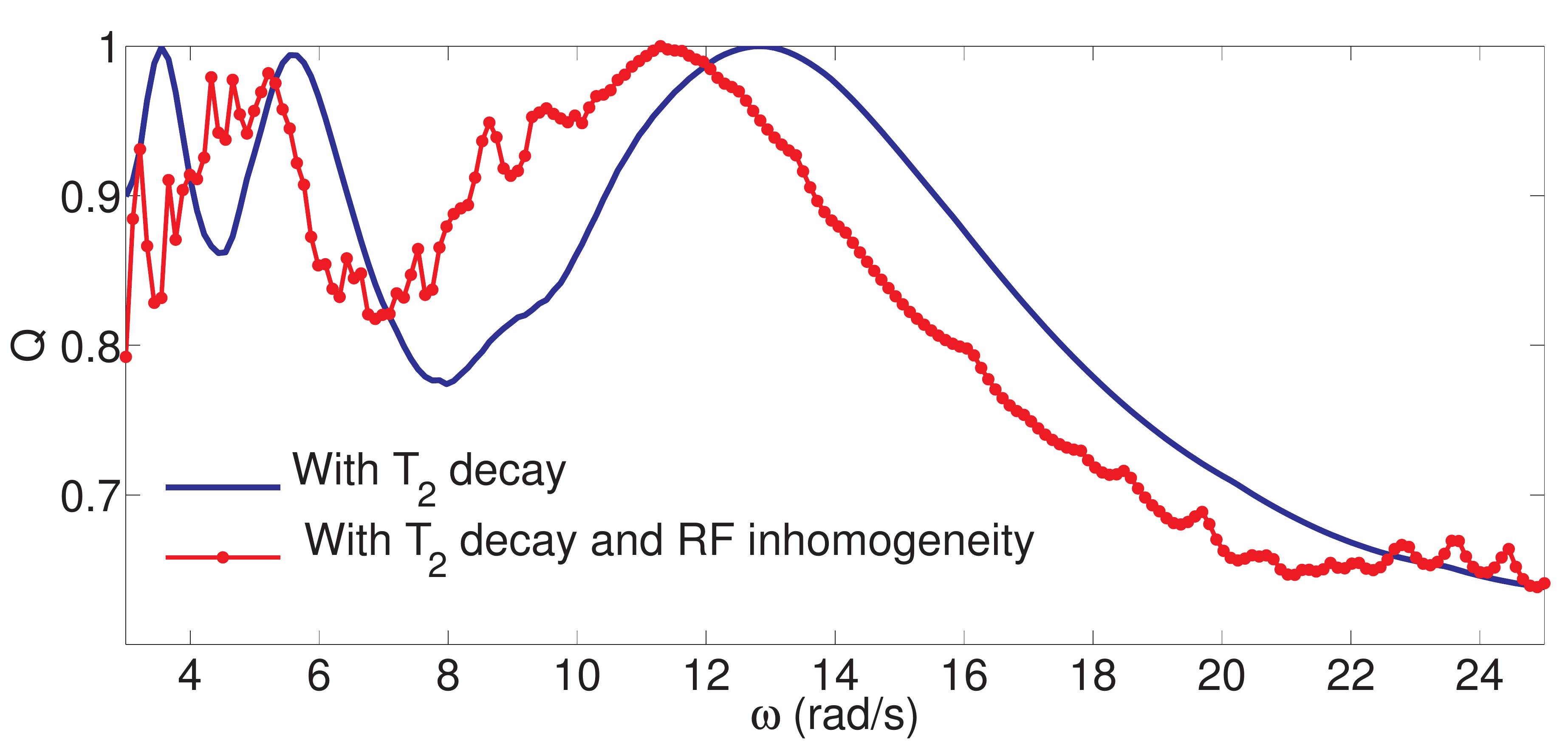}
\caption{ Numerical simulation for $Q$ vs $\omega$ by incorporating errors for $h_0 = 5\pi$ and ${\cal J} = h_0/20$ starting from an initial state $\rho(0) = \sum_i^3 X_i/2$. 
}
\label{q_rfa}
\end{center}
\end{figure}
is 
adversely affected by the errors with no sign of oscillations in $m^x(t)$. 

Similarly, Fig. \ref{q_rfa} shows how the effect of RF  inhomogeneity changes the freezing points. While the plots with only $T_2$ decay still captures the response correctly, the plots with $T_2$ decay and  RF  inhomogeneity show an erratic response. This
indicates that the adverse effects caused by pulses imperfections is much worse than that due  to the decoherence
effects. 

\textit{Thus experimental implementation of this phenomenon demands for an efficient control technique that is robust against RF  inhomogeneities.
}

\subsection{Overcoming the challenge \label{sec_grape}}
In order to circumvent the above problem, we used an optimal control algorithm called GRadient Ascent Pulse Engineering (GRAPE) \cite{grape}. This algorithm generates robust, high fidelity amplitude and phase modulated RF pulses. 

Consider an $n$- qubit NMR system defined by the Hamiltonian:
\begin{equation}
H(t) = H_0+\sum_{k=1}^m u_k(t) H_k,
\end{equation}
where $H_0$ is the internal Hamiltonian, $H_k$ is the RF Hamiltonian and $u_k(t)$ correspond to the RF amplitudes that can be changed and controlled. Let $\rho(0)$ be the initial state and the evolution of the state $\rho(t)$ under this Hamiltonian is given by
\begin{equation}
\dot{\rho}(t) = -i[H(t),\rho(t)].
\end{equation}
The GRAPE algorithm aims to find the optimal $u_k(t)$ values  that take the state $\rho(0)$ to $\rho(T)$ by maximizing the performance function $\Phi_0$ between the the final state $\rho(T)$ with the desired target state $C$. The performance function is given by
\begin{equation}
\Phi_0 = \langle C|\rho(T)\rangle = \mathrm{Tr}[C^\dagger\rho(T)].
\end{equation}

Let the total time be discretized into $N$ equal steps, each of duration $\Delta t=T/N$ and the control amplitudes $u_k$ are assumed to be constant in each time step. The propagator corresponding to the $j$th step is given by
\begin{equation}
U_j = e^{-i\Delta t H(t)}.
\end{equation}
The final state is $\rho(T) = U_N\cdots U_1\rho(0) U_1^\dagger\cdots U_N^\dagger$ and $\Phi_0$ becomes
\begin{equation}
\Phi_0 = \langle C|U_N\cdots U_1\rho(0) U_1^\dagger\cdots U_N^\dagger\rangle = \langle \lambda_j|\rho_j\rangle,
\end{equation}
where $\lambda_j = U_{j+1}^\dagger\cdots U_N^\dagger C U_N\cdots U_{j+1}$ is the backward propagated target operator and $\rho_j = U_j\cdots U_1\rho(0) U_1^\dagger\cdots U_j^\dagger$ is the density operator at time $j\Delta t$. 

Let $u_k$ for the $j$th step be $u_k(j)$. Suppose $u_k(j)$ is changed to $u_k(j)+\delta u_k(j)$ where $\delta$ is the small perturbation. 
It was shown in \cite{grape} that $\Phi_0$ could be increased if the change in $u_k(j)$ was such that
\begin{equation}
u_k(j) \longrightarrow u_k(j)+\epsilon \frac{\delta \Phi_0}{\delta u_k(j)},
\label{ujk}
\end{equation} 
where $\frac{\delta \Phi_0}{\delta u_k(j)} = -\langle \lambda_j | i \Delta t[H_k,\rho_j]\rangle$ and $\epsilon$ is the small step size.

The first step of the algorithm is to initialize $u_k(j)$ to guess values. This is followed by calculating $\rho_j$ where $\rho_j = U_j\cdots U_1 \rho_0 U_1^\dagger\cdots U_j^\dagger$ for $j\le N$.
The next step is to calculate $\lambda_j$ starting from $\lambda_N=C$. Using this $\lambda_j$, calculate $\frac{\delta \Phi_0}{\delta u_k(j)}$ and update all the values of $u_k(j)$ according to Eq. \ref{ujk}. With these updated $u_k(j)$, repeat the algorithm starting from the calculation of $\rho_j$ until the desired fidelity is achieved.

\section{Experiments}
\begin{figure}[h]
\begin{center}
\includegraphics[trim = 00mm 0mm 70mm 00mm, clip, width=4cm]{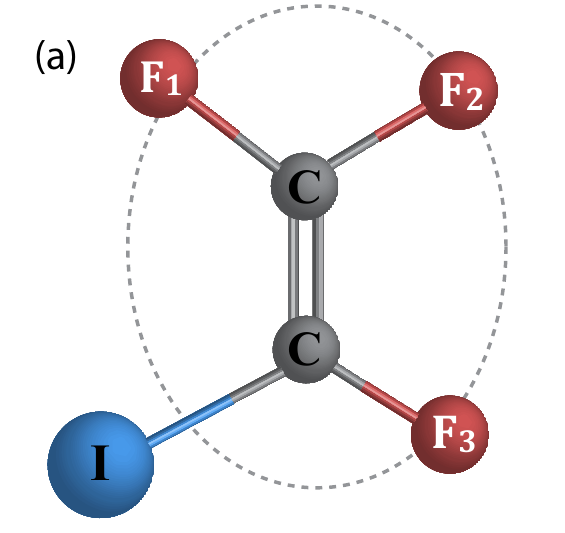}
\includegraphics[trim = 55mm 8mm 00mm 10mm, clip, width=7.0cm]{fr_fig1.pdf}
\caption{Molecular structure of trifluoroiodoethylene. The quantum simulator consists of F$_1$, F$_2$ and F$_3$. Their chemical shifts (diagonal elements) and the scalar couplings (off diagonal elements) in the units of Hz are shown at the right  (Figure from [30]).}
\label{fr_mol}
\end{center}
\end{figure}

We considered the three $^{19}$F nuclear spins in the molecule trifluoroiodoethylene as NMR quantum simulator and its  properties are shown in Fig. \ref{fr_mol}. The molecule is dissolved in acetone-D6 and all the experiments were carried out in Bruker 500 MHz NMR spectrometer at an ambient temperature of 290 K.

\begin{figure}[h!]
\begin{center}
\includegraphics[trim = 20mm 10mm 158mm 165mm, clip, width=8cm]{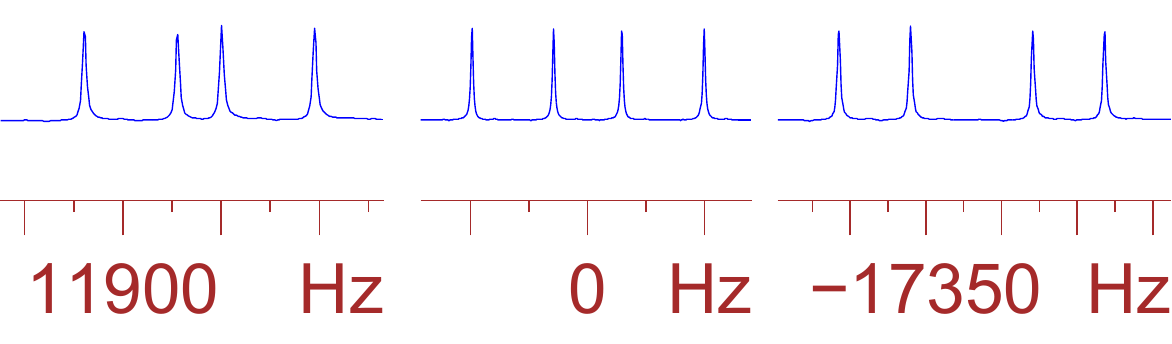}
\caption{ Thermal equilibrium spectra of $F_1$, $F_2$ and $F_3$  (left to right respectively).
}
\end{center}
\end{figure}

The internal Hamiltonian for this NMR system  is given by
\begin{equation}
H_{0} = -\pi \sum\limits_{i=1}^{3}\nu_i Z_i 
+ \frac{\pi}{2} \sum\limits_{\substack{i=1 \\ i<j}}^{3} J_{ij} Z_iZ_{j},
\label{hint}
\end{equation}
where the first term is the Zeeman Hamiltonian and the second term is the spin-spin interaction Hamiltonian. The chemical shifts $\nu_i$ in the rotating frame and the scalar couplings $J_{ij}$ are shown in Fig. \ref{fr_mol} 


\subsection{Quantum simulation}
The basic idea of quantum simulation is to use the NMR simulator along with the external controls to mimic the dynamics of a quantum many-body system evolving under Eq. \ref{H_tic}.
The simulation protocol for the 3-qubit  simulator involves the following main steps:
\begin{enumerate}
\item Initial state  preparation: 

We performed two sets of experiments with different initial states i.e., for  $\rho_1(0) = \sum_i^3 X_i/2$ and for $\rho_2(0) = \sum_i^3 [X_i/2+Z_i\sqrt{3}/2]$. These correspond to the initial transverse magnetization values $m^x(0) = 1 $ and $m^x(0) = 0.5 $ respectively since $m^x(0) = \mathrm{Tr}[\rho(0) (\sum_i^3X_i/2)]$. 

Experimentally, the states $\rho_1(0)$ and $\rho_2(0)$ were prepared by applying RF pulses on the three $^{19}$F spins with rotation angles $\pi/2$ and $\pi/6$ respectively about the $y$-axis.

\item Unitary implementation:

In order to be consistent with the fast drive, we specifically chose the Hamiltonian parameters  as follows: $h_0 = 5\pi$, ${\cal J} = h_0/20$. Note that $\omega\gg 2{\cal J}$, where $2{\cal J}$ is the maximum characteristic frequency of the model system.

Consider  a particular value of $\omega$ in the high frequency regime. For these specific values of $h_0$, ${\cal J}$ and $\omega$, we constructed ${\cal H}(t)$.
We implemented this dynamics by generating the corresponding $U(\tau)$ for a time $\tau = 2\pi/\omega$. Since the terms in ${\cal H}(t)$ do not commute with each other, the simulation under $U(t)$ requires discretization of $t$ into smaller time intervals. We discretized $\tau$ into 11 equal intervals and thus $U(\tau) = U_{11}\cdots U_2 U_1$ where each $U_j = e^{-i {\cal H}(m)m }$ with $m = \tau/11$. Thus the dynamics was realized by implementing $U(\tau)$ $j$ times with $j = 0,1,\cdots, N$ where $N=30$ for a total time of $T = j\tau$. Thus $\rho(0)$ evolves under $U(j \tau)$ as 

\begin{equation}
\rho(j\tau) = U(j\tau)\rho(0)U(j\tau)^\dagger = U(\tau)^j\rho(0)[U(\tau)^\dagger]^j.
\end{equation}

Three important steps contribute in  simulating ${\cal H}(t)$:
The first step is to cancel the evolution under the Zeeman Hamiltonian in Eq. \ref{hint}. The second step is to to realize an effective interaction Hamiltonian with strength ${\cal J}$. The final step is to implement an periodic drive $[-h_0/2]\cos(\omega t)$ about the $x$-axis. All in all, the simulation problem boils down to the realization of 
${\cal H}(t)$ using $H_0$ and external RF controls.  

We realized all the operators $U(\tau)$ by low power GRAPE pulses with durations ranging from a 5ms to to 10ms that were optimized by considering $20\%$ RF inhomogeneity. 
We considered the following RF inhomogeneity distribution: $4.31\%$ of the sample gets an RF inhomogeneity of 0.8, $0.81\%$ of the sample gets an RF inhomogeneity of 0.9, $75.32\%$ of the sample gets no RF inhomogeneity, $8.01\%$ of the sample gets an RF inhomogeneity of 1.1, $4.26\%$ of the sample gets an RF inhomogeneity of 1.2. The GRAPE algorithm first finds an optimized unitary for each of these sample volumes subjected to specific RF inhomogeneity. We calculated the average of these fidelity between the optmized unitary and the target unitary.  The average fidelity for all the pulses  were greater than or equal to 0.99. Fig. \ref{grape_u} shows the GRAPE pulse for a specific unitary as explained in the caption.

\begin{figure}[t!]
\begin{center}
\includegraphics[trim = 0mm 0mm 0mm 0mm, clip, width=12cm]{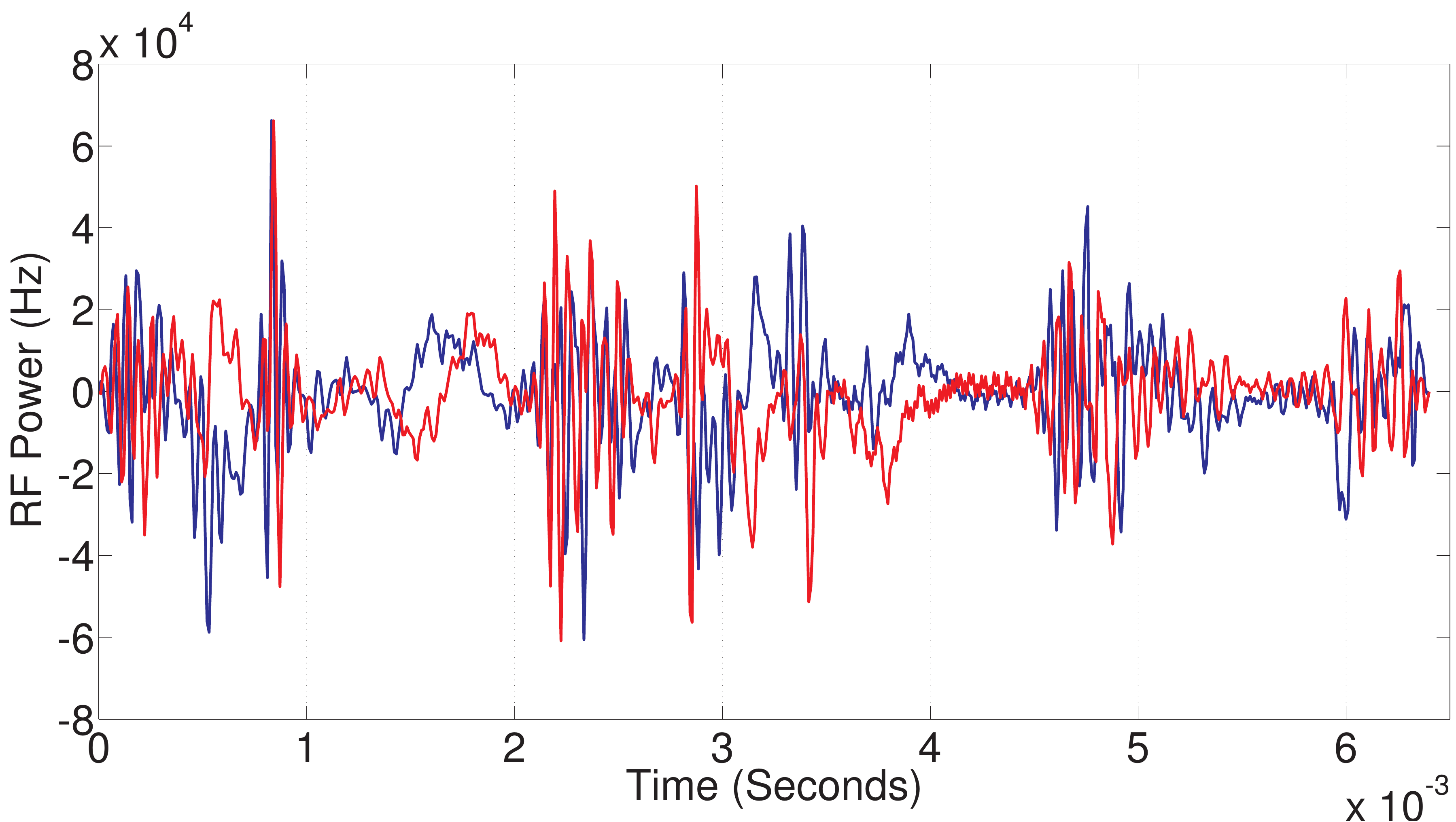}
\caption{The optimized  control field values for $\omega = 5.61$ rad/s as generated by GRAPE for one cycle (corresponding to $U(\tau)$ with $\tau = 2\pi/\omega$).  The blue and red plots correspond to the the x and y components of the control field $u$. 
}
\label{grape_u}
\end{center}
\end{figure}

\item Read-out: 

We measured $m^x(j\tau)$ at regular intervals $t = j\tau$ with $j = 0,1,\cdots, N$ which is given by
\begin{equation}
m^x(j\tau) = \mathrm{Tr}[\rho(j\tau) \sum_i^3X_i/2],
\label{fr_m}
\end{equation}
and hence $Q$ becomes
\begin{equation}
Q = \frac{1}{N+1}\sum_{j=0}^{N} m^x(j\tau).
\label{q}
\end{equation}

\end{enumerate}


\section{Results}
\subsection{Raw experimental results} 
\begin{figure}[b!]
\begin{center}
\includegraphics[trim = 20mm 10mm 158mm 20mm, clip, width=6cm]{fr_fig4.pdf}
\includegraphics[trim = 0mm 19mm 0mm 14mm, clip, width=1.45cm]{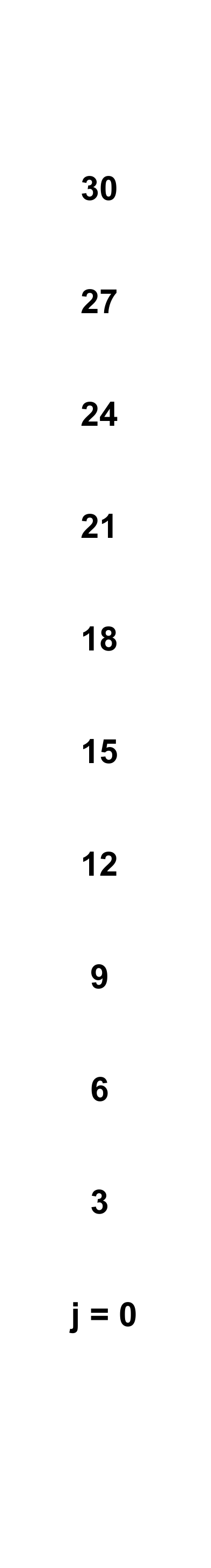}
\includegraphics[trim = 158mm 10mm 20mm 20mm, clip, width=6cm]{fr_fig4.pdf}
\caption{ $^{19}$F spectra for $\omega = 24.54$ rad/s (left) and $\omega   = 5.61$ rad/s (right) corresponding to non-freezing and freezing case respectively starting from $m^x(0)=1$. Here $j= t/\tau$ indicates the number of unitary operations on $^{19}$F spins starting from $j=0$ that corresponds to the equilibrium case. 
}
\label{fr_fig4}
\end{center}
\end{figure}

Fig. \ref{fr_fig4} shows the experimental $^{19}$F spectra corresponding to two different drive frequencies.  The spectra are observed at different instants of time corresponding to Eq. \ref{fr_m} for certain $j$ values.  Starting from the thermal equlibrium state (bottom spectra) with $j=0$, the evolution of the magnetization for different $j$ is indicated in the figure. The left column correspond to the non-freezing case with $\omega = 24.54$ rad/s  and the right column correspond to the freezing case with $\omega = 5.61$ rad/s. While the spectra at the left oscillate with $j$, the spectra at the right remains the same for all $j$. However an overall decay can be seen in both the cases. This decay is due to the phenomenon of decoherence, transverse relaxation of the nuclear spin and RF inhomogeneities. Here the term `raw' refers to the direct experimental results without incorporating any corrections by using numerical processing.

The sum of the area under the spectra of all the three $^{19}$F spins is proportional to the  magnetization $m^x(t)$. The evolution of $m^x(t)$ for different freezing ($\omega = 5.61$ and  $12.88$ rad/s) and non-freezing ($\omega = 8.40$ and $24.54$ rad/s) cases is shown in Fig. \ref{fr_fig2}.  The solid circles indicate the raw experimental results and the solid line indicates the numerical simulation. The decay in magnetization as explained with reference to Fig. \ref{fr_fig4} is reflected in the solid circles.  This decay is corrected by processing the experimental results and is explained in the below subsection.

\subsection{Inverse decay \label{inv_dec}}
In any physical set-up, the interaction of the system qubits with the environment is unavoidable. This leads to the loss of  coherence and is called as decoherence (see chapter \ref{chap5}). The decoherence time scales are characterized by $T_2$ values and theses values for the nuclei  $F_1$, $F_2$, $F_3$ are found to be $2.8$ s, $3.1$ s, $3.3$ s respectively. 

The  decoherence process, longer pulse implementation times, lower fidelity gates and RF  inhomogeneities result in the decay of magnetization as seen in Fig. \ref{fr_fig4}, despite the fact that the gates were optimized for shorter durations with fidelities above $99\%$ and the pulses were made robust even in the presence of $20\%$ RF  inhomogeneity.

 Fig. \ref{fr_m2} shows the evolution of $m^x(t)$ for freezing and non-freezing case. The dots are the experimental $m^x(t)$ values  and it can be noted that the decay corresponds to the decay of amplitude and an overall decay of $m^x(t)$. Below, I will explain the decay model and its correction using inverse decay method using two steps that we implemented in our work.
\begin{enumerate}
\item Suppose $T_d$ is the relaxation time due to all the contributing factors. The decay of $m^x(t)$ is modeled as
\begin{equation}
m^x(t) = \alpha+[\beta+\gamma\cos(ct)]e^{-t/T_d},
\label{decay_model}
\end{equation}
where $\alpha$, $\beta $, $c$ and $T_d$ are the fitting parameters. We calculated these values by fitting the experimental results to Eq. \ref{decay_model}. 
\begin{figure}[h]
\begin{center}
\includegraphics[trim = 0mm 0mm 00mm 00mm, clip, width=14cm]{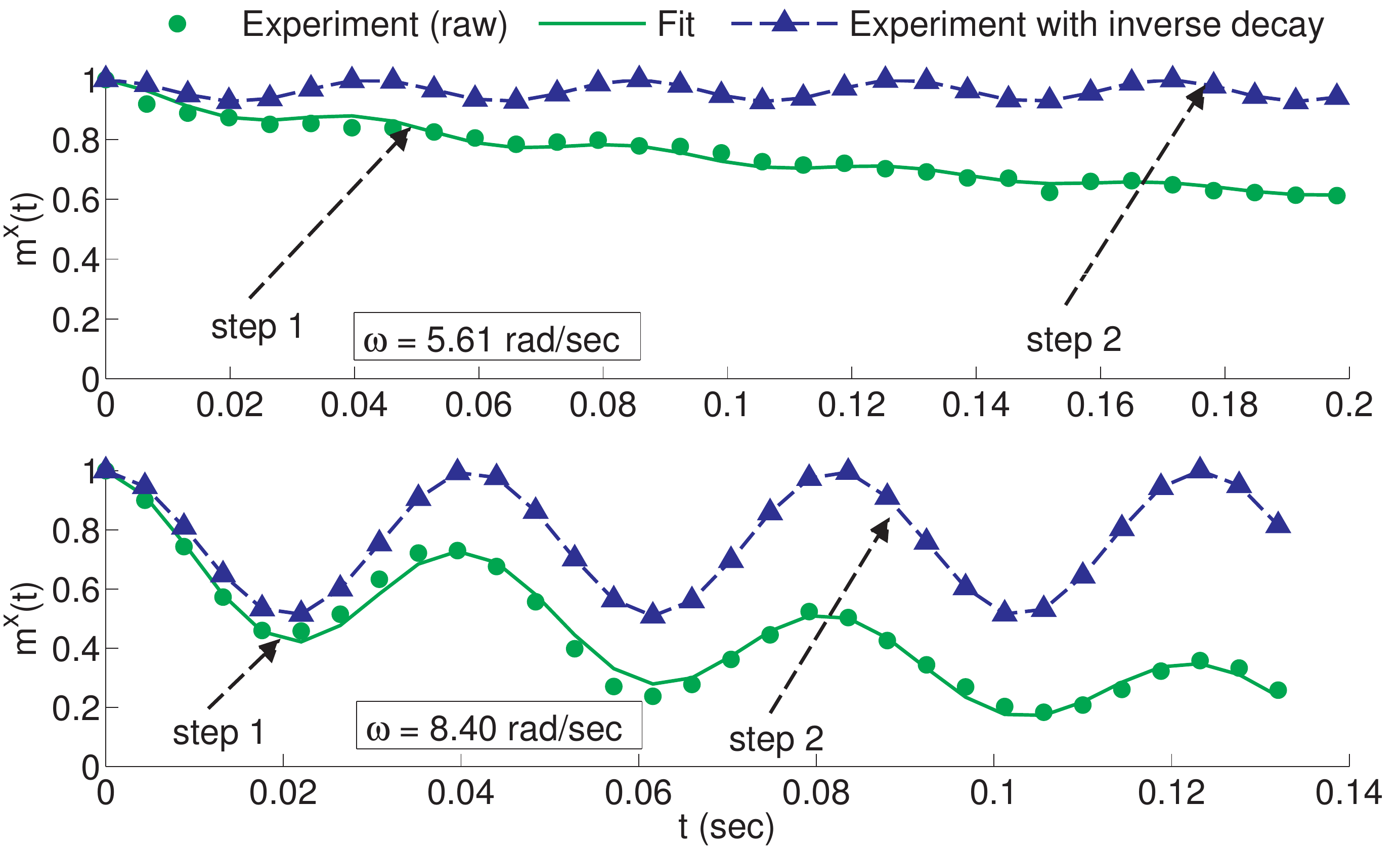}
\caption{Inverse decay method.
}
\label{fr_m2}
\end{center}
\end{figure}
The solid lines connecting the dots in Fig. \ref{fr_m2} correspond to the  fit. 
\item Consider an ideal case where there is no relaxation, i.e., $T_d\to\infty$. In this limit, we processed the evolution of $m^x(t)$ by using the values of $\alpha$, $\beta$ and $c$ that were calculated in step 1. The solid triangles connected by dotted lines indicate the decay corrected $m^x(t)$ values.
\end{enumerate}

\subsection{Theory vs experiments}
In this section, I will show how the experimental results match with the theory. 

\begin{figure}[h!]
\begin{center}
\includegraphics[trim = 30mm 0mm 00mm 00mm, clip, width=15cm]{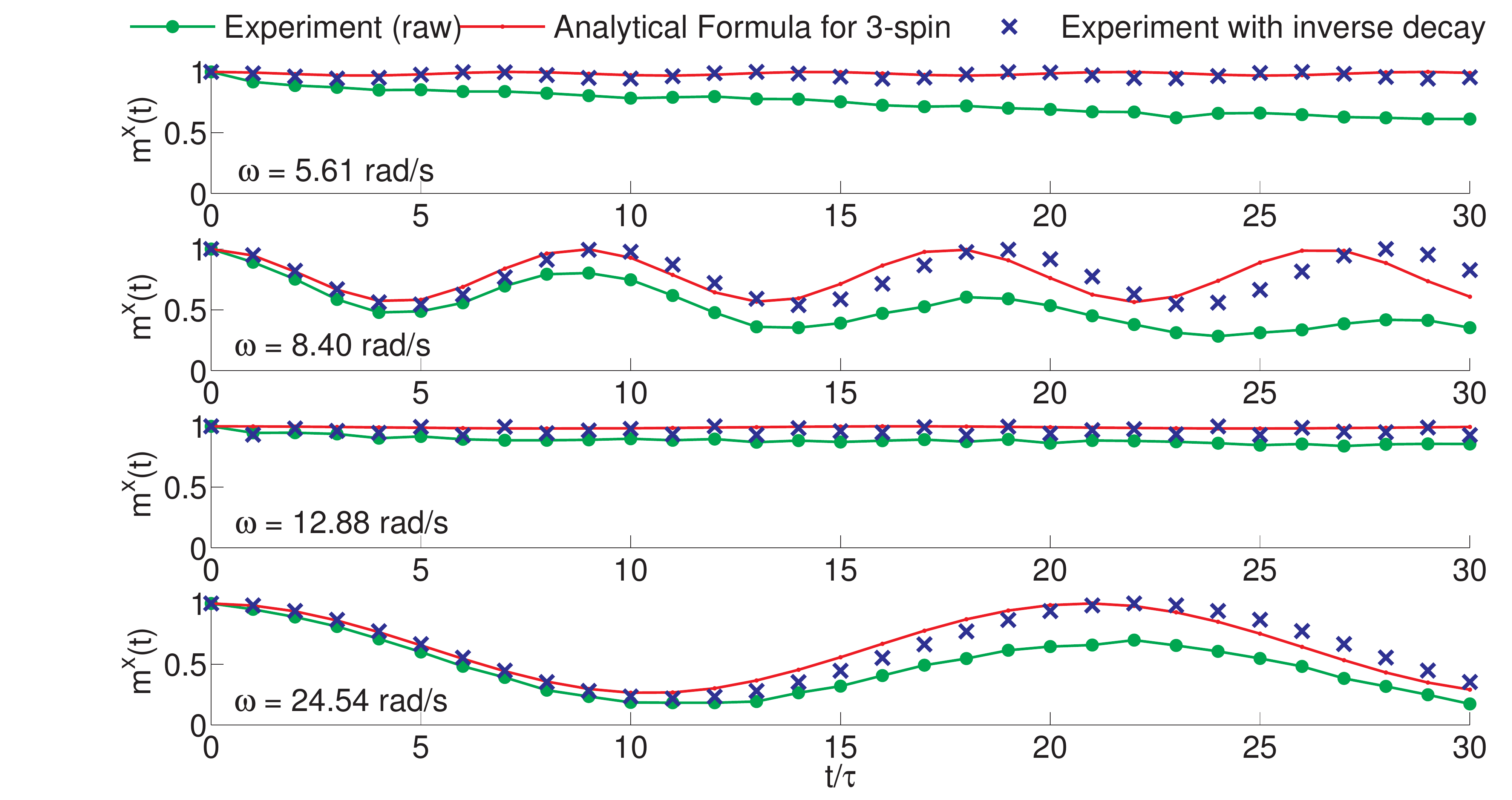}
\caption{Magnetization evolution for various $\omega$ values corresponding to freezing and non-freezing cases. All the plots correspond to an initial magnetization $m^x(0)=1$.
}
\label{fr_fig2}
\end{center}
\end{figure}

Fig. \ref{fr_fig2} shows the behaviour of the magnetization evolution for various $\omega$ values in the high frequency regime. 
$m^x(t)$ plots in each row of Fig. \ref{fr_fig2}  are plotted such that $\omega$ increases from top row to the bottom row.  The first row corresponds to the freezing case. Intuitively, one should have expected freezing of $m^x(t)$ for all the higher $\omega$ values. However, interestingly, as seen in the figure, $m^x(t)$ shows freezing as well as non-freezing (oscillations) response for specific values of $\omega$. Also, the experimental results that are corrected for the decay show similar profiles as that of the theory.

\begin{figure}[t!]
\begin{center}
\includegraphics[trim = 00mm 0mm 215mm 00mm, clip, width=10cm]{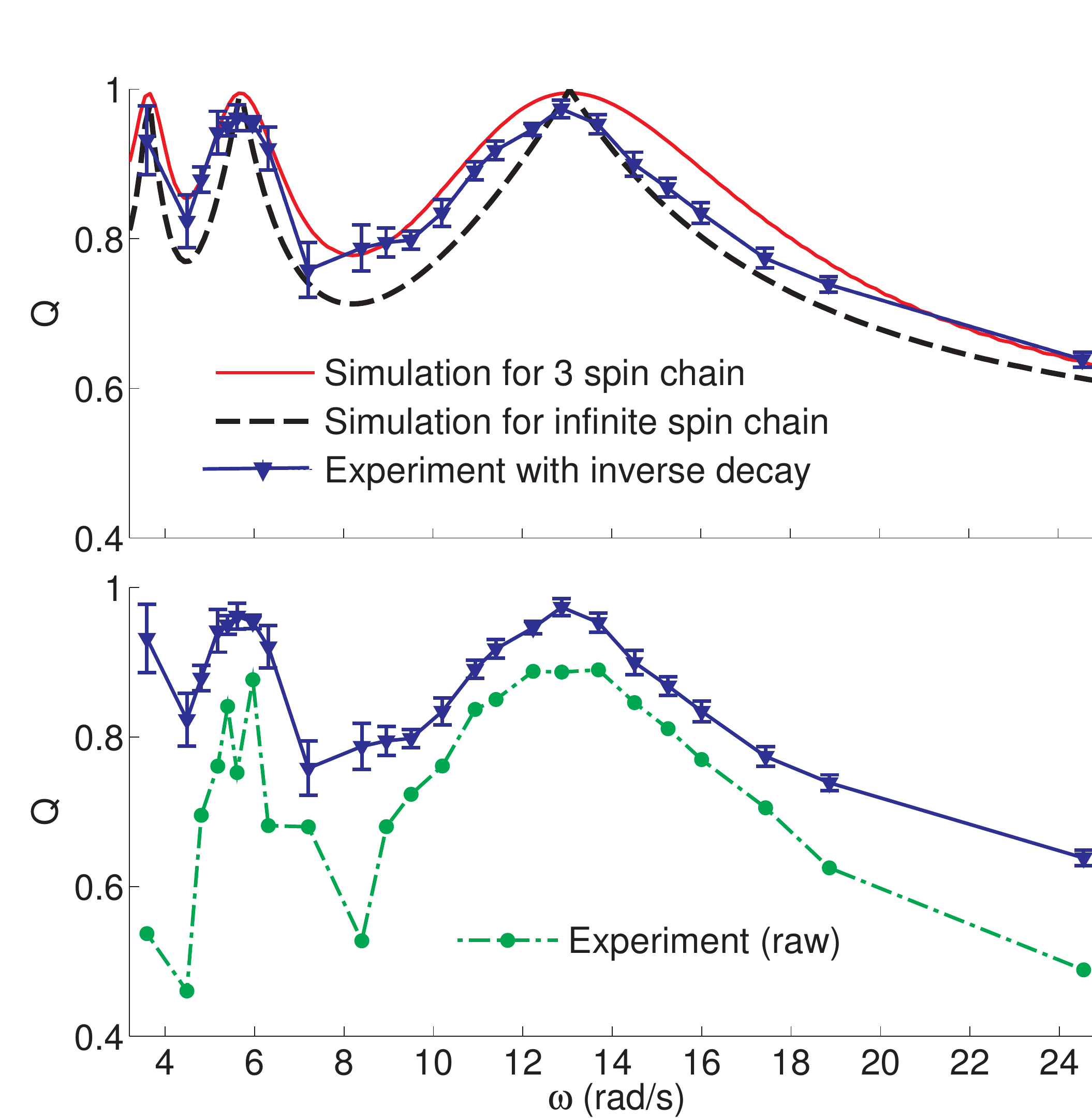}
\caption{$Q$ vs $\omega$ in the high frequency regime for the case of $m^x(t)=1$.  
}
\label{fr_fig3_1}
\end{center}
\end{figure}

Figs. \ref{fr_fig3_1} and \ref{fr_fig3_1a} are the final results of this work that capture the peak valley structure of the response of the one dimensional spin chain that is driven by external field in the high $\omega$ regime. These correspond to the two different initial states with $m^x(0)=1$ and $m^x(0)=0.5$ respectively. The quantity $Q$ is the long time average of $m^x(t)$ and is calculated using Eq. \ref{q}. 
The solid circles are the raw experimental results. Although, the experimental profiles match with the theory, these values are lower than that of the theory due to the relaxation process  as explained in section \ref{inv_dec}. However, the decay corrected experimental results fairly match with the theory. 

As seen in the figure, the striking match between the theory, raw experiments and the decay corrected experimental results reveal the successful demonstration of the dynamical many-body freezing.

\begin{figure}[h!]
\begin{center}
\includegraphics[trim = 216.5mm 00mm 00mm 00mm, clip, width=10cm]{fr_fig3_1.pdf}
\caption{$Q$ vs $\omega$ in the high frequency regime for the case of $m^x(t)=0.5$. 
}
\label{fr_fig3_1a}
\end{center}
\end{figure}

\section{Conclusions and future outlook}

Using a 3-qubit NMR simulator, we demonstrated the first experimental implementation of the phenomenon of DMF. As is proven in \cite{arnab}, the phenomenon of DMF is independent of the system size and thus allowed us to simulate DMF even on a 3-qubit system. The main set up consisted of a quantum many-body system that is driven out of equilibrium by an external periodic drive. Under this set up, we observed the response of the system by systematically monitoring the transverse magnetization. We considered   one dimensional transverse ising spin chain as our model quantum many-body system. By tuning the external drive frequency to some specific values, we observered the non-monotonicity in the response of the system \cite{arnab,freeze2}. We showed that the phenomenon is true for all times and for all states by separately demonstrating the experiments for two arbitrary initial states. Although DMF can be observed with small systems, it would be no less interesting to observe the phenomena in large systems. However, the experimental implementation with larger systems would not be easy. This is because (1) NMR systems are not scalable with the increasing number of qubits (2) Coherent control is difficult. e.g GRAPE algorithm to realize a specific operation might take longer times to converge to a desired fidelity. (3) A system with long $T_2$ value to accommodate the complex unitaries should be considered.

It was theoretically shown that despite the presence of  disorder in the system, the quantum many-body systems under DMF retain long coherence times \cite{arnab2}. This indicates that the freezing points are robust and hence can be used as efficient quantum memories. The future experimental plan is  to simulate this problem to further understand DMF and to see how it can be efficiently used as a novel control technique to manipulate and preserve the states of quantum computers.

\thispagestyle{empty}

\chapter{Pauli Decomposition over Commuting Subsets: Applications in Gate Synthesis, State Preparation, and Quantum Simulations  \label{chap6}}

\section{Introduction}
Quantum devices have the capability to perform several tasks with efficiencies beyond the reach of their classical counterparts \cite{chuang,preskill}.  An important criterion for the physical realization of such devices is to achieve precise control over the quantum dynamics \cite{divincenzo1}. 
The circuit model of quantum computation is based on the  realization of a desired unitary in terms of simpler quantum gates.  However, arbitrarily precise decomposition of a general unitary $U_T$ in the form 
\begin{equation}
U_T = U_m \cdots U_2 U_1,
\label{U}
\end{equation}
is a nontrivial task.
Here each of the $U_j$'s is either of lower complexity or acts on smaller subsystems. 

The decomposition of a unitary operator corresponding to a Hamiltonian ${\cal H} = {\cal H_A}+{\cal H_B}$, where $[{\cal H_A},{\cal H_B}] = 0$, is
trivial, i.e., $U_T = e^{-i{\cal H} t} = e^{-i{\cal H_A} t}e^{-i{\cal H_B} t}$.   When  $[{\cal H_A},{\cal H_B}] \neq 0$, 
one can discretize the time, $\delta = t/m$,
and use the Trotter's formula \cite{trotter}
\begin{equation}
U_T = \left[e^{-i{\cal H_A} \delta}e^{-i{\cal H_B}\delta}\right]^m + {\cal O}(\delta^2)
\label{U_trot}
\end{equation}
 or
its symmetrized form \cite{trotter1}
\begin{equation}
U_T = \left[e^{-i{\cal H_A} \delta/2}e^{-i{\cal H_B} \delta}e^{-i{\cal H_A} \delta/2}\right]^m  + {\cal O}(\delta^3)
\label{U_trots}.
\end{equation}
For a time-dependent Hamiltonian ${\cal H}(t)$, one needs to use Dyson's time ordering operator or the Magnus expansion, and then decompose the time discretized components \cite{sakurai2011modern}.
However, for a given unitary $U_T$, such a decomposition may not be obvious or even after the decomposition, the individual pieces themselves may involve matrix exponentials of non-commuting operators thus failing to reduce the complexity. 

Several advanced decomposition routines have been suggested for arbitrary unitary decomposition. Barenco {\em et al.} have shown that XOR gates along with local gates are universal, and in terms of these elementary gates they have explicitly decomposed several standard quantum operations \cite{barenco}. Tucci presented an algorithm to decompose an arbitrary unitary into single and two-qubit gates using a mathematical technique called CS decomposition \cite{tucci1999rudimentary}. Khaneja {\em et al.} used Cartan decomposition of the semi-simple lie group SU$(2^n)$ for the unitary decomposition \cite{khaneja2001cartan}. A method to realize any multiqubit gate  using fully controlled single-qubit gates using Grey code was given by Vartiainen {\em et al.} \cite{vartiainen2004efficient}.
M\"{o}tt\"{o}nen {\em et al.} have presented a cosine-sine matrix decomposition method synthesizing the gate sequence
to implement a general $n$-qubit quantum gate \cite{mottonen2004quantum}.
 Recently, Ajoy {\em et al.} also developed an ingenious algorithm to 
decompose an arbitrary unitary operator 
using algebraic methods \cite{ashok1}. 
More recently, this method was utilized in the experimental implementation of mirror inversion and other quantum state transfer protocols \cite{kota}.
Manu et al have shown several unitary decompositions by case-by-case numerical optimizations \cite{manu}.

In this work, we propose a general algorithm to decompose an arbitrary unitary upto a desired precision.
It is distinct from the above approaches in several ways. 
Firstly, our method considers generalized rotations of commuting Pauli operators which are more amenable for practical implementations via optimal control techniques.  Secondly, being a numerical procedure, it considers various experimentally relevant parameters such as robustness with respect to fluctuations in the control parameters, minimum rotation angles etc. Besides, the procedure can be extended for quantum circuits, quantum simulations, quantum state preparations, and probably in some cases even for nonunitary synthesis.

The chapter is organized as follows: A detailed explanation of the algorithm for arbitrary unitary decomposition is presented in section \ref{algo}.  Section \ref{pdcs_ex} deals with the applications of the algorithm with explicit demonstrations involving the decomposition of standard gate-synthesis, quantum circuit designs, certain quantum state preparations, and for quantum simulations. Finally the chapter ends with a conclusion in section \ref{pdcs_con}.

\section{Algorithm \label{algo}}
In the following, I will describe an algorithm for Pauli decomposition over commuting subsets (PDCS) for arbitrary unitary operators. Although for the sake of simplicity we utilize 2-level quantum systems,  the protocol applies equally well to any $d$-level quantum systems.

\begin{figure}[b!]
\begin{center}
	\includegraphics[trim=4.5cm 0cm 4.5cm 0cm, clip=true, width=12cm]{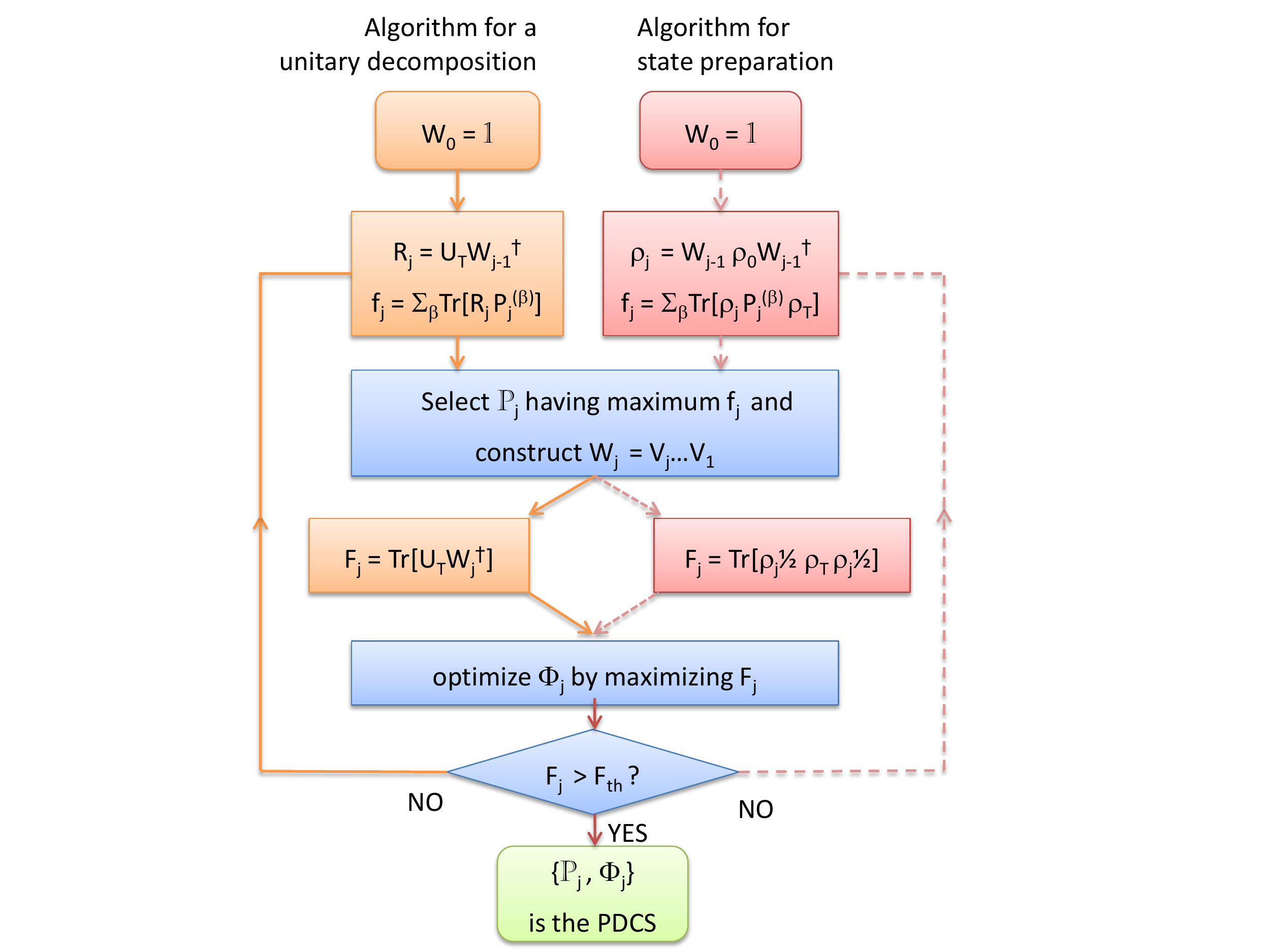}
	\caption{The flowchart describing PDCS algorithm for unitary decomposition (left) and state preparation (right, dashed).
	}
\label{flowchart}
\end{center}
\end{figure}

Let $U_T$ be the desired unitary operator of dimension $N=2^n$ to be applied on an $n$-qubit system.  We seek an $m$-rotor decomposition
$W_m = V_m V_{m-1} \cdots V_1 \approx U_T$,
where the decomposed unitaries $V_j$ with $j\in[1,m]$ have the form
\begin{equation}
	V_j = e^{-i \sum_\beta P^{(\beta)}_j \phi^{(\beta)}_j }.
	\label{Uj}
\end{equation}
Here $\{P^{(\beta)}_j\} \equiv \mathbbm{P}_j$ is a maximally commuting subset of $n-$ qubit Pauli operators,  $\{\phi^{(\beta)}_j\}\equiv \Phi_j$ is the set of corresponding rotation angles, and the index $\beta$ runs over the elements of $\mathbbm{P}_j$.
In general, for an $n$-qubit case, a maximal commuting subset $\mathbbm{P}_j$ can have at most $N-1$ elements. 
The fidelity $F_m$ of the decomposition, defined by
\begin{equation}
F_m = \langle U_T \vert W_m \rangle = \vert \mathrm{Tr}[U_T^\dagger W_m]/N \vert ,
\label{fd}
\end{equation}
should be larger than a desired threshold $F_\mathrm{th}$.

The flowchart for the PDCS algorithm is shown in Fig. \ref{flowchart}.
I will now describe an algorithm to build $W_m$ in $m$ steps.  
To begin with, we start with $W_0 = \mathbbm{1}$.
The $j$th step of the algorithm consists of the following processes:
\begin{enumerate}
\item
Calculate the residual propagator 
$R_{j} = U_T W_{j-1}^\dagger$. 
\item Selection of the commuting subset $\mathbbm{P}_j$ having the maximum overlap 
$f_j = \sum_\beta \mathrm{Tr} [R_{j} P^{(\beta)}_j]$
with the residual unitary $R_j$.
\item Setting up the decomposition $W_j$
and numerically optimizing the rotation angles $\{\Phi_1,\cdots,\Phi_j\}$ by maximizing the fidelity $F_j = \langle W_j \vert U_T\rangle$, where $W_j = V_j \cdots V_1$.
\end{enumerate}
These steps can be iterated upto $m$-steps until the fidelity $F_m > F_\mathrm{th}$ of a desired value is reached.

In general, the solutions to the decomposition may not be unique.  However, it is desirable to attain a decomposition that is most suitable for experimental implementations.  In this regard, we look for solutions with minimum rotation angles $\{\Phi_j\}$, which can be obtained by using a suitable penalty function in step 3 of the above algorithm.

\section{Applications \label{pdcs_ex}}
\subsection{PDCS of quantum gates and circuits}
In this section I will illustrate PDCS of several standard quantum gates. As described in Eq. \ref{Uj}, the $j$th decomposition is expressed in terms of the commuting Pauli operators $\mathbbm{P}_j$ and the corresponding rotation angles $\Phi_j$.  Further, a specific operator $P_j^{(\beta)} \in \mathbbm{P}_j$ can be expressed as a tensor product of single-qubit Pauli operators $X$, $Y$, $Z$, and the identity matrix $\mathbbm{1}$.

\begin{figure}[t!]
\begin{center}
	\includegraphics[trim=0.0cm 17.5cm 17.5cm 0.5cm, clip=true, width=14cm]{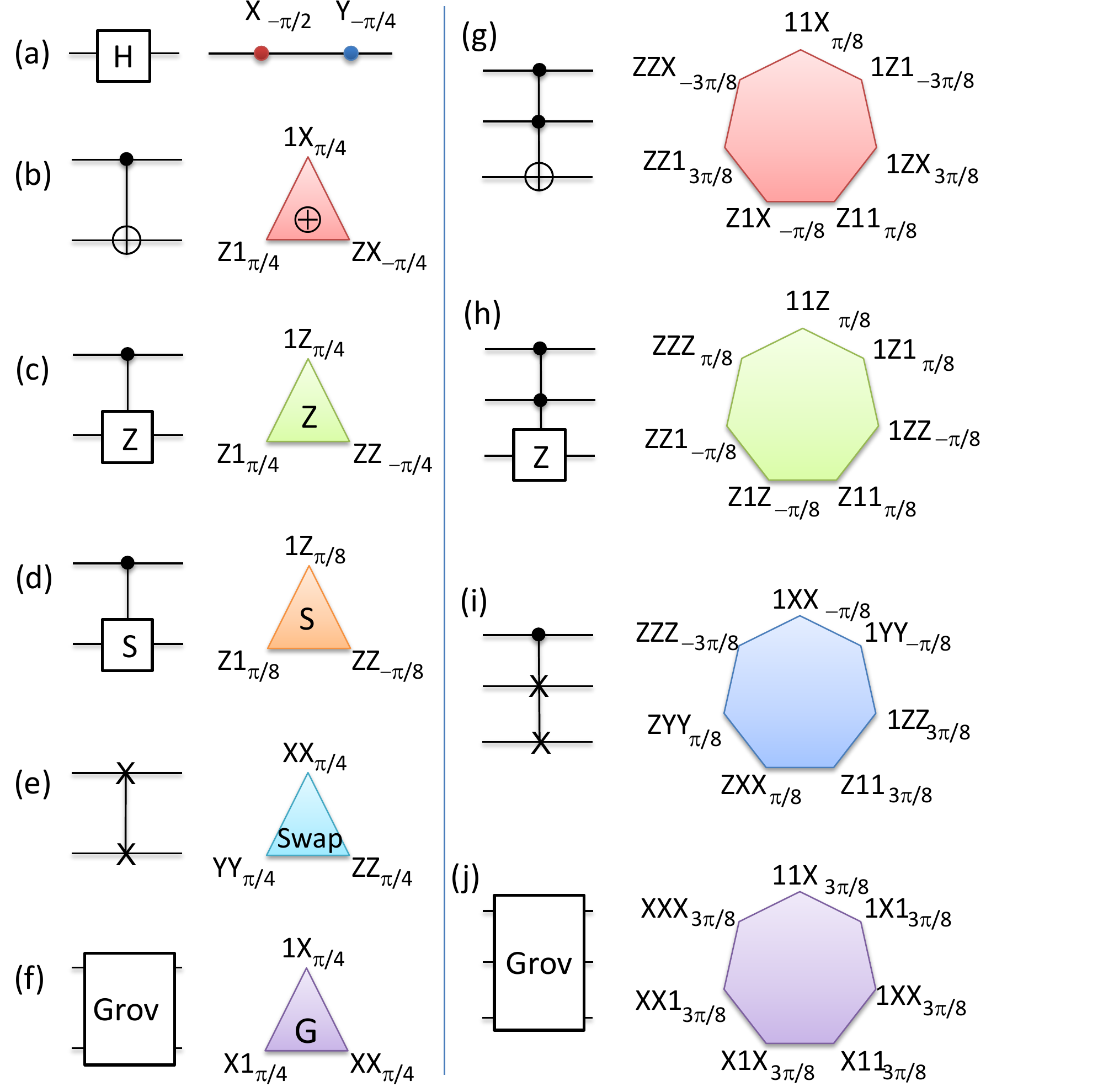}
	\caption{PDCS of some standard quantum gates: (a) single-qubit Hadamard, (b) c-NOT, (c) c-Z,
		(d) c-S, (e) SWAP, (f) 2-qubit Grover iterate, (g) Toffoli, (h) c$^2$-Z, (i) Fredkin, and (j) 3-qubit Grover iterate.  The individual rotors are represented by dots, triangles, and heptagons depending on number of Pauli operators (indicated at the vertices) in each rotor. The corresponding rotation angles are indicated by subscripts.}
		\label{gates1}
\end{center}
\end{figure}

Exact PDCS of several standard quantum gates are shown in Fig. \ref{gates1}.  For a single-qubit Hadamard operation (Fig. \ref{gates1}(a)), we obtain a decomposition with two noncommuting rotations, as is well known \cite{chuang}.  Here $\mathbbm{P}_1=X,~ \mathbbm{P}_2 = Y$ and the corresponding rotation angles $\Phi_1 = -\pi/2, ~ \Phi_2 = -\pi/4$,  are indicated by the subscripts.

In the two-qubit case, the maximal commuting subset can have only three Pauli operators and there are only 15 such subsets. Figs. \ref{gates1}(b-e) describe decompositions of several two-qubit gates namely c-NOT, c-Z, c-S, and SWAP gate. 
Here	$Z = \left[
		\begin{array}{cc}
				1 & 0 \\
				0 & -1
		\end{array}
		\right]$ 
		and
		$S = \left[
		\begin{array}{cc}
		1 & 0 \\
		0 & i
		\end{array}
		\right]$.
It is interesting to note that each of these gates needs a single subset of commuting Pauli operators.  Such a rotation can be obtained by a single matrix exponential and can be thought of as a single generalized rotation in the Pauli space. We refer to such a generalized rotation as a rotor, and since it consists of three operators, we represent it by a triangle. In practice, the individual components of a single rotor can be implemented either simultaneously, or in any order. We find that even a 2-qubit Grover iterate, i.e., $G = 2\outpr{\psi}{\psi}-\mathbbm{1}$, 
where $\ket{\psi} = (\ket{00}+\ket{01}+\ket{10}+\ket{11})/2$, can be realized as a single rotor (Fig. \ref{gates1}(f)).

For three-qubits, the maximal commuting subset can have only seven commuting Pauli operators and there are 135 distinct subsets.  Figs. \ref{gates1}(g-j) describe Toffoli, c$^2$-Z, Fredkin, and 3-qubit Grover iteration respectively.  Again we find that a single heptagon rotor suffices for realizing each of the standard gates.  Similarly, in the case of four-qubits, a maximal commuting subset can have 15 operators and one can verify that a basic gate such as c$^3$-NOT, c$^3$-Z, etc. can be realized by a single rotor.  

\begin{figure*}
	\includegraphics[trim=0.0cm 23.8cm 6cm 1.5cm, clip=true, width=15cm]{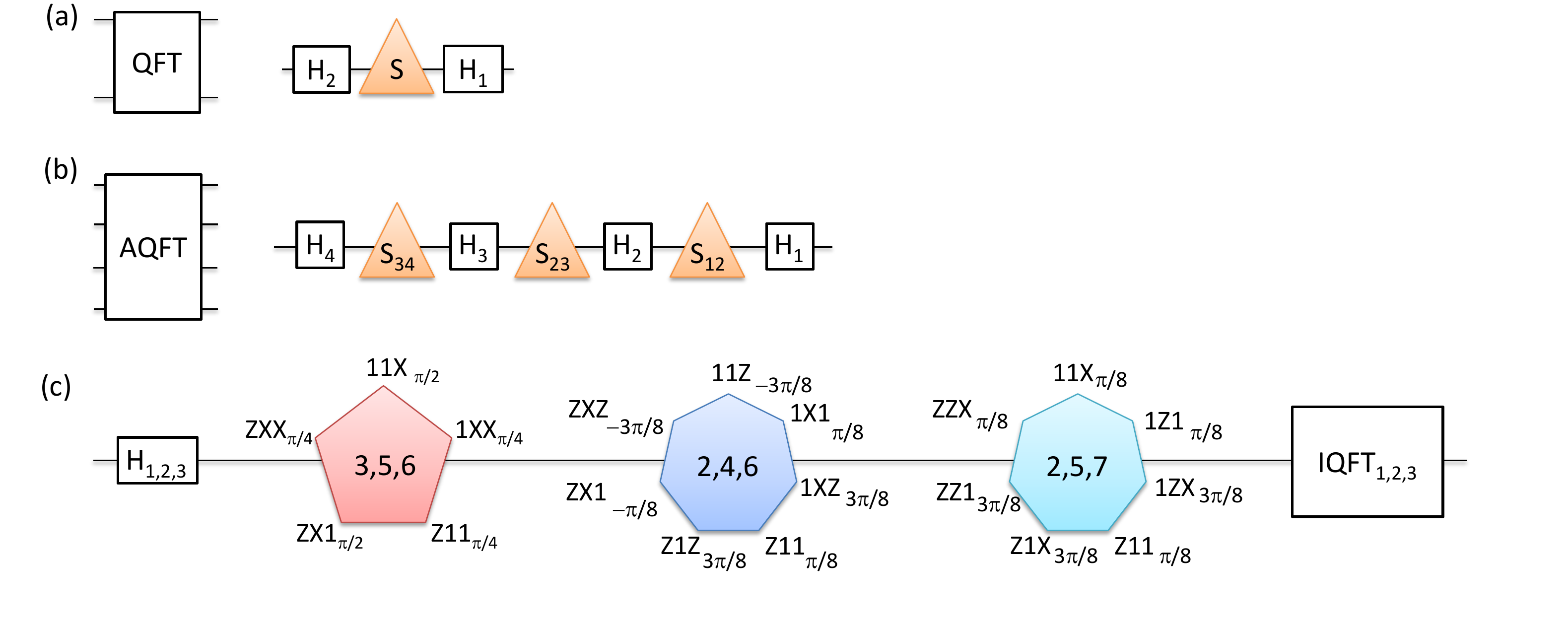}
	\caption{
		PDCS of (a) 2-qubit Quantum Fourier Transform (QFT), 
		(b) 4-qubit approximate QFT (AQFT), and
		(c) 7-qubit Shor's circuit for factorizing 15.
		In (b), each S gate acts on a pair of qubits as indicated by the subscripts.}
		\label{gates2}
\end{figure*}

It is always possible to decompose a multi-qubit quantum circuit in terms of single- and two-qubit gates \cite{barenco}.  
As  examples, we  consider PDCS of a few quantum circuits based on Quantum Fourier Transform (QFT) (Fig. \ref{gates2}). QFT is an important algorithm in quantum computation since it takes only $n^2$ steps to Fourier transform $2^n$ numbers unlike a classical computer that takes $n2^n$ steps for the same.
The two-qubit QFT circuit can be exactly decomposed into three rotors as shown in Fig. \ref{gates2}(a). As another example, PDCS of the 4-qubit approximate QFT (AQFT) circuit \cite{aqft}  results in only single-qubit and two-qubit rotors as shown in Fig. \ref{gates2}(b).  An example based on QFT is shor's circiut that is used to find the prime factors of a given number \cite{shor1}.  PDCS of the 7-qubit Shor's circuit for factoring the  number 15  involves at most three-qubit rotors as shown in Fig. \ref{gates2}(c) \cite{shor}. Although these circuits are mentioned in the respective references, we implemented PDCS algorithm on the corresponding individual operations to aid experimental implementations. In this sense, PDCS of multiqubit quantum gates and quantum circuits is scalable with increasing system size.

\subsection{Quantum state preparation}
\begin{figure}[t!]
\begin{center}
	\includegraphics[trim=2cm 25.0cm 18cm 1.5cm, clip=true, width=12.5cm]{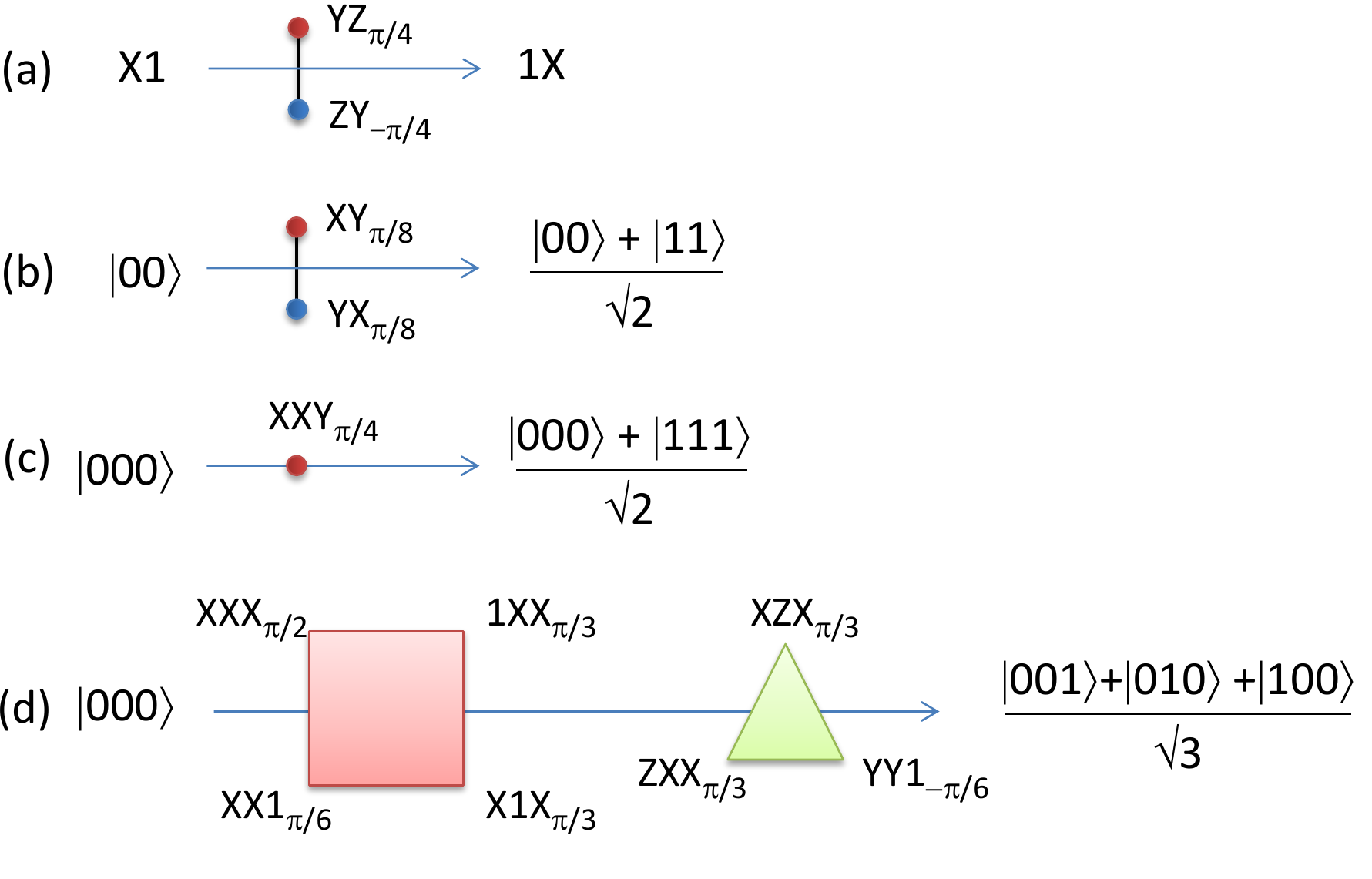}
	\caption{PDCS of some state to state transfers: (a) polarization transfer (INEPT) and (b-d) preparation of Bell, GHZ, and W states respectively starting from pure states.}

	\label{pdstate}
\end{center}
\end{figure}

Here the goal is to prepare a target state
$\rho_T$ starting from a given initial state
$\rho_0$.  In general the unitary operator, connecting the initial and target states, itself is not unique.  The procedure is similar to that described in the previous section, and is summarized in the flowchart shown in Fig. \ref{flowchart}. Here 
the selection of commuting Pauli operators $\mathbbm{P}_j$ is based on the overlap 
$f_j = \sum_{\beta} \mathrm{Tr}[\rho_j P_j^{(\beta)} \rho_T]$,
where $\rho_j = W_j \rho_0 W_j ^\dagger$ is the intermediate state after $j$th decomposition.  As explained before, we select
the commuting subset $\mathbbm{P}_j$ having the maximum overlap $f_j$ and optimize the phases $\Phi_j$ by maximizing the Uhlmann fidelity 
\begin{eqnarray}
F_j = \langle \rho_j \vert \rho_T \rangle =  \mathrm{Tr}[\rho_j^{1/2}\rho_T\rho_j^{1/2}].
\end{eqnarray}
Again, $m$ iterations are carried out until  $F_m \ge F_\mathrm{th}$ is realized.

Fig. \ref{pdstate} displays PDCS of some standard state to state transfers.  The polarization transfer in a pair of qubits (popularly known as INEPT \cite{cavanagh}) requires a single rotor having a pair of bilinear operators (Fig. \ref{pdstate}(a)).
The preparation of a Bell and GHZ states respectively from $\ket{00}$ and $\ket{000}$ states also require a single rotor (Fig. \ref{pdstate}(b-c)).  However, the preparation of a three-qubit W-state is somewhat more elaborated, and requires two rotors (Fig. \ref{pdstate}(d)).  Although these decompositions are not unique, it is possible to optimize them based on the experimental conditions.
Here one can notice that although a maximal commuting subset can have up to $N-1$ elements, it is often possible to decompose a multi-qubit operation over a smaller commuting subset.

\subsection{Quantum simulation}

Utilizing controllable quantum systems to mimic the dynamics of other quantum systems is the essence of quantum simulation \cite{fey82}. Various quantum devices have already demonstrated quantum simulations of a number of quantum mechanical phenomena (for example, \cite{suter2005,nori,atoms1,freeze2}). 
An important application of the decomposition technique described above is in the experimental realization of quantum simulations.
To illustrate this fact,  we experimentally carryout  quantum simulation of a three-body interaction Hamiltonian using a three-qubit system. While such a
Hamiltonian is physically unnatural, simulating such interactions has interesting applications such as in quantum state transfer \cite{suter_3body_statetransfer}.  Specifically,
we simulate the dynamics under the Hamiltonian
\begin{eqnarray}
{\cal H}_\mathrm{S} = X\mathbbm{1}\mathbbm{1}+\mathbbm{1}X\mathbbm{1}+\mathbbm{1}\mathbbm{1}X + J_{123} ZZZ,
\label{H_3body}
\end{eqnarray}
where $ J_{123}$ is the three-body interaction strength.  
A slightly different three-body Hamiltonian simulated earlier by Cory and coworkers \cite{tseng} consisted of only the last term.  The presence of other terms which are noncommuting with the 3-body term necessitates an efficient decomposition of the overall unitary.

\begin{figure}[h!]
\begin{center}
	\includegraphics[trim=4cm 8.5cm 2cm 4.5cm, clip=true, width=10.5cm]{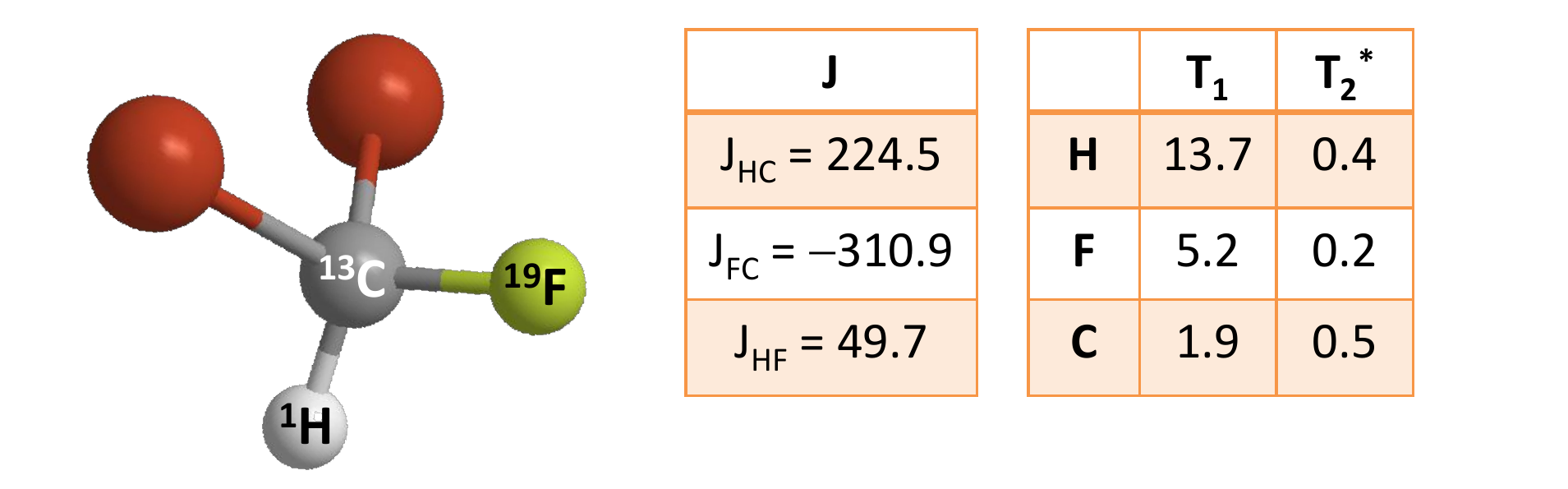}
	\caption{Dibromofluoromethane consisting of three nuclear spin qubits $^1$H, $^{13}$C and $^{19}$F. The tables display the values of indirect spin-spin coupling constants ($J$) in Hz, and the relaxation time constants ($T_1$, and $T_2^*$) in seconds.}
	\label{hfc}
\end{center}
\end{figure}

We use three spin-1/2 nuclei of dibromofluoromethane (Fig. \ref{hfc}) dissolved in acetone-D6 as our three-qubit system.  All the 
experiments are carried out on a 500 MHz Bruker NMR spectrometer at an ambient temperature of
300 K.  In the triply rotating frame at resonant offsets, the internal Hamiltonian of the system is given by
\begin{eqnarray}
{\cal H}_\mathrm{int} = 
\left[ J_\mathrm{HF}ZZ\mathbbm{1} +
J_\mathrm{HC}Z\mathbbm{1}Z +
J_\mathrm{FC}\mathbbm{1}ZZ\right]\pi/2,
\end{eqnarray}
and the values of the indirect coupling constants $\{J_\mathrm{HF},J_\mathrm{HC},J_\mathrm{FC}\}$ are as in Fig. \ref{hfc}.  This internal Hamiltonian along with the external control Hamiltonians provided by the RF pulses are used in the following to mimic the three-body Hamiltonian in Eqn. \ref{H_3body}.

The traceless part of the thermal equilibrium state of the 3-qubit NMR spin system is given by
$\rho_{eq} = (Z\mathbbm{11}+\mathbbm{1}Z\mathbbm{1}+\mathbbm{11}Z)/2$ \cite{cavanagh}.  The initial state $\rho(0) = (X\mathbbm{11}+\mathbbm{1}X\mathbbm{1}+\mathbbm{11}X)/2$ is prepared by applying a $90^\circ$ RF-pulse  about $Y$.  
The goal is to subject the three-spin system to an effective three-body Hamiltonian $H_S$ and monitor the evolution of its state
$\rho(t) = U_S(t) \rho(0) U_S(t)^\dagger$, where
$U_S(t) = e^{-i{\cal H}_St}$.
We choose to experimentally observe the  transverse magnetization 
\begin{eqnarray}
M_x(k \tau) = \mathrm{Tr}[\rho(k \tau)(X\mathbbm{11}+\mathbbm{1}X\mathbbm{1}+\mathbbm{11}X)/2]
\end{eqnarray}
for $ J_{123}=5$ Hz at discrete time intervals $k \tau$, where $k = \{0,\cdots,20\}$ and $\tau=0.8$ s.

\begin{figure}[t!]
\begin{center}
	\includegraphics[trim=6.5cm 8.2cm 4.5cm 5cm, clip=true, width=10.5cm]{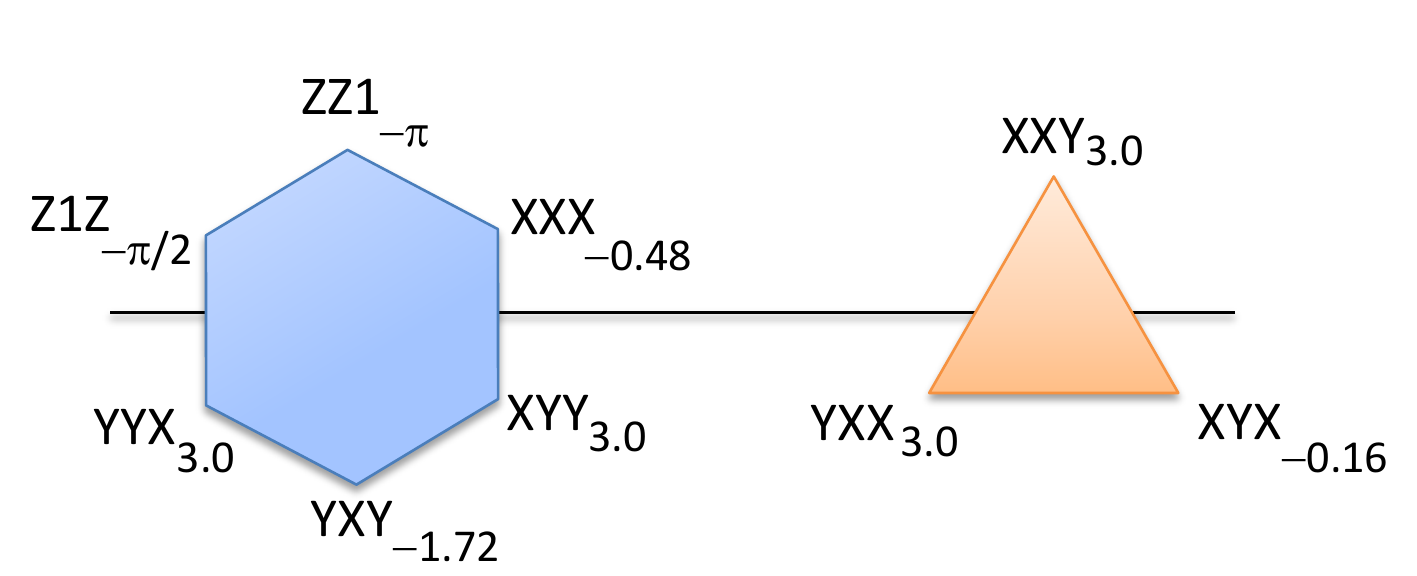}
	\caption{PDCS of 
		$U_s(\tau) = \exp(-i{\cal H}_s\tau)$ for $J_{123}=5$ Hz and $\tau = 0.8$ s.}
	\label{U_3body}
\end{center}
\end{figure}

\begin{figure}[b!]
\begin{center}
 	\includegraphics[trim=0.2cm 8.9cm 0cm 0cm, clip=true, width=12cm]{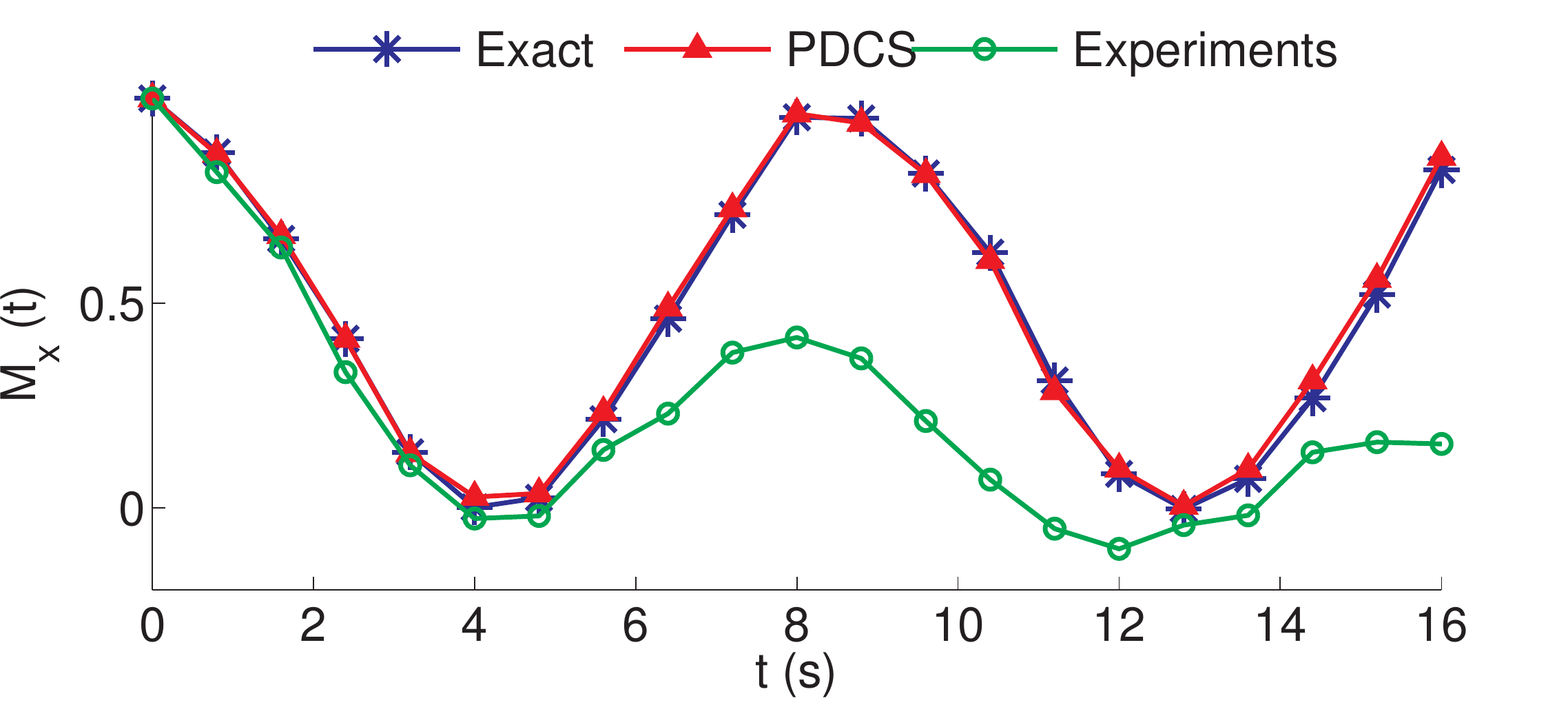}
 	\caption{Magnetization vs time. The first point is the initial state magnetization. The second and the subsequent points correspond to the application of the operator $U_s$ for k times with $k=1 \cdots 20$ respectively. The decay in the experimental results are due to the decoherence and other experimental imperfections. 
 	}
 	\label{simu}
\end{center}
 \end{figure}

The PDCS of $U_S$ shown in Fig. \ref{U_3body} consists of two rotors: a hexagon followed by a triangle.  We utilized bang-bang (BB) control technique for generating each of the two rotors \cite{gaurav}.  The duration of each BB-sequence was about 7 ms and fidelities were above 0.98 averaged over a 10\% inhomogeneous distribution of RF amplitudes.  The results of the experiments (hollow circles) and their comparison with numerical simulation of the PDCS (triangles) and exact numerical values (stars) are shown in Fig. \ref{simu}.  The first experimental data point was obtained after a simple 90 degree RF pulse and was normalized to 1.
The $k^{\mathrm{th}}$ point is obtained by $k$ iterations of the BB-sequence for $U_S(\tau)$. While the experimental curve displays the same period and phase as that of the simulated curve, the steady decay in amplitude is mainly due to decoherence and other experimental imperfections such as RF inhomogeneity and  nonlinearities of the RF channel.

\begin{figure}[t!]
\begin{center}
	\includegraphics[trim=3cm 5cm 3cm 5cm, clip=true, width=10cm]{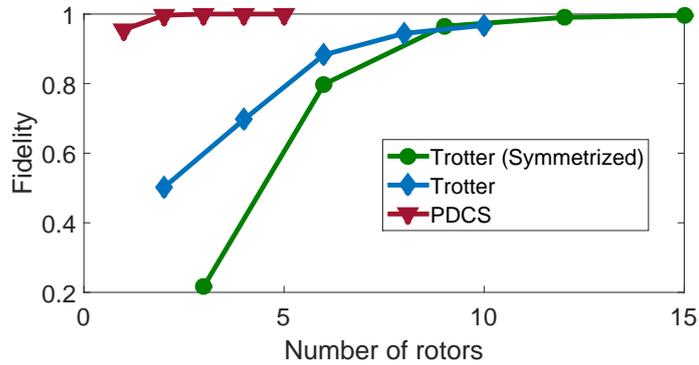}
	\caption{Fidelity ($F_m$) vs number ($m$) of rotors for three different types of decompositions of $U_s$. As seen, the decompositions corresponding to PDCS converge faster towards fidelity 1.  Here the fidelity is with respect to the rotors and $U_s$. 
	}
	\label{comptrot}
\end{center}
\end{figure}

To compare the efficiency of PDCS with that of Trotter decomposition (in Eq. \ref{U_trot} and \ref{U_trots}), we calculate the fidelities ($F$) of the decomposed propagator with the exact propagator
$U_S(\tau)$ as a function of number $m$ of rotors (see Fig. \ref{comptrot}).  It can be observed that, with increasing number of rotors, PDCS fidelity converges faster than the Trotter.

\section{Conclusions and future outlook \label{pdcs_con}}
In this work we proposed a generalized numerical algorithm based on Pauli decomposition over commuting subsets (PDCS). The aim of the algorithm is to decompose an arbitrary target unitary into simpler unitaries, referred to as rotors.  Each rotor consists of only commuting subset of Pauli operators. These rotors are optimized to be robust against experimental errors by minimizing the rotation angles and by considering other control errors.  Thus apart from providing an intuitive and topological representation of an arbitrary quantum circuit, the method is also useful for its efficient physical  realization.

We demonstrated the robustness and efficiency of the decomposition using numerous examples of quantum gates and circuits. It is interesting to note that several standard quantum gates correspond to single rotors.     
We also discussed the applications of  PDCS in quantum state-to-state transfers and illustrated it using several examples. 

Another important application of PDCS is in  quantum simulations.  As an 
example, we described the quantum simulation of a three-body interaction.  We used PDCS algorithm to decompose the unitary corresponding to such a Hamiltonian and found it to be more efficient than Trotter decomposition
in terms of the number of rotors. Further, we have demonstrated the quantum simulation by experimentally monitoring the evolution of magnetization using a three qubit NMR system.  The experimental results matched with the numerical simulations upto a decay factor arising  predominantly due to decoherence. 

 The decompositions can be  made robust against noise.   This can be done by incorporating the noise in the algorithm wherein the first decomposition optimizes against this noise and the next iteration considers this operator to optimize the next operator, again in the presence of noise.  Further, this algorithm can also be generalized to decompose nonunitary operators.  We also believe such unitary decomposition strategies combined with sophisticated optimal control techniques will greatly assist in efficient quantum control.

\thispagestyle{empty}
\part{Non-unitary Control}
\thispagestyle{empty}

\chapter{Engineered decoherence: Characterization and suppression  \label{chap5}}

\section{Introduction \label{intro_dec}}

Quantum devices  which are perfectly isolated from their environment follow  unitary dynamics wherein the purity of the density operators are preserved throughout the state evolution. Although this  ideal case  is strongly desired in the field of quantum computation and communication, in practice, no quantum device is perfectly isolated from its environment. This leads to  inevitable interactions between a quantum system and the environment which ultimately entangles the two. For sufficiently large times and for large environmental size, the evolution of quantum system becomes non-unitary leading to an irreversible information transfer from the quantum system to the environment. 

The most common information losses correspond to the coherence decay, also known as phase decoherence, and  energy dissipation   of the quantum systems. 
The decay constants $T_1$ and $T_2$ are associated with energy dissipation and decoherence processes respectively, that are borrowed from NMR terminology \cite{cavanagh}. In general, $T_1>T_2$ which implies that the quantum systems lose phase information faster than their  energy. Hence in any quantum information protocol it is important to implement the gates within the time scale of $T_2$. The very fact that the phenomenon of decoherence has been a severe threat to the physical realization of a quantum computer  has lead towards several theoretical and experimental studies on decoherence
\cite{cory_dec,zurek_dec,poyatos1996quantum,myatt2000decoherence,paz1999quantum,paz2001quantum,viola_dec}. 

In this chapter, I will explain our work that deals with the understanding of phase decoherence and is organized into three parts \cite{dec}: 
\begin{enumerate}

\item We experimentally simulated artificial phase decoherence. Although, in practice, one does not have any control over the environment, emulation of decoherence gives a direct control over it. By systematically controlling the environment one can study its effects on the system coherences. 
\item We suppressed the induced decoherence using standard dynamical decoupling (DD) sequences. The simultaneous competition between the DD sequences and the decoherence process might give insight into the decoherence process and ways to improve DD sequences.

\item We characterized the amount of the induced decoherence in the system qubits using noise spectroscopy (NS) and quantum process tomography (QPT).  NS gives the frequency distribution of the noise and QPT gives the entire   information of the noise process for a specific noise frequency. 
\end{enumerate}

We implemented the above steps using a 2-qubit NMR simulator. The model considered one qubit as a system qubit and the other as an environment qubit. Additional decoherence, apart from the inherent decoherence, was induced using random classical fields on the environment qubit \cite{cory_dec}. We studied the effect of controlled noise on the system qubit. In the following sections, I will introduce to decoherence model, DD sequences, NS, QPT,  experiments and results.

\section{Decoherence models}
In this section, I will explain the phase decoherence model with $ZZ$ type system-environment interaction. 
\subsection{Zurek's decoherence model \label{zurek_model}}
This model was given by Zurek \cite{zurek} and is explained below.

Consider an $n$ qubit composite system  consisting of two subsystems. One qubit is considered as the system of interest and the rest of the qubits are considered as the environment. The total Hamiltonian and the corresponding unitary operator are respectively given by
\begin{equation}
H_{SE} = \sum_{j=2}^n J_{1j} Z_1 Z_j \qquad \mathrm{and} \qquad U_{SE}(t) = e^{-iH_{SE}t}.
\end{equation} 
 Here $J_{1j}$ is the coupling between the system (represented by subscript $1$) and the environment (represented by subscript $j$). Zurek showed that such a Hamiltonian with $ZZ$ type system-environment interaction leads to phase decoherence. 

Let the combined system start with a separable state:
\begin{equation}
|\psi(0)\rangle_{SE} = |\psi(0)\rangle_{S} \otimes |\psi(0)\rangle_{E}.
\end{equation} 
Here the pure state $|\psi(0)\rangle_{S} = a|0\rangle_{1}+b|1\rangle_{1}$ with $|a|^2+|b|^2=1$ is the system state and $|\psi(0)\rangle_{E} = \otimes_{j=2}^n(\alpha_j|0\rangle_{j}+\beta_j|1\rangle_{j})$ with $|\alpha_j|^2+|\beta_j|^2=1$ is the environment state.

The evolution of $|\psi(0)\rangle_{SE}$ under the $U_{SE}$ entangles the system and the environment as below:

\begin{align}
|\psi(t)\rangle_{SE} &= U_{SE}(t)|\psi(0)\rangle_{SE}\nonumber  \\
&= a|0\rangle_1 \otimes_{j=2}^n(\alpha_j e^{-iJ_{1j}t}|0\rangle_{j}+\beta_j e^{iJ_{1j}t} |1\rangle_{j}) \nonumber\\
&+ b|1\rangle_1 \otimes_{j=2}^n(\alpha_j e^{iJ_{1j}t}|0\rangle_{j}+\beta_j e^{-iJ_{1j}t} |1\rangle_{j}).
\end{align}
The corresponding density operator is given by
$\rho_{SE}(t) = |\psi(t)\rangle_{SE} \langle\psi(t)|_{SE}$ and the system density operator $\rho_S(t)$ is obtained by tracing out the environment subsystem from $\rho_{SE}(t)$, i.e., $\rho_S(t) = \mathrm{Tr}_E[\rho_{SE}(t)]$. The quantity that we are interested in is the coherence part of the density operator. As was already mentioned  in Eq. \ref{rho_pop_coh},  the off-diagonal term $\rho_S^{01}$  encodes the coherence information and this matrix element in $Z$ basis is given by
\begin{align}
\rho_S^{01}(t) & =  _1\langle 0|\rho_{SE}(t)|1\rangle_1 \nonumber \\
&= ab\cdot  \Pi_{j=2}^n(|\alpha_j|^2 e^{-2iJ_{1j}t}+|\beta_j|^2 e^{2iJ_{1j}t})\nonumber \\
&= ab\cdot z(t),
\end{align}
where $\{|0\rangle_1, |1\rangle_1\}$ are the basis states of the system qubit and $z(t)$ is called as the decoherence factor. As seen from the above equation, $|z(t)|\to 0$ implies the decay in the coherences of the initial system state $|\psi(0)\rangle_S$ after time $t$.  Further, it can also be noted that irreversible decoherence can occur when $n\to\infty$, i.e., when the environmental size is large.

\subsection{Simulation of decoherence \label{artificial_dec}}

As was already discussed, 
Zurek's decoherence model requires large environmental size for irreversible phase damping. However, one question in experimental realization is whether one can still simulate the same process using only finite sized environment.  
In this section, I will give a brief review of the methods given by Teklemariam {\em et al.} \cite{cory_dec} to emulate artificial decoherence even when the environment size is finite. Such finite sized decoherence simulation allows for the direct control over the environment that can be easily implemented in laboratory with a goal to study the decoherence process.

The model given by Teklemariam {\em et al.}  differs from the Zurek's model as follows: Suppose, the dimension of the Hilbert space  of the quantum system is $2^n$. The model considers its interaction with environment described by a maximum Hilbert space dimension of $2^{2n}$. This greatly restricts the size of the environment for very small $n$ but favors experimental studies on decoherence.  Further, in order to mimic infinite sized environment and to induce irreversible phase damping on the system qubits, this model uses additional stochastic classical fields on the environment.

For the sake of simplicity, consider a two qubit system-environment model initially in the product state,
\begin{equation}
\rho_{SE}(0) = \rho_S(0) \otimes \rho_E(0),
\end{equation}
Initially the composite system is assumed to be a closed system and the total Hamiltonian is given by
\begin{equation}
{\cal H}=\pi(\nu_SZ_{S}+\nu_E Z_{E}+\frac{J}{2}Z_{S}Z_{E}),
\label{Ham}
\end{equation}
where $\nu_S$ and $\nu_E$ are the resonant frequencies of the system (S) and the environment (E) qubits respectively, and $J$ is the strength of the coupling  between the two. We consider the Hamiltonian in the rotating frames where $\nu_S=\nu_E=0$.
The state $\rho_{SE}(0)$ evolves under the propagator $U(T)$ for  a total time $T$ which is given by
\begin{equation}
U(T)=e^{-i{\cal H}T}
\label{U_dec_se}
\end{equation}
that entangles $S$ and $E$ as was discussed in section \ref{zurek_model}. 

Suppose,  $E$ is perturbed by random  classical fields without externally disturbing $S$. These perturbations are called as kicks and each kick operator $K_m$ corresponds to the rotation of $E$ with an arbitrary rotation angle $\epsilon_m$ about $y$-axis. For the $m^{\mathrm{th}}$ kick, we have $K_m = \mathbbm{I}_S\otimes e^{-i\epsilon_m Y_E}$ where $\mathbbm{I}_S$ is the Identity on the system and $\epsilon_m$ is chosen randomly between $[-\alpha, \alpha]$ with $\alpha$ being a small angle. 

The kicks are assumed to be  instantaneous with the kick rate $\Gamma = k/T$ where $k$ is the total number of kicks. Under this action, Eq. \ref{U_dec_se}  is modified to incorporate its dependency on the random angles $\epsilon_m$ and is given by
\begin{equation}
U_k(T)= K_k U(\delta) K_{k-1} U(\delta) \cdots K_1 U(\delta); \qquad \delta = \frac{T}{k}.
\end{equation}
A state $\rho_{SE}(0)$ evolves under this operator as $\rho_{SE}(T) = U_k(T)\rho_{SE}(0) U_k(T)^\dagger$ and the system and environment states are given by $\rho_S(T) = \mathrm{Tr}_E[\rho_{SE}(T)]$ and $\rho_E(T) = \mathrm{Tr}_S[\rho_{SE}(T)]$ respectively. 

An ensemble realization
over many random $\epsilon_m\in[-\alpha,\alpha]$
leads to an average behavior represented by
\begin{equation}
\bar{\rho}_s(T) = \int_{-\alpha}^{\alpha}\frac{d\epsilon_k}{2\alpha}\cdots\int_{-\alpha}^{\alpha}\frac{d\epsilon_1}{2\alpha} \mathrm{Tr}_E[U_k\rho_S(0)U_k^\dagger].
\end{equation}
Teklemariam {\em et al.} showed that \cite{cory_dec}, 
\begin{equation}
\bar{\rho}_S(T) = \sum_{r,s=0,1}D_{rs}(k,T)\rho_S^{rs}(0)|r\rangle\langle s|,
\end{equation}
with $ |r\rangle$, $|s\rangle \in \{|0\rangle,|1\rangle\}$  being the eigenstates of $Z_S$ and $D_{rs}(k,T)$ is the decoherence factor which is given by
\begin{equation}
D_{rs}(k,T) = \mathrm{Tr}_E[{\cal O}^k(\rho_E(0))].
\label{coryt2}
\end{equation} 
Here ${\cal O}$ is the superoperator that is neither trace preserving nor Hermitian and its action is defined as ${\cal O}(\rho_E) = cV_K \rho_E V_K + d  Y V_K \rho_E V_K Y$ with $V_K = e^{-i\pi J \delta Z_E/2}$, $c+d=1$, $c-d=\gamma$, and $\gamma = \sin(2\alpha)/(2\alpha)$. This indicates that for a specific value of $\epsilon$, $J$ and $\gamma$, one can simulate a unique type of phase decoherence.   

Teklemariam {\em et al.} showed that for smaller $\epsilon$'s and for lower $\Gamma$, the decoherence rate $1/T_2$ was proportional to $\Gamma$. However, for certain $\Gamma$ value, $1/T_2$ saturated and for a much higher $\Gamma$ value, $1/T_2$ decreased exponentially with $\Gamma$. Thus the former case correspond to the decoherence inducing effect while the latter case corresponded to noise decoupling effect \cite{ernst1966nuclear}. This latter case was not explored in our work due to the experimental limitations considering the fact that the pulses with very high kick rates could damage the RF-coils in NMR setup. 

\section{Suppressing Decoherence}
Preserving the qubit information  against noise is one of the most crucial steps in quantum information processing. Different techniques have been developed to suppress decoherence like dynamical decoupling (DD) \cite{cpmg,uhrig}, quantum
error correction \cite{qec}, use of robust approaches such as adiabatic quantum computation \cite{farhi}, or encoding quantum information in decoherence-free subspaces \cite{dfs}. In this chapter, I will explain two standard DD techniques that are utilized in our work in order to suppress the inherent as well as induced artificial decoherence. One of the major advantages of this technique is that, unlike
the other techniques, DD does not require extra qubits and it
can be combined with other quantum gates leading to fault
tolerant quantum computation \cite{ng2011combining,zhang2014protected}. 

I will first explain a way to suppress the static noise, a technique known as Hahn echo sequence. However, in general, the noise is time-dependent. I will explain two standard DD techniques for suppressing time-dependent noise, namely CPMG and Uhrig DD sequences.  All these techniques are explained in the case of NMR setp-up. 

\subsection{Hahn Echo \label{hahn_echo}}
\begin{figure}[h]
\begin{center}
\includegraphics[trim = 10mm 10mm 20mm 10mm, clip, width=14cm]{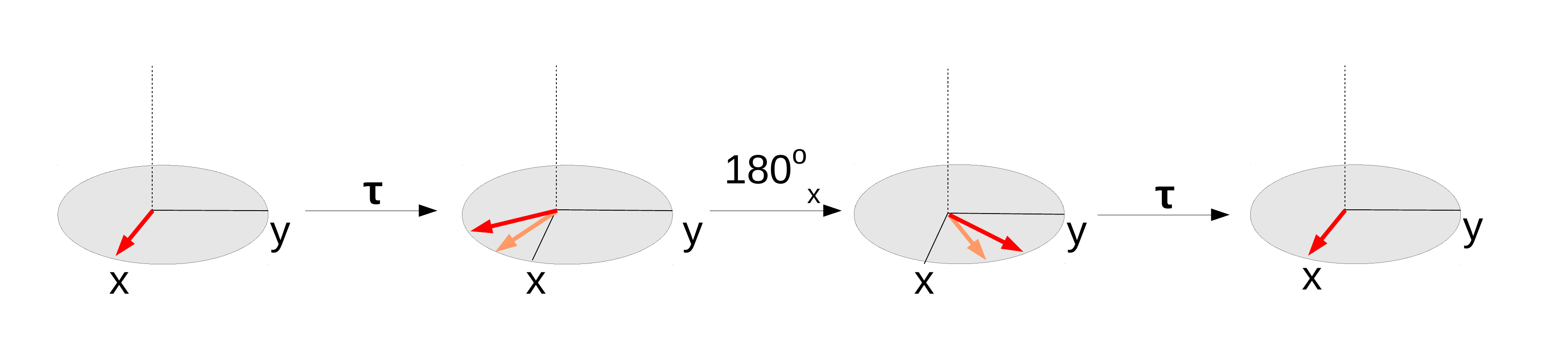}
\caption{Evolution of the net magnetization (indicated by arrows) under the Hahn echo sequence. The dotted arrows represents slow precessing spins and the solid arrows represents fast preseccing spins.  In this case, the precession of the magnetization about the $z$-axis is assumed to be clockwise.
}
\label{hahn}
\end{center}
\end{figure}

A technique to suppress time-independent noise in a single qubit was given by Hahn \cite{hahn}. Suppose the static magnetic field $B_0$ has a spatial inhomogeneity (NMR set-up). This small change in $B_0$ changes the larmor frequencies of the nuclei and hence different nuclei experience different larmor frequencies. However the desired scenario is the case wherein all the nuclei  precess with the same larmor frequency. In order to achieve this, Hahn gave a sequence as shown in Fig. \ref{hahn}.

 The effect of the Hahn echo pulse sequence is explained as follows: The spins evolve freely for a time $\tau$ during which  different nuclei will pick up different larmor frequencies in the presence of $B_0$  inhomogeneity. The nuclei fan-out with a range of larmor frequencies. The slow moving components are represented by dotted arrows and the fast moving components are represented by solid arrows. A $\pi$ pulse about the $x$-axis rotates the spins and  during the free evolution for time $\tau$, the faster moving components catch-up with the slower moving components. Finally, at time $2\tau$, all the spins are along the $x$-axis. 

\subsection{CPMG DD sequence \label{cpmg}}
\begin{figure}[h]
\begin{center}
\includegraphics[trim = 30mm 120mm 20mm 32mm, clip, width=12cm]{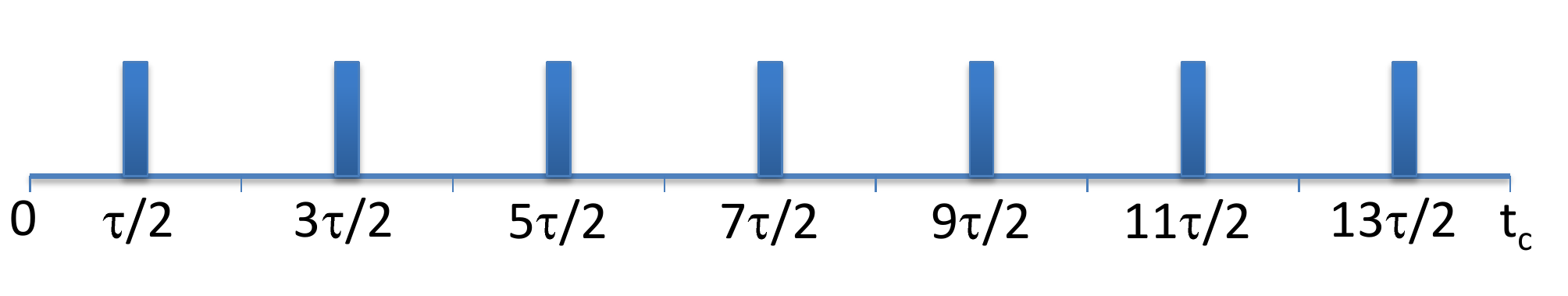}
\caption{CPMG DD pulse sequence for a cycle time of $t_c$ and for $N=7$ where $N$ is the number of $\pi$ pulses in one cycle. The solid bars indicate the $\pi$ pulses that are applied on the system qubit. $\tau$ indicates the duration between the consecutive  $\pi$ pulses.
}
\label{dec_dd_cpmg}
\end{center}
\end{figure}

The term CPMG refers to Car-Purcell-Meiboom-Gill, named after the people who came up with a decoherence suppression technique when the noise is time-dependent \cite{cpmg}. This method is similar to Hahn echo  except that 
CPMG DD consists of a train of equidistant
$\pi$ pulses that are applied on the system qubit as shown in Fig. \ref{dec_dd_cpmg}.   The $\pi$ pulses are applied at regular intervals $\tau$.

A CPMG sequence with a $\tau$ value  much shorter than the noise correlation time can suppress the corresponding noise. 
In general, the
smaller the value $\tau$, the larger the
bandwidth of noise that is suppressed, and thus increases
the efficiency of DD. 

It is important to note that the phases of $\pi$ pulses are chosen such
that the initial state is stationary under the pulses, so that
the DD sequence is robust against pulse errors. In other words, if the  magnetization just before the CPMG sequence is about $x$-axis then the $\pi$ pulses in Fig. \ref{dec_dd_cpmg} are applied about the $x$-axis and vice versa.

\subsection{Uhrig DD sequence}
\begin{figure}[h]
\begin{center}
\includegraphics[trim = 30mm 90mm 20mm 72mm, clip, width=12cm]{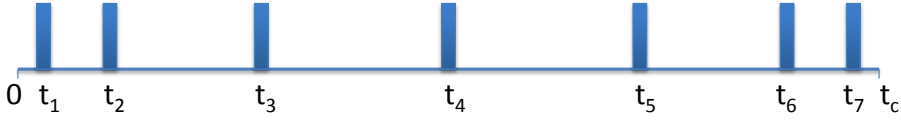}
\caption{UDD pulse sequence for a cycle time of $t_c$ and for $N=7$ where $N$ is the number of $\pi$ pulses in one cycle. The solid bars indicate the $\pi$ pulses that are applied on the system qubit.}
\label{dec_dd_udd}
\end{center}
\end{figure}

Uhrigh DD (UDD) is another technique  to suppress low-frequency noise \cite{uhrig}. Unlike CPMG DD, here the $\pi$ pulses are not equidistant but the $\pi$ pulse spacing is given by
\begin{equation}
t_j = t_c \sin^2 \left[\frac{\pi j}{2(N+1)}\right],
\label{t_udd}
\end{equation}
where $N$ is the total number of $\pi$ pulses, and $t_c$ is the cycle time. Fig. \ref{dec_dd_udd} shows the pulse sequence  for $N=7$. 

\section{Characterizing decoherence}
In this section, I will show how  decoherence can be characterized using two techniques, i.e. by NS and QPT. NS gives the  noise information in the qubit for different noise frequencies. Recently, NS has emerged to be of particular interest in quantum information processing due to its use in optimizing the DD sequences \cite{biercuk,biercuk2011dynamical,pan}. QPT gives the entire process. In our work, the  process is decoherence process at particular noise frequency. This technique also quantifies the type of induced noise, e.g. bit flip or phase flip. 

\subsection{Noise spectroscopy \label{dec_ns}}
NS gives the frequency distribution of the noise which essentially contains the information about qubit noise content. Yuge {\em et al.} \cite{yuge} and Alvarez {\em et al.} \cite{suter_ns} independently proposed the method to experimentally measure the noise spectrum. Noise spectrum is defined by the quantity $S(\omega)$ which is a function of noise frequency $\omega$. 

\begin{figure}[h]
\begin{center}
\includegraphics[trim = 50mm 80mm 00mm 00mm, clip, width=13cm]{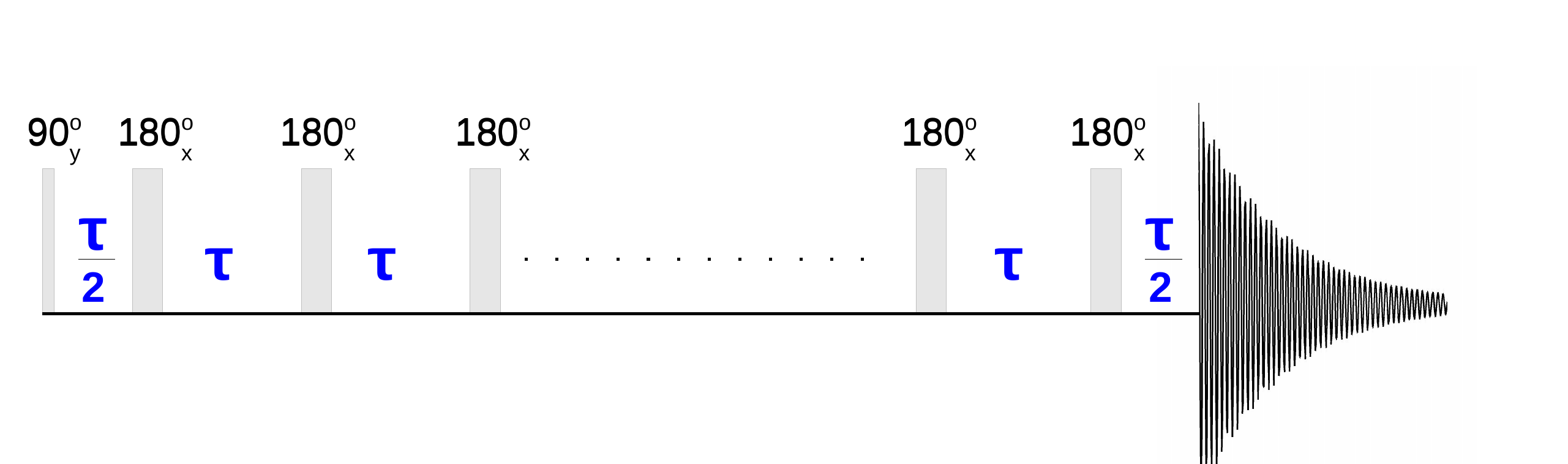}
\caption{Pulse sequence to measure $S(\omega)$. 
}
\label{sd_ns}
\end{center}
\end{figure}

We utilize the method given by Yuge {\em et al.}. Fig. \ref{sd_ns} shows the pulse sequence to measure $S(\omega)$. 
This sequence is basically CPMG sequence and is used to measure the decay constant $T_2$. 
By varying $\tau$, one can have a distribution of $T_2$ values. Further, it was shown  that \cite{yuge}
\begin{equation}
S(\omega) \approx \frac{\pi^2}{4T_2(\omega)},
\end{equation}  
where $\omega = \pi/\tau$.
Thus by scanning a range of $\omega$ and by measuring $T_2(\omega)$, one can obtain $S(\omega)$.

\subsection{Quantum process tomography \label{dec_qpt}}
QPT is a technique to reconstruct the entire quantum process \cite{chuang}. 

Consider a quantum operation $\mathcal{E}$ which transforms the initial state $\rho$ to a final state $\rho'$ as follows:
\begin{equation}
\rho' = \mathcal{E}(\rho).
\end{equation}
$\mathcal{E}$ can be any process which can either be unitary or non-unitary. 
The goal of quantum process tomography is to determine  $\mathcal{E}$ \cite{chuang,qpt1}.

Suppose, 
\begin{equation}
\mathcal{E}(\rho) = \sum_j E_j \rho E_j^\dagger; \qquad \mathrm{where} \quad E_j = \sum_m e_{jm} \tilde{E}_m.
\end{equation}
Here $\tilde{E}_m$ are the fixed set of operators and $e_{jm}$ are the complex numbers. Hence
\begin{equation}
\mathcal{E}(\rho) = \sum_{mn} \tilde{E}_m \rho \tilde{E}_n^\dagger \chi_{mn}; \qquad \chi_{mn} = \sum_j e_{jm} e^{\ast}_{jn}.
\end{equation}
Thus one can see that for a fixed set of operators $\tilde{E}_j$, one needs to determine the coefficients of $\chi$. This is known as $\chi$ matrix representation. 

After some algebra, one can deduce that
\begin{equation}
\sum_{mn} \beta^{mn}_{pq} \chi_{mn} = \lambda_{pq}
\end{equation} 
where,
\begin{itemize}
\item $\lambda_{pq} = \mathrm{Tr}[\rho^{'}_p \rho_q]$.
\item $\beta_{pq}^{mn} = \mathrm{Tr}[\tilde{E}_m\rho_p \tilde{E}_n \rho_q]$.
\end{itemize}

In order to calculate $\lambda_{pq}$, one needs to know the final state $\rho'$. This characterization of $\rho'$ is done by using quantum state tomography (QST) \cite{chuang,qst1,qst2}.
For the sake of simplicity, I will briefly explain this procedure for a single qubit case.

As already mentioned in equation \ref{density1}, a single qubit density operator has the form
\begin{equation}
\rho = \frac{1}{2}(\mathbbm{I}+\sum_i r_i \sigma_i); \qquad \mathrm{where} \quad
r_i = \langle \sigma_i\rangle = \mathrm{Tr}[\rho \sigma_i]. 
\nonumber
\end{equation}
Here, $\sigma_i\in\{X, Y, Z\}$. For the spin operators $\mathbbm{I}/2$, $X/2$, $Y/2$ and $Z/2$, it follows that
\begin{equation}
\rho = \frac{\mathrm{Tr}[\rho]\mathbbm{I}+\mathrm{Tr}[X\rho]X+\mathrm{Tr}[Y\rho]Y+\mathrm{Tr}[Z\rho]Z}{2}.
\end{equation}

Thus characterizing $\rho$ involves the measurements of the average value of the operator corresponding to  $X$, $Y$ and $Z$ which are given by  $\mathrm{Tr}[X \rho]$, $\mathrm{Tr}[Y \rho]$ and $\mathrm{Tr}[Z \rho]$ respectively. 


\section{Experiments and results}

\begin{figure}[h]
\begin{center}
\includegraphics[trim = 00mm 20mm 240mm 20mm, clip, width=4.5cm]{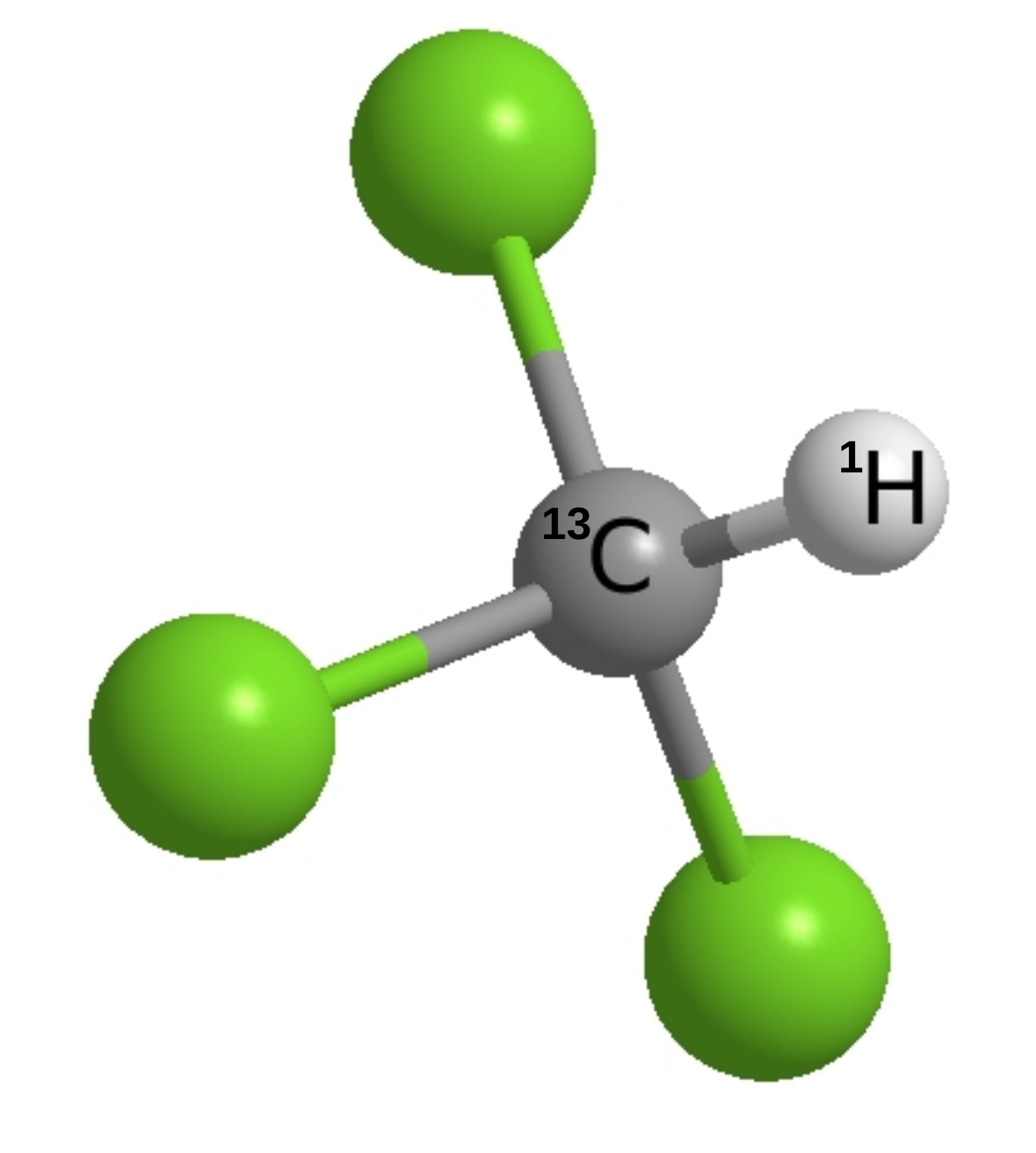}
\caption{$^{13}$C$^1$HCl$_3$ as NMR quantum simulator. The resonance offsets of $^1$H and $^{13}$C are $104.7$ Hz and $0$ Hz respectively. The $J$-coupling between the two is $209.4$ Hz. The $T_1$ for $^1$H and $^{13}$C is $4.1$s and $5.5$s respectively, and $T_2$ for the same is $2.9$s and $0.8$s respectively.
}
\label{dec_mol}
\end{center}
\end{figure}

The experiments were carried out on $^{13}$C$^1$HCl$_3$ molecule dissolved in CDCl$_3$ at an ambient temperature of $300$ K. The nuclei $^{13}$C and $^1$H form the two qubit system. The molecule and its properties are shown in Fig. \ref{dec_mol}. 

We chose $^{1}$H as our system qubit and $^{13}$C as our environment qubit.  
Here, I will explain the three parts of our experimental work, i.e., introducing artificial phase decoherence in $^1$H by randomly perturbing $^{13}$C, Suppressing the 
decoherence in $^1$H and finally characterizing the decoherence process that is induced in $^1$H.

The NMR Hamiltonian is similar to Eq. \ref{Hamn} and is given by
\begin{equation}
{\cal H}=\pi(\nu_HZ_{H}+\nu_CZ_{C}+\frac{J}{2}Z_{H}Z_{C}),
\label{Hamn}
\end{equation}
where $\nu_H$ and $\nu_C$ are the chemical shifts of the system ($^1$H) and the environment ($^{13}$C) qubits respectively, and $J$ is the scalar coupling.

We prepared  $^1$H qubit in the initial state
$\rho_H(0) = \mathbbm{I}_H/2+p_H X_{H}$ by applying a $\pi/2$ pulse about the $y$-axis on the thermal equilibrium state $\mathbbm{I}_H/2+p_H Z_{H}$. Here $p_H \sim 10^{-5}$ is the spin polarization. Also $^{13}C$ qubit was prepared in the  initial thermal equilibrium state $\rho_C(0) = \mathbbm{I}_C/2+p_C Z_{C}$ where $p_C \sim 10^{-5}$ . 

\subsection{Emulation of decoherence}
\begin{figure}[h]
\begin{center}
\includegraphics[trim = 10mm 25mm 10mm 20mm, clip, width=14cm]{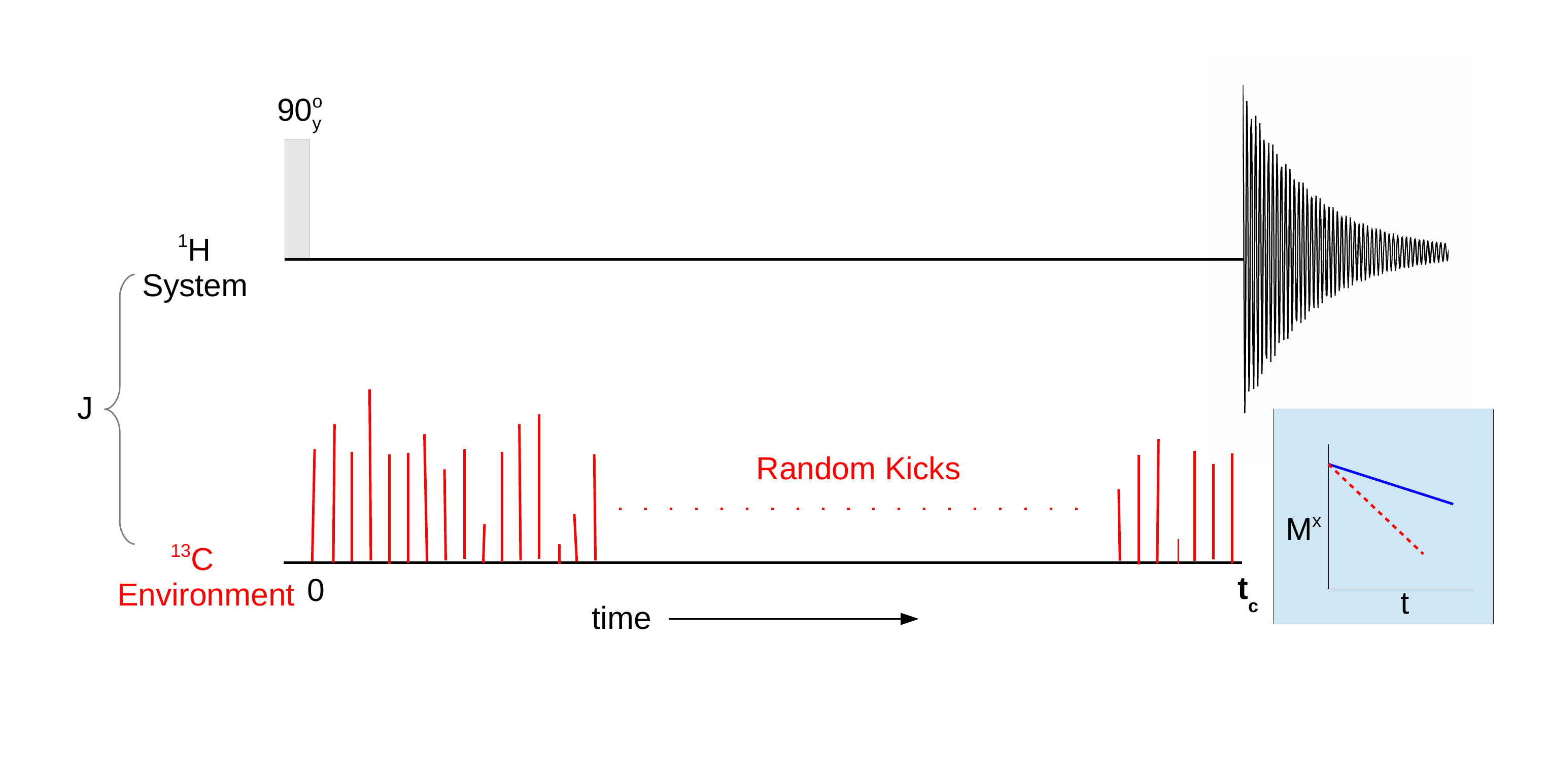}
\caption{Method to introduce artificial decoherence.  The filled bar on the system qubit corresponds to the RF pulse with rotation angle $90^\circ$ about $y$-axis. This pulse prepares the system qubit in the required initial state. The vertical lines on the environment qubit are the random kicks applied for time $t_c$. The inset in the lower right corner represents the expected magnetization decay wherein the solid line corresponds to inherent decay and the dotted line corresponds to the inherent decay as well as decay due to the artificial decoherence.
}
\label{dec_kicks1}
\end{center}
\end{figure}

Evolution under the action of kicks as explained in section \ref{artificial_dec} was realized by perturbing the $^{13}C$ qubit. 
These kicks were RF-pulses with random rotation angles and random phases. The experimental realization of artificial decoherence as explained in \cite{cory_dec} is shown in Fig. \ref{dec_kicks1}.

We performed different sets of experiments with kicks corresponding to $\epsilon_m \in  [0, 1^\circ]$, $[0, 2^\circ]$ and random phases between $0$ and $2\pi$, while allowing the $^1 H$ qubit to evolve freely. A large number of random pulses (with random angles and phases) were numerically generated and were then fed to the spectrometer to implement the consecutive pulses. These random pulse applications over many realizations of $\epsilon_m$ and phases were used to emulate decoherence.

\begin{figure}[h]
\begin{center}
\includegraphics[trim = 00mm 0mm 00mm 00mm, clip, width=10cm]{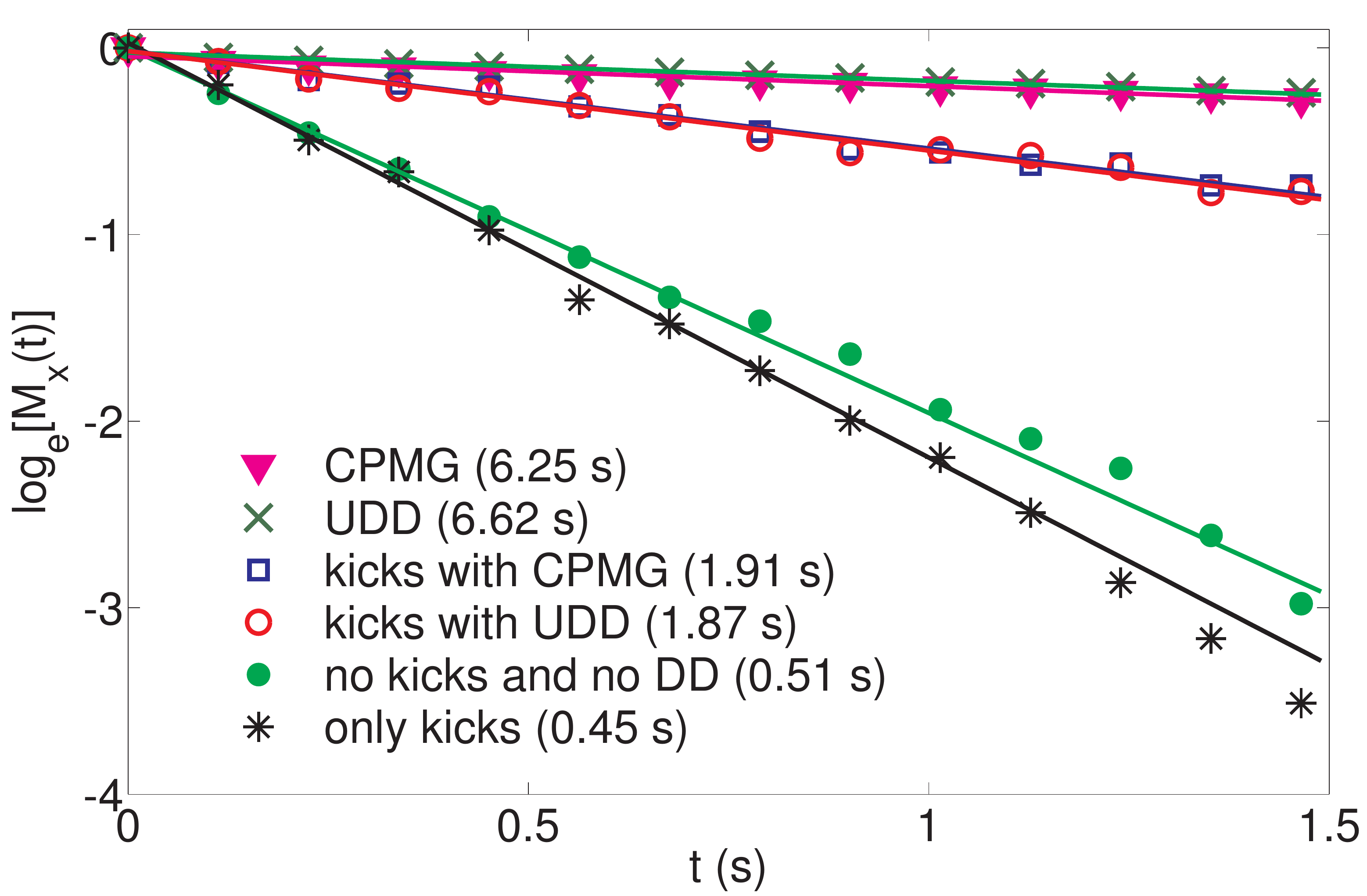}
\caption{ Decay of $M_x(t)$ under various cases. The numbers in the legend represent the $T_2$ values for the corresponding cases. The kick parameters are $\epsilon_m=[0^\circ,1^\circ]$, $\Gamma=25$ kicks/ms, and $t_c = 22.4$ ms and $\tau  =3.2$ ms. (Figure from [36]) 
}
\label{dec_mx}
\end{center}
\end{figure}

Decoherence is observed by measuring the transverse magnetization (averaged over many experiments with random kick angles and phases) 
$M_x(m t_c) = \mathrm{Tr}[\rho_H(m t_c) X_H]$ after $m$ cycles each of duration $t_c$ with $m=0,1,\cdots,N$ where $N$ is total number of cycles. 
Fig. \ref{dec_mx} shows the results of the experiment for $\epsilon_m \in [0^\circ,1^\circ]$ and $\Gamma = 25$ kicks/ms (indicated by stars).  
As can be seen from the figure, the decay of magnetization is higher than that without kicks (indicated by filled circles) indicating that we introduced additional decoherence apart from the natural relaxation processes. 

\subsection{Suppression of decoherence}
\begin{figure}[h]
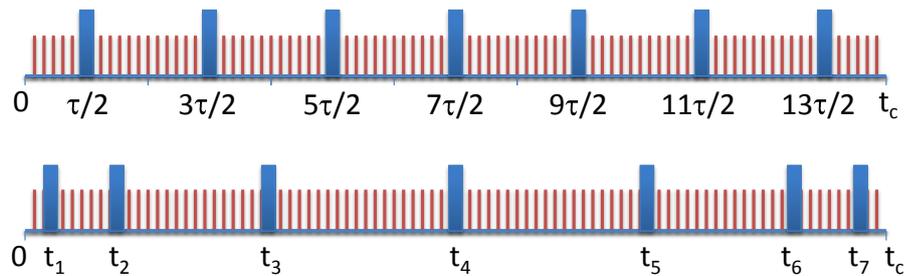

\begin{center}
\includegraphics[trim = 30mm 50mm 20mm 102mm, clip, width=12cm]{dec_dd_seq.pdf}
\includegraphics[trim = 30mm 20mm 20mm 140mm, clip, width=12cm]{dec_dd_seq.pdf}
\caption{The top and bottom figures correspond to the CPMG and Uhrig DD pulse sequence in the  presence of kicks  for a cycle time of $t_c$ and for $N=7$. The solid bars indicate the $\pi$ pulses that are applied on the system qubit and the vertical lines indicate the kicks on the environment qubit. 
}
\label{dec_dd_cpmgk}
\end{center}
\end{figure}

After emulating decoherence in $^1$H, we suppressed it by using CPMG DD and UDD sequences.  Fig. \ref{dec_dd_cpmgk} shows the pulses sequences for implementing CPMG and Uhrig DD in the presence of kicks. 

\begin{figure}[h!]
\begin{center}
\includegraphics[trim = 20mm 20mm 20mm 10mm, clip, width=14cm]{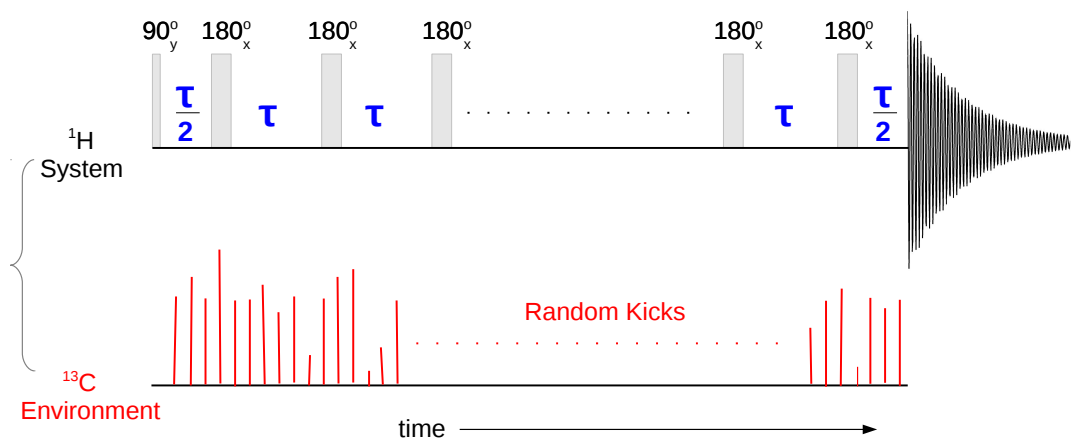}
\caption{Measuring NS in the presence of kicks. The pulses on $^1$H is basically CPMG sequence to measure $T_2$. 
}
\label{sd_ns1}
\end{center}
\end{figure}

While the $\pi$ pulses were applied on $^1$H, the kicks were applied on $^{13}$C simultaneously. The experimental results as shown in Fig. \ref{dec_mx} (indicated by open circles and boxes) show that the DD sequences were successful in suppressing decoherence even in the presence of kicks. 
It may be noted that detailed comparative studies of CPMG and UDD under natural relaxation processes have been studied elsewhere 
\cite{ashok,soumyaudd}

\subsection{Characterization of decoherence}
As the last step, we characterized decoherence using NS and QPT. The way to measure $S(\omega)$ as given in section \ref{dec_ns} but in the presence of kicks on $^{13}$C is shown in Fig. \ref{sd_ns1}. $T_2$ of $^1$H for each $\tau$ is obtained by fitting the experimental magnetization values to the decay model given by $M_x(t)  = M_x(0) e^{-t/T_2}$, where $M_x(0)$ is the initial transverse magnetization. By vayring $\tau$, we measured $T_2(\omega)$ where $\omega=\pi/\tau$ for various kick parameters.

\begin{figure}[t!]
\begin{center}
\includegraphics[trim = 00mm 0mm 00mm 00mm, clip, width=10cm]{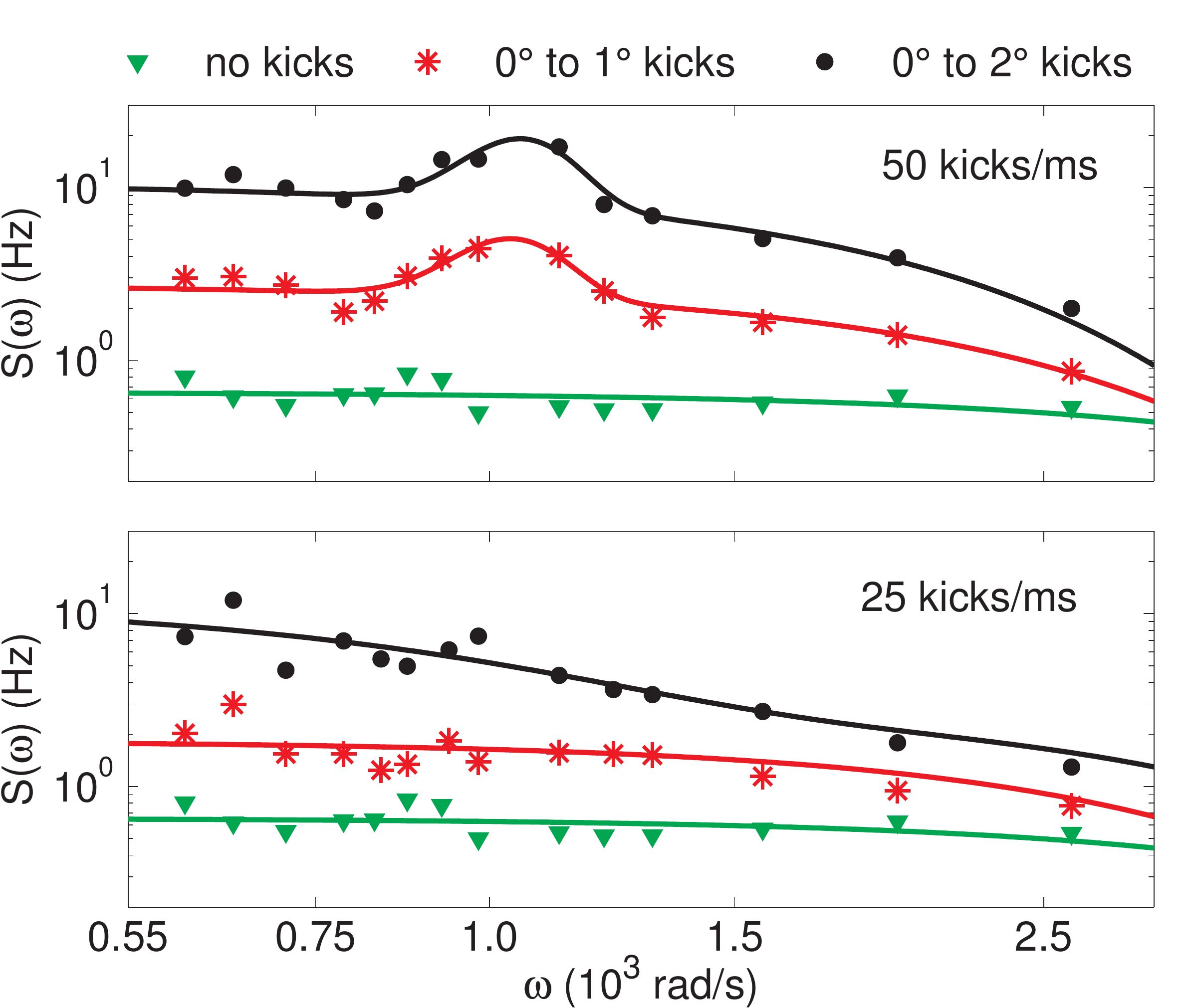}
\caption{Spectral density distribution for various kick parameters. 
}
\label{sd_new}
\end{center}
\end{figure}

Fig. \ref{sd_new} shows the noise spectral density distribution for various kick parameters. For comparison, we have also measured $S(\omega)$ in the absence of kicks (indicated by filled triangles). As expected, the $S(\omega)$ plot in the presence of kicks has higher values than that due to the inherent decay indicating  that the effect of kicks is to induce noise.  Generally, the noise spectra for the inherent noise has a Gaussian profile \cite{cowan} and the results agree in the case of $S(\omega)$ of inherent decay. However, an interesting characteristic
features in the noise spectral density at higher kick-rate (50 kicks/ms) was observed.  Similar
features were earlier observed by Suter and co-workers due to a decoupling
sequence being applied on environment spins \cite{suter_ns}.
\begin{figure}[b!]
\begin{center}
\includegraphics[trim = 10mm 0mm 0mm 00mm, clip, width=8cm]{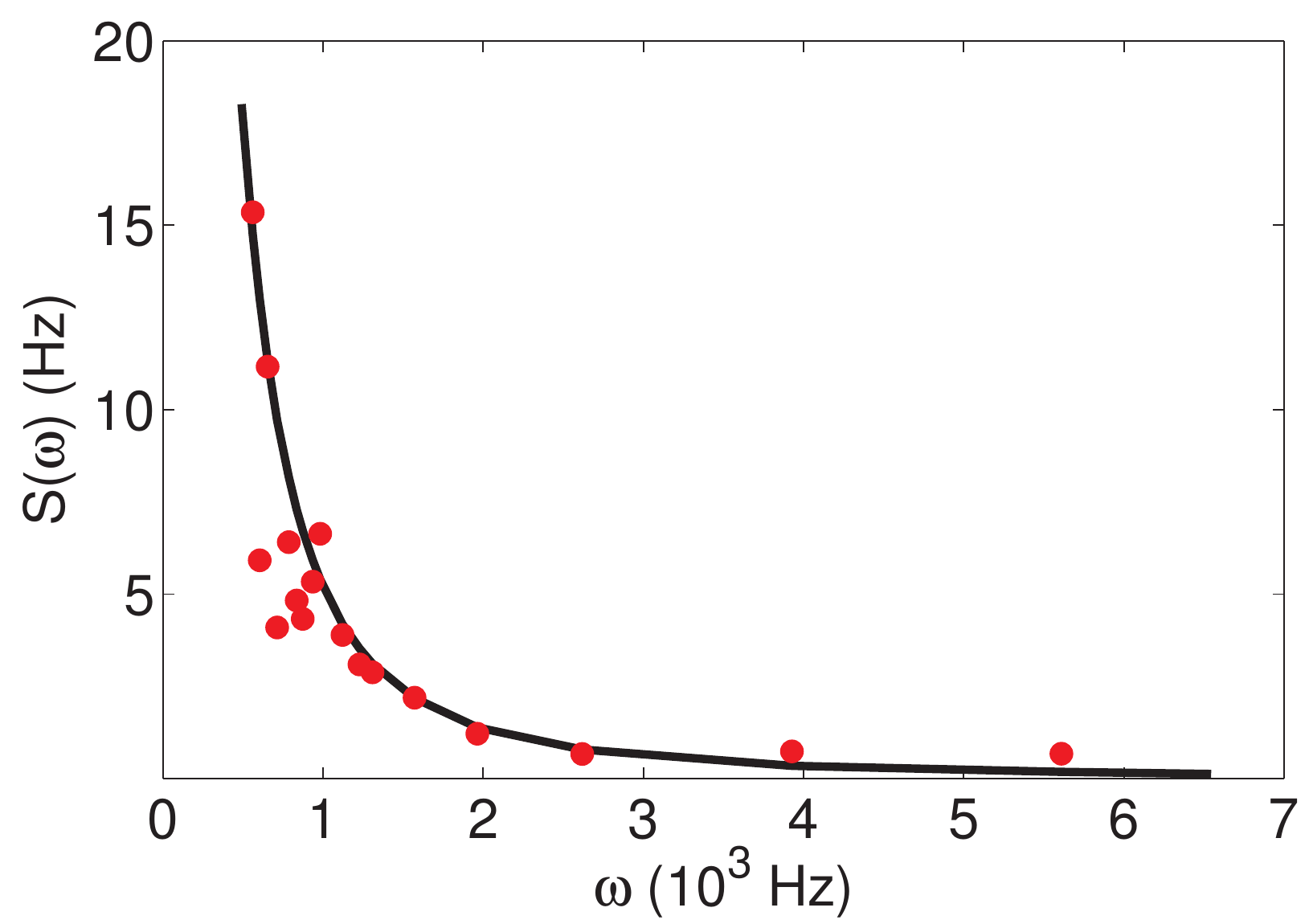}
\caption{ The experimental $S(\omega)$ (dots) vs the theoretical $S(\omega)$ (solid curve) corresponding to the kick parameters $\Gamma = 25$ kicks/ms and 
 $\epsilon_m \in [0^\circ,2^\circ]$. }
\label{sd_new_th}
\end{center}
\end{figure}

Fig. \ref{sd_new_th} shows the comparison  between the theoretical $S(\omega)$ using the methods given by Teklemariam {\em et al.} and Yuge {\em at el.} vs the experiemental $S(\omega)$ for  kick-rate of $25$ kicks/ms
and kick-angles in the range $0$ to $2$ degrees.  
To obtain the experimental $S(\omega)$ due to kicks alone, we
subtracted the intrinsic spectral density of the system qubit (with no kicks) 
from the total spectral density with kicks. 
A fair agreement between the numerically simulated curve and the experimental data
confirms the relevance of the model in low $\Gamma$ regime as given in \cite{cory_dec}.

We also characterized the induced phase decoherence by QPT and the general protocol was given in section \ref{dec_qpt}. The single qubit QPT protocol consisted of the initial preparation of the four states as follows: $\rho_j = |\psi_j\rangle \langle\psi_j|$, with $|\psi_j\rangle \in \{ |0\rangle, |1\rangle, (|0\rangle+|1\rangle)/\sqrt{2}, (|0\rangle-i|1\rangle)/\sqrt{2} \}$. The fixed set of operators   $\tilde{E}_p$ were chosen from the set $\{E, X, -iY, Z \}$,  where $E$ is the identity matrix and $X$, $Y$, $Z$ are the Pauli matrices. The goal is to obtain the $\chi$ matrix which corresponds to kick induced noise process.

\begin{figure}[b!]
\begin{center}
\includegraphics[trim = 10mm 0mm 10mm 20mm, clip, width=13cm]{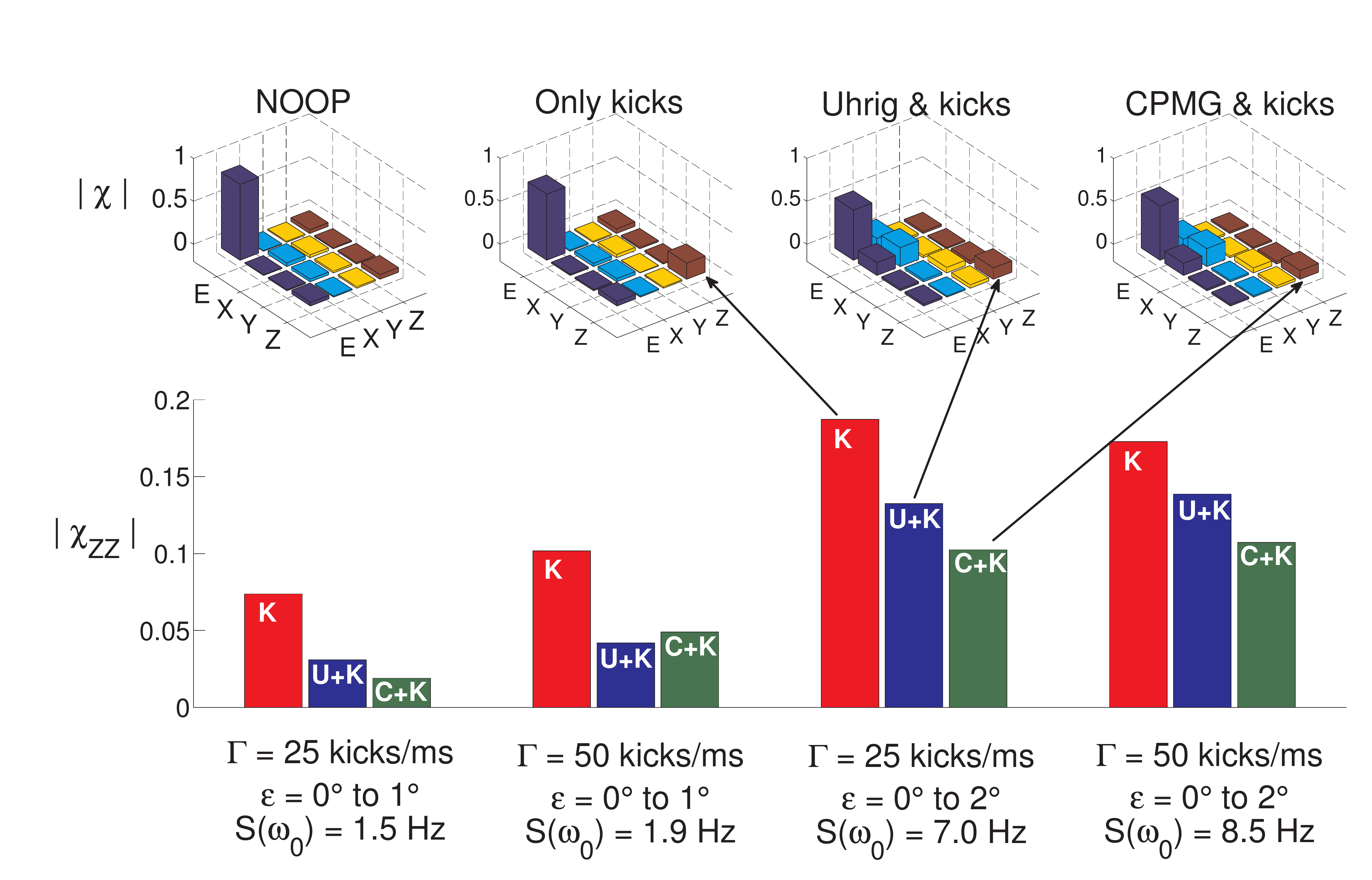}
\caption{ QPT for various kick parameters. The top row corresponds to the absolute value of $\chi$ matrix elements. The bottom corresponds to the enlarged version of $\chi_{zz}$ components. (Figure from [36])
}
\label{dec_tomo}
\end{center}
\end{figure}

Fig. \ref{dec_tomo} shows the experimental QPT of the phase decoherence process for various kick parameters.
 The top figures correspond to the entire $\chi$ matrix expressed in the basis of $\{E, X, -iY, Z \}$. Among these figures are QPT of NOOP which is an Identity operator, i.e., no kicks and no DD, only kicks, UDD with kicks and CPMG DD with kicks. As evident, the Identity process has only $EE$ component while the QPT in the presence of kicks has additional components in the $\chi$ matrix. As was already explained in section \ref{zurek_model}, the system-environment interaction of the type $ZZ$ gives rise to phase decoherence which corresponds to $ZZ$ components in the figure, thus indicating that the decoherence that we induced and studied was indeed phase decoherence. The extra components like $EX$, $XE$ and $XX$ arise due to the  nonidealities in the
$\pi$
pulses in the DD sequences that
introduce
NOT
operations.
All in all, we see that the DD sequences were successful in suppressing the decoherence with CPMG out-performing UDD due to the flat spectral density profile as shown in Fig. \ref{sd_new} \cite{ashok}.

\section{Conclusions and future outlook}
One of the main challenges in the physical realization of quantum computers is the phenomenon of decoherence. Our  work addressed this phenomenon and  is a step towards understanding phase decoherence. We simulated decoherence using the method given by Teklemariam {\em et al.} \cite{cory_dec} which involved the process of perturbation of the environment qubits by random classical fields. Using this approach we were able to vary the noise parameters in a controlled way. We showed that the amount of decoherence induced was related to the kick parameters - stronger kick angles realized over many such realizations with higher kick rates applied for  longer times proved to be effective to emulate decoherence which is also intuitive. We showed that we introduced additional decoherence apart from the inherent decoherence. After introducing decoherence, our main contribution was in suppressing and characterizing decoherence. Suppression of decoherence was achieved by standard DD sequences, i.e., CPMG and UDD.   Characterization of decoherence was done using NS and QPT. While the NS gave the  spectral density distribution, QPT revealed the entire phase decoherence acted on the system qubit. Besides, CPMG outperformed UDD as expected from the broad spectral density profiles revealed by NS. 

We believe that characterizing noise by the above methods not only provide an important platform for designing new robust optimized DD sequences, but also furnish insights into the origin of quantum noise.



               
\thispagestyle{empty}

\cleardoublepage
\bibliographystyle{IEEEtran}
\bibliography{reference} 
 
\end{document}